%% file: 2014arxiv_v14_2.tex
\documentclass[12pt,a4paper,twoside,openright]{report}%
\usepackage{amsmath}%
\usepackage{amsfonts}%
\usepackage{amssymb}%
\usepackage{graphicx}

\usepackage[T1]{fontenc}
\usepackage[utf8]{inputenc}
\usepackage{authblk}

\usepackage[top=1.25in, bottom=1.25in, left=1.25in, right=1.25in]{geometry}      
\usepackage{pdflscape}                    

\usepackage{chngcntr}

\usepackage{fancyhdr}
\usepackage{subcaption}

\usepackage[rightcaption]{sidecap}

\usepackage[nottoc,notlot,notlof]{tocbibind}

\usepackage{caption}


\newcounter{qcounter}

\DeclareGraphicsExtensions{.png,.pdf,.jpg}

\counterwithin*{equation}{chapter}
\numberwithin{equation}{chapter}

\newcommand{\oldref}[1]{}

\begin{document}

\include{2014arxiv_v14_3_title}        
\tableofcontents
\newpage

\include{2014arxiv_v14_3_0f}           
\include{2014arxiv_v14_2_0i}            
\include{2014arxiv_v14_2_1}

\include{2014arxiv_v14_2_2}

\include{2014arxiv_v14_2_3}

\include{2014arxiv_v14_2_4}

\include{2014arxiv_v14_2_5}             
\include{2014arxiv_v14_2_6}

\include{2014arxiv_v14_3_7}             
\include{2014arxiv_v14_3_8}             
\include{2014arxiv_v14_3_9}             
\include{2014arxiv_v14_2_9c}

\fancyhf{}
\fancyhead[LE,RO]{Bibliography}
\fancyhead[RE,LO]{L. Bilbao, L. Bernal, F. Minotti}
\fancyfoot[RE,LO]{Vibrating Rays Theory}
\fancyfoot[LE,RO]{\thepage}
 
\renewcommand{\headrulewidth}{1pt}
\renewcommand{\footrulewidth}{1pt}

\bibliography{2014arxiv_v14_2}{}          
\bibliographystyle{unsrt}

\end{document}

%% file: 2014arxiv_v14_3_title.tex

\title{Vibrating Rays Theory}

\author[1]{Luis Bilbao}
\author[2]{Luis Bernal}
\author[1]{Fernando Minotti}
\affil[1]{INFIP, UBA-CONICET, and Departamento de Fisica, Facultad de Ciencias Exactas y
Naturales, Universidad de Buenos Aires, Argentina}
\affil[2]{Departamento de Fisica, Facultad de Ciencias, Universidad de Mar del Plata, Argentina}

\renewcommand\Authands{ and }

\date{\today}
\maketitle

\begin{abstract}

The present work is aimed to explain why we started to consider Vibrating Rays
Theory (VRT) as a viable representation of nature, and to elaborate some of
its consequences. We first note that we have kept the probably unsuitable term
``vibrating rays'' as homage to its insightful introducer: Michael Faraday.
Certainly, the image of rays or ``protrusions'' emanating from an electric
charge is not a very palatable one for a contemporary physicist. The term is
used in this work only as a reference to a complex, and as yet not studied,
possible means of interaction among particles. In 1846 Faraday
\oldref{[R1]}\cite{faraday1846thoughts} introduced
the concept of vibrating rays, in which an atom is conceived as having rays
that extend to infinity and move with it. According to this point of view,
electromagnetic radiative phenomena correspond to vibration of these rays,
which propagate at speed $c$ relative to the rays (and the atom). Although a
discussion on this subject might seem to be out-of-date, there are many
reasons that justified this work. The first reason is based on the fact that
the constancy of the speed of light, irrespective of the source movement, has
not been demonstrated experimentally in a conclusive way. In fact, only
ballistic emission theories (see, for example, Ritz theory 
\oldref{[R2]}\cite{ritz1908recherches}) can be discarded by the experimental results 
(for example, Brecher \oldref{[R3]}\cite{brecher1977speed} or Alvager et al. \oldref{[R4]}\cite{alvager1964measuring} both compare their results against a ballistic theory). 
The second reason is based on the fact that study of radiometric data from spacecrafts
indicates the existence of different kinds of anomalous Doppler residuals. In
1998 Anderson et al. \oldref{[R5]}\cite{anderson1998indication} reported an anomalous 
acceleration towards the Sun
obtained from the analysis of the Doppler data of the Pioneer 10/11
spacecraft. The anomaly was inferred from a small, blueshift Doppler residual
obtained as the difference between measured and modeled values. Besides this
term (that may be a thermal effect) there is an annual term, and in a much
shorter time scale, there is a diurnal term. Also, for the Pioneer 11
there was a fast increase in the anomaly right at Saturn encounter.
More recently the Doppler frequency data of different spacecrafts before 
and after the closest approach to Earth have shown an unexpected frequency shift, 
which had been called flyby anomaly \oldref{[R6]}\cite{anderson2008anomalous}. 
Further, a range disagreement has been measured
between active and passive reflection \oldref{[R7]}\cite{antreasian1998aiaa}. 
We will show that the above mentioned anomalies exhibit a signature of VRT. 
The third reason is related to the time definition in a rotating frame. According 
to Special Relativity (SRT) there is not a unique way to assign a time, 
whereas under VRT no contradictions are present. And finally, that VRT is 
compatible with all known experiments on electromagnetism and light
propagation. In the present work we will (1) explain how VRT should be
interpreted, and why past experiments were misinterpreted (for example, by the
use of the motion of a mirror image as if it were a real source, when
according to SRT itself both produce different results), (2) show the
characteristics of VRT that are present in spacecraft anomalies, (3) give a
possible theoretical model (including the possible presence of longitudinal
waves), and (4) describe results of an experiment designed to
distinguish between VRT and SRT models.
\end{abstract}


%% file: 2014arxiv_v14_3_0f.tex


\chapter*{Preface}\label{ch:f}

\addcontentsline{toc}{chapter}{Preface}

\fancyhf{}
\fancyhead[LE,RO]{Preface}
\fancyhead[RE,LO]{L. Bilbao, L. Bernal, F. Minotti}
\fancyfoot[RE,LO]{Vibrating Rays Theory}
\fancyfoot[LE,RO]{\thepage}
 
\renewcommand{\headrulewidth}{1pt}
\renewcommand{\footrulewidth}{1pt}

Almost 170 years ago Michael Faraday gave an improvised talk at the Royal
Society presenting what he described as ``thoughts on ray vibrations''\oldref{[R1]}
\cite{faraday1846thoughts}. His ideas were intended to do away with the ether in favor
of lines connecting the particles; light being the vibrations of these lines
or rays. These thoughts were not developed further into a more precise
formulation, probably because of the following developments of electromagnetic
field theory by Maxwell, happening shortly afterward. That these ideas are
potentially fruitful can be seen, for instance, in \oldref{[R8]}\cite{mansfield2011faraday}, 
where Maxwell equations are derived as statistical averages over the states of those lines,
employing techniques of string theory.

The purpose of the present notes is to show that Faraday's thoughts,
appropriately interpreted, can be the base of an alternative approach to
electrodynamics. We consider that this approach has not been properly studied
so far, and cannot be dismissed with the actual experimental evidence,
especially considering its far reaching consequences. The first question that
comes to mind is: why bother to develop alternatives to a well established and
widely accepted theory as Maxwell's electrodynamics? A possible answer is
that, as we endeavor to show in the following pages, there are some more or
less subtle indications that Maxwell theory may not be absolutely right in all
its predictions and, consequently, some basic postulates of Special Relativity
Theory (SRT) may not be fully justified. In particular, the postulate of the
velocity of light being independent of the motion of its source, on the one
hand, appears not to be experimentally verified so far in a correct, univocal
way for macroscopic sources and, on the other hand, seems to be challenged by
particular aspects of various spacecraft anomalies. As we will detail, the
conception of the speed of light being dependent on the motion of its source
(even after being emitted) seems to be indicated by those aspects of the
anomalies, and by some recent experiments.

Moreover, important epistemological issues about the foundations of the 
electromagnetic theory have recently
been raised by Natiello and Solari \cite{natiello2019construction}, indicating 
that some of the implicit assumptions of the theory are contradictory. Also,
from the historical point of view, the authors point out the 
unjustified lack of consideration of alternative theories that were developed at 
the time of Maxwell's work, which are free of those contradicting assumptions.  

It is important to mention that the light speed dependence on the source
motion that we consider is based on Faraday's idea of vibrating rays
(Vibrating Rays Theory, or VRT, as we will denominate it), and corresponds to
the rays, or lines, being carried along with the source (in fact being an
intrinsic aspect of it), so that light moves with constant speed $c$ at all
instants relative to its source, irrespective of the motion of the latter.
This is in sharp contrast with ballistic type of theories like Ritz', in which
light is detached from the source after emission, being affected by the source
motion only at that moment. The idea is not new, as it was put forward before 
by others. In the 19th century it was most clearly and explicitly expounded by 
C. Neumann, for the case of delayed action-at-a-distance interactions, in his 
short answer \oldref{[Neumann1869]}\cite{neumann1869mathematische} 
to R. Clausius' criticism \oldref{[Clausius1869]}\cite{clausius1869philosophical} 
to one of his previous works \oldref{[Neumann1868]}\cite{neumann1868tubingen}, 
and during the 20th century by Dingle \oldref{[R9]}\oldref{[R10]}\cite{dingle1960doppler,dingle1960dopplerb}, 
and Moon and Spencer \oldref{[R11]}\oldref{[R12]}\cite{moon1956establishment,moon1989universal}.

However, we will develop it further to show a series of important points:
\begin{list}{(\roman{qcounter})}{\usecounter{qcounter}}
\item particular aspects of the Pioneer and flyby anomalies are reproduced in detail, 
\item contrary to SRT, light motion in rotating systems is shown to be free of
conceptual problems with the VRT approach, 
\item an experiment is proposed to measure the speed of light with sources 
in different states of motion, and its results are shown 
that seem to favor VRT over SRT, and 
\item a complete electrodynamic theory is derived from Faraday's ideas,
which includes Weber's theory \oldref{[R13]}\cite{weber1846electrodinamica} 
as its non-radiative part, and which incorporates radiation with very similar 
properties to those in Maxwell's theory.
\end{list}


%% file: 2014arxiv_v14_2_0i.tex


\chapter*{Introduction}\label{ch:i}

\addcontentsline{toc}{chapter}{Introduction}

\fancyhf{}
\fancyhead[LE,RO]{Introduction}
\fancyhead[RE,LO]{L. Bilbao, L. Bernal, F. Minotti}
\fancyfoot[RE,LO]{Vibrating Rays Theory}
\fancyfoot[LE,RO]{\thepage}
 
\renewcommand{\headrulewidth}{1pt}
\renewcommand{\footrulewidth}{1pt}

This work is a preliminary draft intended to summarize the basic properties of
VRT, and to discuss possible measurements and a theoretical description of
VRT. Although VRT might seem to be out-of-date, there are many reasons that
justify its study. The main one is that VRT was not properly considered, nor
properly tested.

The tracking of spacecraft shows many differences between the measured and the
modeled Doppler frequencies, although models are very complete. Both the
Pioneer anomaly \oldref{[R5]}\cite{anderson1998indication} and the flyby 
anomaly \oldref{[R6]}\cite{anderson2008anomalous} refer to small residuals of the
differences between the measured Doppler frequencies and the modeled ones.
Although they are very small the problem is that they exibit a non-random
pattern indicating failures on the model. For example, according to the
temporal variation of this residual, the Pioneer anomaly has a main term, an
annual term, a diurnal term and a term that appears during planetary encounters.

Radiometric data from the Pioneer 10 and 11 spacecraft indicate an apparent,
constant skewing between the predicted and observed Doppler shifts. This
offset has been attributed to a possible acceleration of of 
$8\times 10^{-10}$ m/s$^{2}$ directed toward the Sun. The anomaly was
inferred from a small, blueshift Doppler residual obtained as the difference
between measured and modeled values. A more detailed paper 
\oldref{[R14]}\cite{turyshev2005study} determined a
value of $(8.74\pm 1.33)\times 10^{-10}$ m/s$^{2}$ for the anomalous
acceleration. Besides this term, there is an annual term with corresponding
mean amplitude of 12 mHz, and in a much shorter time scale, a diurnal term 
is also reported.
Further, for the Pioneer 11 spacecraft right at Saturn encounter, when the
craft passed into an hyperbolic escape orbit, there was a fast increase in the
anomaly, which afterward settled down to its canonical value 
\oldref{[R15]}\cite{turyshev2005route}.

Studies of radio Doppler data of six spacecraft flybys (namely, Galileo, NEAR,
Cassini, Rosetta and Messenger) show an anomalous Doppler shift between pre-
and post-encounter data. The same inconsistency is observed when the ranging
data is differenced. Anderson et al. \oldref{[R6]}\cite{anderson2008anomalous} 
give an empirical prediction formula that fits six flybys. The anomaly is found to 
depend on the declination of the incoming and outgoing asymptotic velocity 
vectors and on the tangential speed of Earth.

It should be clarified that a few years ago an explation of the Pioneer
anomaly was published \oldref{[R16]}\cite{turyshev2012support}, 
that many people have taken as the end of the
dispute. However, it is only a very specific solution that only applies to the
main term of the Pioneer spacecraft, but left unresolved many other anomalies,
including those of the spacecrafts Cassini, Ulysses and Galileo; the annual
term; the diurnal term; the increases of the anomaly during planetary
encounters; the flyby anomaly; and the possible link between all these (is
hard to think that there are so many different causes as mentioned anomalies).
For all this, we believe that the issue can not be closed.

Since the effects cannot be explained by previously known physics or
spacecraft properties, a possibility exists that the used relativistic Doppler
formula could be wrong. Mbelek \oldref{[R17]}\cite{mbelek2008special} claims 
that unaccounted transverse Doppler
shift may explain the flyby anomaly. Mbelek wrongly thought that only first
order Doppler was used. However, in satellite Doppler tracking light
propagation is correct to order of $c^{-2}$ \oldref{[R18]}\cite{anderson2005study}.
As we will show below the problem is the opposite: that is, the presence of 
second order terms in the relativistic Doppler modeling, that seems to be 
absent in the radiometric data, are the cause of the anomalies.

The Pioneer 10/11 communication systems use S-band Doppler frequencies (2.113
GHz up and 2.295 GHz down). Phase coherency with the ground transmitters is
maintained by means of an S-band transponder with the 240/221 frequency
turnaround ratio. Other spacecrafts use X-band frequencies. The essential
point of the Doppler data is that the signal is actively reflected, i.e. that
the downlink signal is provided by an onboard spacecraft transmitter. While in
SRT there is no difference between passive and active reflection, VRT produces
different results. We will show that considering that difference it is
possible to understand both anomalies.

Further, during the flyby of the NEAR spacecraft in 1998, its trajectory was
measured near the point of closest approach with two radars (Millstone and
Altair) of the Space Surveillance Network (SSN), and compared to the
trajectory from the Deep Space Network (DSN) \oldref{[R7]}\cite{antreasian1998aiaa}.
As for the range, the two
measurements should match within a meter-level accuracy (the resolution is 5 m
for Millstone and 25 m for Altair), but measurements show a maximum difference
of about 1 km, ie more than 100 times larger than the accuracy of the
equipment used (see figure 10 of \oldref{[R7]}\cite{antreasian1998aiaa}). 
Moreover, this difference depends linearly with time but with different slopes 
for the two radars used. No explanation has been given for this discrepancy.

As a matter of fact, the range difference, $\delta R$, is well fitted by
\begin{equation}
\delta R=-\frac{\mathbf{R\cdot v}}{c}  \tag{I.1}\label{I.1}
\end{equation}
where $\mathbf{R}$ is a vector range pointing from the spacecraft to the
radar, $\mathbf{v}$ the spacecraft velocity relative to the radar, and $c$
the speed of light.  It reproduces the (almost) linear dependence with 
time during the measured interval, and the two different slopes for 
Millstone and Altair stations due to their different locations (on Section
\ref{sec:6.3} we will develop further this subject).

Actually, since the range is calculated from the time of flight of the
signals, the validity of the above fit means that the speed of electromagnetic
waves (microwave) of the DSN and SSN travel at different speeds. Specifically,
from  \oldref{(I.1)} (\ref{I.1}) follows that the waves of the SSN travel at $c$ 
(relative to the radar) while those of the DSN travel at $c$ plus the projection 
of the speed of the spacecraft into the direction of the beam (of course, within 
this first order fit, it would also be mathematically equivalent to assume that 
the speed of the DSN waves is $c$, while those of the SSN is $c$ minus the 
projection of the speed of the spacecraft into the direction of the beam). 
This is in sharp contrast with the Second Postulate of SRT.

In view of the above result one may ask: 
\begin{enumerate}
\item {If the velocity of electromagnetic waves depends on the speed 
of the source, why wasn't this observed in the past?;} 
\item {Are there simultaneous measurements of the speed of light from
different moving macroscopic sources (not moving images) with different
velocities?;}
\item {Since ballistic (emission) theories are ruled out (see, for
example, DeSitter\oldref{[R19]}\oldref{[R20]}
\cite{desitter1913ein,desitter1913uber},  Brecher\oldref{[R3]}
\cite{brecher1977speed} and Alvager et al.\oldref{[R4]}
\cite{alvager1964measuring}), how else could the speed of light 
depend on that of the source?;}
\item {How is it possible that
there is a first order difference (in $v/c$) in the speed of light compared to
the Second Postulate while at the same time, there are many experiments on
time dilatation that are consistent with SRT to second order in $v/c$ (see, for
example, \oldref{[R21]}\cite{botermann2014test})?;}
\item {Assuming that surveillance radars are well calibrated, the
measured range difference would be due to a defect in the DSN system, and
therefore should be present in all spacecraft tracking. Is it possible that
the Pioneer and flyby anomalies are the manifestation of this problem?.}
\end{enumerate}

To the best of our knowledge there is no known experimental work that
simultaneously measures the speed of light from two different sources (not
images), or that simultaneously measures the speed of light and that of the
source. For example, in the work by Alvager et al. 
\oldref{[R4]}\cite{alvager1964measuring} the speed of light is
measured at a later time (\~ 200 ns) after the radiation is produced, while the
speed of the source is estimated (it was not measured) at the time of emission
of the radiation. There is no measurement nor estimation of the speed of the
source at the time of the detection of the light. It is also important to note
that measurements involving moving images produce different results than those
produced by mobile sources. Therefore, to ensure the independence of the speed
of light with the source, it is mandatory to have two sources with different
movements. The experiences mentioned above\oldref{[R3]}\oldref{[R4]}\oldref{[R19]}\oldref{[R20]}
\cite{brecher1977speed,alvager1964measuring,desitter1913ein,desitter1913uber} 
only rule out ballistic theories in which radiation maintains the speed of the source
adquired at the time of emission, but do not rule out the present theory (VRT)
based on Faraday's ideas \oldref{[R1]}\cite{faraday1846thoughts}.

We will further show that many optical effects and experiments, like Doppler
shifts, Fresnel drag and light aberration, are fully compatible with VRT. In
particular, we show that VRT allows consistently interpreting the delays
observed in the Sagnac experiment, while at the same time being compatible
with the experiments that indicate the isotropy of the velocity of light in
rotating frames.

Also, we will show that VRT can be developed into a complete electrodynamic
theory, which presents both, instantaneous and delayed action at a distance.
It incorporates Weber's electrodynamics \oldref{[R13]}\cite{weber1846electrodinamica}, 
together with electromagnetic
propagation and even radiation damping, and is derived from a fully Galilean
covariant action. It is interesting that, as was shown by others \oldref{[R22]}\cite{schrodinger1925die}, 
if a similar instantaneous action at a distance is assumed for the gravitational
interaction, a fully Machian theory emerges with the action of the distant
universe giving rise to inertial effects that could even account for mass
increments with velocity similar to those described by SRT, and in which time
dilation can be interpreted as a retardation effect linked to mass changes of
the system constituents. In this respect it can be understood why, while
having differences relative to SRT at first order in $v/c$, VRT formalism can
accomodate second order effects predicted by SRT.

Finally, an actual experiment is presented to measure the one-way difference
of the speed of light between two sources with different velocities. Results
of the experiment are shown, and suggestions for improving the
experiment are discussed.

The present work is divided into nine chapters. All chapters are somewhat
independent and can be read separately. Chapters 1 to 5 are of an introductory
level included mainly to review some historical key experiments from VRT
perspective. In Chapter 1, we present the basic considerations on Vibrating Rays
Theory (VRT) based on the ideas of Faraday and Dingle. In Chapter 2, 3 and 4 the
basic concepts are applied to Aberration, Doppler and Fresnel drag phenomena.
In Chapter 5, we will show the Sagnac experiment and the definition of time in
rotating frames from both SRT and VRT point of view. In Chapter 6 we will show
that the difference between VRT and SRT Doppler formulas allows interpreting
well the spacecraft anomalies. In Chapter 7 the compatibility between GPS
measurements and VRT is described. In Chapter 8, a complete electrodynamic theory
based on VRT is presented. Finally, in Chapter 9 results from an experiment aimed
to distinguish between SRT and VRT are presented.


%% file: 2014arxiv_v14_2_1.tex


\chapter{Basic considerations on the Vibrating Rays Theory of
electromagnetism (VRT)}\label{ch:1}

\fancyhf{}
\fancyhead[LE,RO]{1. Basic considerations}
\fancyhead[RE,LO]{L. Bilbao}
\fancyfoot[RE,LO]{Vibrating Rays Theory}
\fancyfoot[LE,RO]{\thepage}
 
\renewcommand{\headrulewidth}{1pt}
\renewcommand{\footrulewidth}{1pt}

\section{Introduction}\label{sec:1.1}

The aim of this study is to give a basic description of VRT and to show that
in the past there was no experiment designed to test VRT. Further, the
experimental demonstration of the Second Principle of Special Relativity
Theory (SRT) does not contradict VRT. On the internet one can find a very
complete description of the experimental basis of SRT \oldref{[R23]}
\cite{roberts2014relativity}. Despite the many
indirect experiments that are compatible with this principle, after more
than a century of relativity there is no direct experiment that compares the
speed of propagation of light produced by sources moving with different
speeds. By source we refer to macroscopic sources within the meaning of the
proposal of Dingle \oldref{[R9]}\cite{dingle1960doppler}.

Terms like ballistic theory, or corpuscular theory refer to the model in
which light from a moving source has a velocity equal to the sum of the
velocity of light from a stationary source and the velocity of the source
itself. After Ritz \oldref{[R2]}\cite{ritz1908recherches} the previous sentence is 
interpreted with the
velocity of the source at the instant of \textit{emission}. Light is then
emitted like a bullet. It is thus usually granted that light retains the
source speed at time of \textit{emission} (in a similar way a bullet retains
the gun speed). As Moon et al. quoted \oldref{[R11]}\cite{moon1956establishment} 
``\textit{this concept is based on
the usual idea that once a pulse of radiation has left the source, this
pulse has independent existence and is unaffected by subsequent motion of
the source. On the contrary, the theory to be described here allows the
radiation to remain in some way coupled to its source even after it has been
emitted}.'' As it will be shown below VRT cannot be considered a ``ballistic
theory'' since radiation remains coupled to the source. Therefore
experiments that contradict ballistic theories are not necessarily a
demonstration against VRT.

We will quote some key experiments in this direction. Usually it is granted
that the most definite evidence against emission theories are probably the
works of Brecher \oldref{[R3]}\cite{brecher1977speed} and Alvager et al.
\oldref{[R4]}\cite{alvager1964measuring}. Brecher's work is based on
DeSitter's binary star observation \oldref{[R19]}\oldref{[R20]}
\cite{desitter1913ein,desitter1913uber}. Assuming a ballistic emission 
theory, the idea (first advanced by Comstock in 1910\oldref{[R24]}
\cite{comstock1910meeting}) is that the light emitted when a star 
is receding from Earth can be overtaken by the light emitted by the 
same star half period later when the star is approaching Earth. 
DeSitter was the first to make a quantitative evaluation of this 
phenomenon. DeSitter pointed out that distant binary stars should 
exhibit very strange behavior that had not been observed: infinite 
Doppler shifts, multiple images, and apparent variation in magnitude
\oldref{[R19]}\oldref{[R20]}\cite{desitter1913ein,desitter1913uber}. 
He argued that none of the star systems he had studied showed such 
extreme optical effect.

Fox \oldref{[R25]}\cite{fox1962experimental} observed that the passage of 
the light through some material medium
should be taken into account, and therefore, the conclusions of the
Ewald-Oseen extinction theorem\oldref{[R26]}\oldref{[R27]}
\cite{oseen1915uber,ewald1916theorie} can invalidate DeSitter's
conclusions, for the extinction length is evaluated as a few light-year.
Nevertheless, more recent observations in X-ray\oldref{[R3]}
\cite{brecher1977speed} that are not affected by
the extinction lead to the same conclusions as De Sitter (X-ray extinction
length is many orders of magnitude larger than visible light extinction
length). Note that, as it will be shown in Chapter \ref{ch:3}\oldref{3}, 
according to VRT the speed of light in a transparent medium depends 
\textit{on both} the refraction index \textit{and} the speed of the source, 
thus the concept of extinction length must be reformulated under VRT.

In most papers about the binary star problem%
\oldref{[R3]}\oldref{[R19]}\oldref{[R20]}\oldref{[R24]}\oldref{[R25]}\oldref{[R28]}
\cite{brecher1977speed,desitter1913ein,desitter1913uber,comstock1910meeting,fox1962experimental,fox1965evidence}
a ballistic theory of light is implicit, i.e., the assumption that the
velocity of light depends on the source-velocity, in the same way that the
velocity of a bullet depends on that of the gun. In other words, that the
light retains the velocity of the source at the \textit{time of emission}.
Probably the idea is based on Ritz \oldref{[R2]}\cite{ritz1908recherches} ``ballistic'' 
theory where electric charges constantly emit infinitely small, fictitious particles 
in all directions with a radial velocity $c$ with respect to the source at the
\textit{time of emission}. On the contrary, VRT allows the radiation to
remain coupled to its source even after it has been emitted.

From VRT point of view, DeSitter's problem is associated with the
mean velocity of light between emission and reception rather than
instantaneous velocity. Ritz theory is related to uniformly moving sources
where the mean and the instantaneous speed coincide. As it will be shown
below the \textit{mean speed of light} is related to the \textit{mean speed
of the source}, therefore Ritz theory is not applicable to this problem.
According to Dingle \oldref{[R9]}\cite{dingle1960doppler} ``\textit{the velocity of light 
and of the source must, on the relativity principle, be related to some standard 
of rest, and deSitter tacitly chose the Earth as such standard. Strictly speaking, 
this phenomenon lies outside Ritz's considerations, for the relative velocity of
the Earth and star is not constant, and its variation is an essential part
of the test. But if we generalise the postulate of relativity, which Ritz
accepted for uniform motions, to motions of all kinds, then we would expect
the velocity of light with respect to its source to remain constant, and not
its velocity with respect to an arbitrarily chosen body such as the Earth.
In that case the phenomena cited by deSitter would be quite consistent with
the ballistic theory}.'' Therefore in order to test VRT it is necessary to make 
a simultaneous laboratory comparison of the velocities of light from stationary 
and moving sources.

\section{About the tests of the independence of the speed of light with the 
source movement}\label{sec:1.i}

One century ago Comstock wrote \oldref{[R24]}\cite{comstock1910meeting}: 
``\textit{The assumption that the velocity of light depends on that of the 
source, so far as the author is aware, has never been properly examined. 
This is strange, but is explainable as a natural result of the complete trust 
which has been put for years in the concept of the ether}.''
In a similar way, Stewart \oldref{[R29]}\cite{stewart1911second} said: 
``\textit{We may then say that the
results of the relativity principle are due to a generalization of a law of
mechanics (the first postulate) and to the assumption that the velocity of
light is independent of the velocity of the source. This assumption has been
generally accepted on account of our concept of the ether as a fixed medium
filling all space}'' .

Surprisingly, after one century of SRT we found that the constancy of the
speed of light, irrespective of the source movement, has not been
demonstrated experimentally. Different authors claimed for such kind of
experiments as can be seen from the following few cites along the 20th
century.

In 1912: ``\textit{A definite experimental decision between the relativity
theories of Ritz and Einstein is a matter of the highest importance}'' \oldref{[R30]}
\cite{tolman1912some}.

In 1960: ``\textit{The most urgent need is, of course, an experiment to
determine whether, in fact, light from relatively moving bodies travels at a
single velocity through space}.'' And further, ``\textit{that would at once
settle the emission between the two forms of the velocity of light postulate
which distinguish the theories, in a manner that would carry conviction to
physicist. At the same time, more general considerations belonging to the
philosophy of science are relevant in order that the experimental results
shall be not only accepted but understood}'' \oldref{[R10]}\cite{dingle1960dopplerb}.

In 1962: ``\textit{Nevertheless if one balances the
overwhelming odds against such an experiment yielding anything new against
the overwhelming importance of the point to be tested, he may conclude that
the experiment should be performed}'' \oldref{[R25]}\cite{fox1962experimental}.

In 1993: ``\textit{There has not, in the past, been a case
of applying the theory of Relativity to well defined macroscopic bodies with
well defined velocities}'' \oldref{[R31]}\cite{beckmann1993philosophy}.

According to Dingle \oldref{[R9]}\cite{dingle1960doppler} 
``\textit{an experiment is perfectly conceivable in
which two bodies, A and B, relatively at rest at a distance X apart, carry
clocks which are synchronized by light-signals. A third body, C, moving at
high velocities with respect to A and B, travels from the former to the
latter, and at the instant of its coincidence with A, both A and C send out
light pulses towards B. Einstein's theory is built on the assumption that
these signals would reach B at the same instant; on the ballistic theory the
signal from C would arrive first. Even without an actual measurement of the
velocities, the correctness of Einstein's postulate could be tested by the
fact of their simultaneous or successive arrival. It is very greatly to be
hoped that such an experiment as thus will be attempted if possible.}'' Note
that it would not be necessary to make a precise measurement of the time of
travel. As we will show below there was no experiment properly designed to
compare the speed of light of different sources at different speeds. An
experiment in this direction will be shown in Chapter \oldref{9}\ref{ch:9}.

Although VRT is radically different from SRT, it is not easy to design an
experiment that brings out the differences between them. This can be viewed
using the experiments on ether detection performed in the past. From a
mathematical point of view, VRT, which states that the speed of light is
constant relative to the source, can be compared to an ether fixed to the
source. Thus, experiments to detect the motion of the ether should also be
used for detecting the motion of the source. In the nineteenth century, in a
series of papers, Veltmann and Potier (see the work of revision of Newburgh
\oldref{[R32]}\cite{newburgh1974fresnel}) came to the conclusion that 
absolute motion with respect to the ether
is undetectable using first order methods, that is, to measure some changes
in the deviation of prisms or the time taken for light to pass through a
certain thickness of dense transparent medium with the orientation of the
apparatus. By combining Fresnel's theory with Fermat's principle of least
time, Potier showed that to the first order in $v/c$ absolute motion with
respect to the ether is undetectable by optical means. Moreover it means
that interference phenomena are unaffected by motion. For example, let $ABC$
and $AB^{\prime }C$ be the two paths determined by an interferometer in
going from $A$ to $C$. For simplicity consider the entire apparatus encased
in the medium $M$. If the medium were at rest, a fringe pattern would
result. Since motion of the medium increase the time travel for each path by
the same amount, the phase difference between the two rays arriving at $C$
would be unchanged by the motion. Therefore no fringe shift would occur. The
same is valid for VRT since speed of light is constant relative to the
preferred frame of the source. Note that only a difference will appear with
more than one source moving at different speeds. Unfortunately, no such
experiment has been performed yet.

Therefore VRT, as SRT does, will give null results in the case of Arago's
deflection of light star through a prism, Airy's experiment to measure
stellar aberration with a telescope filled with water, Hoek interferometer
\oldref{[R33]}\cite{hoek1868determination}, Mascart interferometer 
\oldref{[R34]}\oldref{[R35]}\cite{mascart1872lumiere,mascart1874lumiere2}, 
and Michelson interferometer with unequal arms. None of them can detect 
the speed of the ether, thus cannot detect the speed of a source.

As an example, consider the attempt in this direction performed by Majorana
in 1919 \oldref{[R36]}\cite{majorana1919experimental}. The experiment 
was based upon a Michelson interferometer (with arms of different length) 
using electric arcs located on the edge of a turntable as a source, having a 
tangential speed of about 80 m/s. Majorana measured fringe shift when the 
source went from rest to speed $v$. Results matched those predicted by SRT. 
However, they were consistent with VRT as well.

\section{On the movement of sources and images}\label{sec:1.ii}

Notice that the movement of a source or the movement of an image produces
different results. As an example, consider the twin paradox. According to
SRT the traveling twin will return younger than the twin that remained on
Earth, therefore asymmetric behavior will result for a moving source of
light relative to a stationary one. Now, consider a moving mirror experiment
where the traveling ``twin'' is the image produced by the mirror. Upon
return of the mirror, the traveling ``twin'' (i.e. the image) will exhibit
the same age as that of the person in front of the mirror. In other words,
the image does not exhibit an asymmetric aging as an actual twin does.
Associating age with the number of beats of a source it follows that in the
interference between two coherent sources, moving one of them back and forth
will give a different interference pattern than the original one, while in
an interferometer, moving a mirror back and forth, the same interferometric
pattern is recovered. This can also be seen by integrating the relativistic
Doppler shift for a moving source and for a moving image (in Chapter \oldref{4}\ref{ch:4}  
we will describe further the Doppler effect). 

Consider, first, that at time there
are two coherent sources ($A$ and $B$), with proper frequency $f_{0}$, at
rest at position $x_{1}$. A detector, located at position $x^{\prime }$,
collects light from both sources. While source $A$ remains at rest, source 
$B$ starts to move relative to $A$ and the detector. The intensity at the
detector will start to vary later at time $t_{1}^{\prime }$,
\begin{equation}
t_{1}^{\prime }=t_{1}+\frac{x^{\prime }-x_{1}}{c}  \label{1.1}
\end{equation}

After arriving at position $x_{2}$ at time $t_{2}$ ($t_{2}^{\prime }$ as
seen by the detector) the source $B$ is stopped and remains at rest. The
number of fringes, $\Delta N_{S}$, counted by the detector between $
t_{1}^{\prime }$ and $t_{2}^{\prime }$ is
\begin{equation}
\Delta N_{S}=-\frac{x_{2}-x_{1}}{\lambda _{0}}+f_{0}\int_{t_{1}}^{t_{2}}
\left( 1-\frac{1}{\gamma \left( t\right) }\right) dt  \label{1.2}
\end{equation}
where
\begin{equation}
\gamma \left( t\right) =\frac{1}{\sqrt{1-v^{2}\left( t\right) /c^{2}}}
\label{1.3}
\end{equation}
$\lambda _{0}$ is the proper wavelength of the source, and $v\left( t\right)
$ is the speed of the source.

Now consider an interferometer with a stationary source and a moving plane
mirror in one arm. After moving the mirror from $x_{1}$ to $x_{2}$, the
number of fringes, $\Delta N_{M}$, counted by the detector is
\begin{equation}
\Delta N_{M}=-\frac{2\left( x_{2}-x_{1}\right) }{\lambda _{0}}  \label{1.4}
\end{equation}
(the factor 2 comes from the fact that the displacement of the image of a
plane mirror is twice the displacement of the mirror itself).

Clearly, when $x_{1}=x_{2}$ (that is, the source and the mirror return to
the original position) it holds that $\Delta N_{S}\neq 0$ and $\Delta
N_{M}=0$. With the moving mirror the original pattern is recovered while
with the moving source it is not. Note that the difference is difficult to
measure, for it is second order in the velocity. Therefore, a moving image
experiment cannot be considered as a moving source experiment.

It should be noted that many of the experiments, considered as moving
sources experiments were in fact experiments on moving images of stationary
sources, produced by transparent media or movable mirrors, see, for example,
\oldref{[R37]}\oldref{[R38]}\cite{babcock1964determination,beckmann1965test}. 
Even the famous experiment of Ives and Stilwell \oldref{[R39]}\cite{ives1938experimental} 
suffers from this defective interpretation, since it uses a moving source and its image,
rather than two different sources. Therefore, to test the possible validity
of VRT, it is mandatory to use differently moving (preferable macroscopic)
sources, or, alternatively, to measure the speed of light and the speed of
the source at the \textit{same instant}. To the best of our knowledge no
experiment has been performed in the past, either using two different moving
sources or measuring the speed of the light and of the source at the same
time.

Another surprising result is that VRT can also explain all known
experimental result that led to SRT. The advantage of VRT over SRT, is that
VRT may explain the spacecraft anomalies, and gives no contradiction in
rotating frames.

\section{VRT: Basic hypothesis}\label{sec:1.2}

Vibrating Rays Theory (VRT) is based on the works by Faraday \oldref{[R1]} 
\cite{faraday1846thoughts}, and, more
recently, Dingle \oldref{[R9]}\oldref{[R10]}\cite{dingle1960doppler,dingle1960dopplerb} 
and Moon et al. \oldref{[R11]}\oldref{[R12]}\cite{moon1956establishment,moon1989universal}. 
The basic idea is that the speed of light, $c$, \textit{at any instant} must be 
constant with respect to the source \textit{at the same instant}. The source 
sends out a spherical disturbance which expands at velocity $c$, and the center 
of this sphere is always at the source, no matter how the source moves. The
previously mentioned ``ballistic'' theory, on the contrary, assumed that the
center of the sphere is where the source would be if the latter continued to
move uniformly at the velocity it had at the instant of emission.

Faraday introduced the concept of vibrating rays, in which a charge has rays
that extend to infinity and move with it. Electromagnetic phenomena are the
vibration of these rays and propagate with speed $c$ relative to the rays,
which in turn can move with respect to the observer.

During a Friday evening discourse at the Royal Institution in 1846, Faraday
presented ideas which he called ``thoughts on ray vibrations'' \oldref{[R1]} 
\cite{faraday1846thoughts}. In
Faraday's view, the ``ultimate atoms'' of matter are centers of force only,
and do not have ``\textit{a definite form and a certain limited size... That
which represents size may be considered as extending to any distance to
which the lines of force of the particle extend: the particle indeed is
supposed to exist only by these forces, and where they are, it is.}'' Light
and such vibrations ``\textit{occur in the lines of force which connect
particles, and consequently masses of matter, together... I do not perceive
in any part of space, whether (to use the common phrase) vacant or filled
with matter, anything but forces and the lines in which they are exerted...
The view which I am so bold as to put forth considers, therefore, radiation
as a high species of vibrations in the lines of force which are known to
connect particles and also masses together. It endeavours to dismiss the
aether, but not the vibrations... The aether is assuming pervading all
bodies as well as space: in the view now set forth, it is the forces of the
atomic centers which pervade (and makes) all bodies, and also penetrate the
space.}'' Of course, today the word ``atom'' may be replaced by ``charge'',
i.e. the source of radiation.

Suppose that each source is the origin of ``lines of force'' proceeding
outwards in all directions. No relative motion is possible between a source
and its rays: if we regard the source as moving, its rays move
instantaneously with it. A light pulse from the source consists of a wave
traveling along a ray with an invariable velocity $c$ (invariable with
respect to the source and to the ray). Therefore its velocity with respect
to a detector would be the resultant of $c$ and the relative velocity of
receiver and source \oldref{[R10]}\oldref{[R11]}\oldref{[R12]}
\cite{dingle1960dopplerb,moon1956establishment,moon1989universal}.

The main idea is that the source itself is a privileged system where light
propagates in vacuum at a constant speed $c$ (to avoid confusion it is
designated by $c$ the speed of light in vacuum in the source own system).
The difference between these ideas and the theory of Ritz is evident in the
case where the source does not move with constant velocity. According to
Ritz the propagation speed of light is $c$ with respect to the source in the
time of \textit{emission}. This speed does not change even if the source
changes its state of motion. Instead, according to VRT the speed of light
will be $c$ relative to the source at all times, regardless of its state of
motion.

In other words, the \textit{instantaneous} velocity measured by a detector
at rest in an inertial system will be $c$ according to SRT; and, $c+v\left(
t_{r}\right) $ where $t_{r}$ is the velocity of the source at the epoch of
\textit{reception} (not emission) according to VRT.

The latter assumption can be extended to a \textit{non-rotating} reference
frame where the source is at rest, independently of its motion (i.e.,
whether or not it is an inertial system). Of course, the validity of the
previous sentence is limited to phenomena described in the present work. It
does not intend to be valid over all scales present in the universe.

In order to solve a problem for light propagation, a non-rotating reference
system fixed to the source should be used (preferred system). In this system
light propagates at $c$. Note that if in this system acceleration can be
neglected, then both VRT and SRT give the same results. Thus interferometric
experiments where solely a stationary (or uniformly moving) source is used
cannot be used to distinguish between VRT and SRT.

Consider a source and a detector that move at arbitrary velocities. Let $K$
be the proper, non-rotating system of the source. At time $\tau $ (emission
time) a pulse is emitted from the source located at $\mathbf{\xi }$, that
arrives at time $t$ (time of reception) to a moving detector instantaneously
located at $\mathbf{x}\left( t\right) $. Then, the traversed distance, $
\mathbf{x}\left( t\right) -\mathbf{\xi }$, and the duration of the journey, $
t-\tau $, fulfill
\begin{equation}
\mathbf{x}\left( t\right) -\mathbf{\xi }=c\left( t-\tau \right) \mathbf{\hat{n}}  \label{1.5}
\end{equation}
where $\mathbf{\hat{n}}$ is the unit vector in the direction of the
trajectory (which coincides with the direction of the beam in $K$).

Now, consider a system $K^{\prime }$ fixed to the detector at time $t$. In
this system the detector is at rest at $\mathbf{x}^{\prime }$ while the
source moves according to $\mathbf{\xi }^{\prime }\left( t \right) $.
Then a \textit{Galilean transform} gives
\begin{equation}
\mathbf{x}^{\prime }-\mathbf{\xi }^{\prime }\left( t \right) =\mathbf{x}
\left( t\right) -\mathbf{\xi }  \label{1.6}
\end{equation}

Thus, according to a Galilean observer in $K^{\prime }$ the path of the
light started at $\mathbf{\xi }^{\prime }\left( \tau \right) $ and ended in $
\mathbf{x}^{\prime }$, while the duration of the trip is the same as
measured in $K$, that is, $t-\tau $. Therefore, the velocity of the
propagation at $t$ in $K^{\prime }$ is
\begin{equation}
c^{\prime }\mathbf{\hat{n}}^{\prime }\mathbf{=}\frac{\mathbf{x}^{\prime }-\mathbf{\xi }^{\prime }\left( \tau \right) }{t-\tau }\mathbf{=}c\mathbf{\hat{n}}+\mathbf{\bar{v}}  \label{1.7}
\end{equation}
where $\mathbf{\hat{n}}^{\prime }$ is the unit vector in the direction of
the trajectory (which differs from the direction of the trajectory in $K$,
the angle between $\mathbf{\hat{n}}$ and $\mathbf{\hat{n}}^{\prime }$ is the
aberration, see Chapter \ref{ch:2} 2), and
\begin{equation}
\mathbf{\bar{v}=}\frac{\mathbf{\xi }^{\prime }\left( t\right) -\mathbf{\xi }^{\prime }\left( \tau \right) }{t-\tau }  \label{1.8}
\end{equation}
is the \textit{mean speed} of the source between emission and reception in
system $K^{\prime }$. Note that trajectory depends on the reference system
(in fact, aberration is the angle between the trajectory in two different
systems), while beam direction is invariant under a Galilean transform.

In conclusion, VRT predicts that the \textit{instantaneous speed} of light
is $c+v\left( t_{r}\right) $ where $v\left( t_{r}\right) $ is the
instantaneous speed of the source at \textit{time of detection}, while the
\textit{mean speed} of light over a given time interval is $c+\bar{v}$ where
$\bar{v}$ is the \textit{mean velocity} of the source during the considered
time interval.

As a consequence of this, VRT predicts:

\begin{list}{(\arabic{qcounter})}{\usecounter{qcounter}}
\item In the proper, non-rotating system of the source the light
propagates with constant velocity $c$ in vacuum.
\item The standard wave equation is not invariant under a Galilean
transformation. It can only be applied in the proper frame of the source, in
a similar way that a mechanical wave equations can only be used in the
system attached to the body in which it propagates. In Chapter \oldref{8}\ref{ch:8} 
a Galilean invariant wave equation will be introduced.
\item The travel time of a pulse is equal in all systems and can be
calculated as the distance between the source and detector, both measured at
\textit{time of reception}, divided by $c$. This is immediately obvious if
one imagines the source to be stationary with the detector moving.
\item The average speed of light in any system in a given time interval
is the sum of $c$ plus the average speed of the source during the same time
interval.
\end{list}

According to the previous interpretation, DeSitter (and Brecher) arguments
vanish, particularly for fast rotating, distant stars. In this problem what
counts is the total travel time, that is, the mean speed rather than the
instantaneous speed. If light remains coupled to the source, then also the
mean speed of the source should be used. A fast rotating and distant star
will exhibit a almost null mean speed (relative to the center of rotation),
in accordance with the observations.

In addition to these objections also the following objections are invalid:

\begin{list}{(\alph{qcounter})}{\usecounter{qcounter}}
\item ``\textit{If a radiant star moves across our field of vision,
light given off by differently-moving atoms in its atmosphere should take
different amounts of time to reach us. Since the retreating atoms would have
a red Doppler shift, and the approaching ones a blue Doppler shift, the
passing star might be expected to appear as a rainbow streak.}'' This is
false since what counts in this problem is the mean velocity of the source
(i.e. the atoms) between the epoch of emission and the epoch of reception.
Light from any star will take many years to reach us. Since the average
velocity of any atom in a star over a period of many years approaches zero
(relative to the star), then, light from red or blue Doppler shift will take
the same time to reach us. Therefore no \textit{rainbow} effect will be
observed.
\item ``\textit{Similarly, if a radiant star is eclipsed, one might
expect the eclipsing shadow to appear to intercept different colours of
Doppler-shifted light in sequence - the eclipse might appear to have
coloured fringes.}'' False. As before, light coming from different atoms
from a star will take almost the same amount of time to reach us since their
mean velocity over large periods is very close to zero relative to the star.
\item ``\textit{For the case of a double-star system
seen edge-on, light from the approaching star might be expected to travel
faster than light from its receding companion, and overtake it. If the
distance was great enough for an approaching star's fast signal to catch up
with and overtake the slow light that it had emitted earlier when it was
receding, then the image of the star system should appear completely
scrambled.}'' False. According to VRT what counts is the
average speed of the stars during the time interval from emission to
reception. For far distant double stars the mean orbital velocity approaches
zero relative to their center of mass. Therefore $c^{\prime }\approx c$ thus
no anomalies should be observed.
\end{list}


%% file: 2014arxiv_v14_2_2.tex


\chapter{Aberration}\label{ch:2}

\fancyhf{}
\fancyhead[LE,RO]{2. Aberration}
\fancyhead[RE,LO]{L. Bilbao}
\fancyfoot[RE,LO]{Vibrating Rays Theory}
\fancyfoot[LE,RO]{\thepage}
 
\renewcommand{\headrulewidth}{1pt}
\renewcommand{\footrulewidth}{1pt}

The phenomenon, known today as ``stellar aberration,'' which was published
by Bradley in 1729 \oldref{[R40]}\cite{bradley1729letter}, refers to the 
north-south shift of the passage of a
star through the meridian, as measured along the year. It has maximum
amplitude (in the case of a star perpendicular to the ecliptic) of about
41'' between their northernmost and southernmost points. In his original
work, Bradley called it ``alteration of declination.'' The term aberration
was coined later, around 1737 \oldref{[R41]}\cite{bradley1931discovery}.

While Bradley was the first to report an accurate measurement, there were
observations about 50 years earlier. According to Sarton\oldref{[R41]} 
\cite{bradley1931discovery}, in 1671 Piccard observed annual variations 
in the position of the North Star. There were also observations in 1674 by 
Hooke, and between 1689 and 1697 by Flamsteed. Hooke and other 
researchers thought it was a parallax effect.
However, as demonstrated by Cassini and Manfredi, the measured effect
pointed in a direction rotated 90${{}^\circ}$ from the expected one for a
parallax effect, although these researchers did not find the correct
explanation of the phenomenon.

The high precision work by Bradley and S. Molyneux between 1725 and 1727 was
the turning point in advancing to the right interpretation. The two main
observations of Bradley were: (a) that as they crossed the meridian at 6
o'clock in the afternoon, the stars did it at the northern extreme. However,
when they crossed at 6 am the maximum deviation to the south occurred; (b)
the extent of the alteration of the declination was proportional to the sine
of the angle of the star to the ecliptic.

Today, stellar aberration is defined as the apparent motion of celestial
objects around their true location, due to a combination between the speed
of light and Earth's orbital velocity. Terms like ``true location'' or
``fixed stars'' actually refer to the \textit{celestial coordinate system},
and not to a standard Galilean reference frame.

In what follows we will refer to first order aberration, since all
measurements were made within this limit.

The word ``aberration'' is used in at least three different contexts, often
without adequate explanation, namely:

\begin{list}{\arabic{qcounter})}{\usecounter{qcounter}}
\item To refer to the transformation of the direction of propagation of
light in different systems (e.g.\oldref{[R42]}\oldref{[R43]}
\cite{sommerfeld1964lectures,pauli1981theory}). The direction of propagation 
of light depends on the coordinate system and, therefore, the transformation
depends on the relative velocity of those systems. The angle between $
\mathbf{\hat{n}}$ (direction of the trajectory in system $K$) and
 $\mathbf{\hat{n}}^{\prime }$ (direction of the trajectory in system 
 $K^{\prime }$) in \oldref{(1.7)}(\ref{1.7}) is called ``aberration.''
\item To refer to the angle between the direction of light propagation
and the axis of the telescope. In Bradley's figure of page 646 in reference
\oldref{[R40]}\cite{bradley1729letter} it is depicted how a telescope should be 
tilted to admit a ``particle of light'' through a ``tube'' (telescope) in order 
to reach the ``eye'' (detector). In an inertial system the angle, 
$\mathbf{\alpha }$, is given by
\begin{equation}
\mathbf{\alpha =-}\frac{1}{c} \, \mathbf{\hat{n}\times v}_{d}  \label{2.1}
\end{equation}
where $\mathbf{\hat{n}}$ is the direction of the trajectory of light and $
\mathbf{v}_{d}$ is the velocity of the telescope. This expression is valid
to first order in $\mathbf{v}_{d}/c$.
\item To refer to the ``alteration of declination'' as measured by Bradley. 
``Alteration of declination'' is measured as the variation of
the angle between the telescope and a fixed direction (i.e. the Earth axis).
It does not depend on relative speeds or on the coordinate system (i.e. the
calculation in any inertial system will give the same result). We will call
it ``stellar aberration.'' Today's usual
definition of stellar aberration is that bodies are observed in a position
displaced from the true position (in the celestial coordinate system). What
has been demonstrated is that the speed shown in the formula of the stellar
aberration is not the relative velocity between source and detector
\oldref{[R44]}\oldref{[R45]}\cite{phipps1988relativity,eisner1967aberration}. 
This is verified by observation of those binary stars which have a
short period of rotation, and which move with velocities exceeding that of
the Earth around the Sun. If the phenomenon depended on the relative
velocities, then these binary stars should exhibit an anomalous aberration,
which has not been observed.
\end{list}

The angle between the trajectory and the telescope axis is given by 
\oldref{(2.1)}(\ref{2.1}), where the direction of the trajectory is defined as
\begin{equation}
\mathbf{\hat{n}=}\frac{\mathbf{x}_{d}\left( t\right) -\mathbf{x}_{s}\left(
\tau \right) }{\left\vert \mathbf{x}_{d}\left( t\right) -\mathbf{x}_{s}\left( \tau \right) \right\vert }  \label{2.2}
\end{equation}
$\mathbf{x}_{s}\left( \tau \right) $ being the position of the source at
time of emission, $\tau $, and $\mathbf{x}_{d}\left( t\right) $ the position
of the telescope at time of detection, $t$. Consider that the inertial
system $K$ is the solar barycenter system. Therefore the velocity of the
telescope, $\mathbf{v}_{d}$, is the velocity of Earth which changes over the
year. For any star, the direction of propagation of light is almost constant
for any detector located in the Earth orbit (that is equivalent to neglect
the parallax as compared to the stellar aberration). Then, the stellar
aberration can be calculated as the variation of $\mathbf{\alpha }$ with the
orbital velocity of the Earth using \oldref{(2.1)}(\ref{2.1}), and the two main
observations by Bradley are recovered. Clearly, what counts in stellar
aberration is the variation with time of the velocity of the detector in a
given (any) inertial system.

What is wrong is to interpret directly that ``stellar aberration'' as
measured by Bradley corresponds to ``aberration'' (i.e., the angular
difference between the trajectories in two systems). This has generated
controversy and confusion that still survive. Even today some argue that
``stellar aberration'' should depend on the relative velocity\oldref{[R46]}
\cite{kassner2002bradley}. Others argue that the formula of aberration 
(which involves the relative velocity) is only valid in inertial frames, and 
as Earth changes its speed during a year, it cannot be used.

Another point that often creates confusion is the term \textit{beam} (or
\textit{ray}) of light and its relation to the \textit{trajectory} of
photons. The deduction referred to above, involving the line joining the
source (at time of emission) with the detector (at time of reception), can
be linked to the trajectory of a photon, but not to the concept of beam.

When we talk about \textit{beam} we refer to the sequence of all illuminated
regions \textit{at a given instant}. In contrast, trajectory is the points
that a photon occupied in \textit{different moments of time}. A clear
example of the difference can be seen in the book by Fayngold\oldref{[R47]} 
\cite{fayngold2002superluminal} in the case of a rotating source where the 
beam is a spiral and the trajectory is a straight line.

Note also that neither \textit{beam} nor \textit{trajectory} corresponds to
the concept of Faraday's ``vibrating rays,'' for the latter do not rotate.
In the case of uniform motions we define the beam as a straight line joining
the source with the detector both at the time of reception, while the
trajectory of the photon is the line between the source at the time of
emission and the detector at the time of reception, i.e., the direction
of propagation of the information.

\begin{figure}[h!]
\centering
\begin{subfigure}{.5\textwidth}
  \centering
  \includegraphics[scale=0.5]{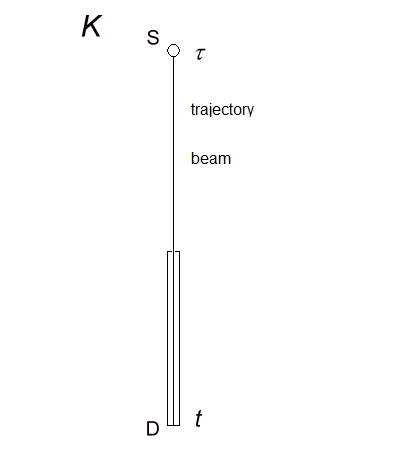}
  \caption{}
  \label{fig:2.1.a}
\end{subfigure}%
\begin{subfigure}{.5\textwidth}
  \centering
  \includegraphics[scale=0.5]{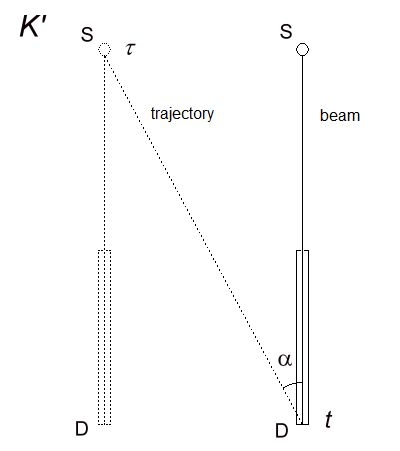}
  \caption{}
  \label{fig:2.1.b}
\end{subfigure}
\caption{\oldref{Figure 2.1.}A vertical beam seen from two different systems. (a) In $K$, the
source $S$ and detector $D$ are stationary, the trajectory of the photon
coincides with the beam. (b) In system $K^{\prime }$, moving with constant
velocity relative to $K$, the trajectory of the photon does not coincide
with the axis of the telescope, there is an angle $\alpha $ between them. In
both cases, $\tau $ refers to the moment of emission and $t$ of detection,
respectively, measured in each system.}
\vspace{-8pt}
\label{fig:2.1}
\end{figure}

As an example, suppose we have a vertical \textit{beam} in a system where
the source is at rest, figure \oldref{2.1a}\ref{fig:2.1.a}. In this system the \textit{beam} 
and the \textit{trajectory} coincide. Viewed from a moving system the 
\textit{beam} will remain vertical, while the \textit{trajectory} of photons will 
be tilted, figure \oldref{2.1b}\ref{fig:2.1.b}. There is ``aberration'' between 
trajectories in $K$ and $K^{\prime }$, regardless of the fact that the telescope 
is vertical in both cases (i.e., no ``stellar aberration'' is measured).
This shows that ``aberration'' is relative,
not because the orientation of the telescope changes relative to a given
fixed direction, but because the orientation of the trajectory of the photon
depends on the reference system.

The ``stellar aberration'' is measured relative to a fixed direction rather
than to the source-detector direction (which depends on the reference
system); the same tilt of the telescope is obtained when analyzed from $K$
or from $K^{\prime }$. This is consistent with measurements made by Kwiek
and Sikorski \oldref{[R48]}\cite{kwiek2002comment}, and explains why the 
mechanism proposed by Sardine \oldref{[R49]}\cite{sardin2001measure}
to measure absolute velocities does not work.

\begin{figure}[h!]
\centering
\begin{subfigure}{.5\textwidth}
  \centering
  \includegraphics[scale=0.5]{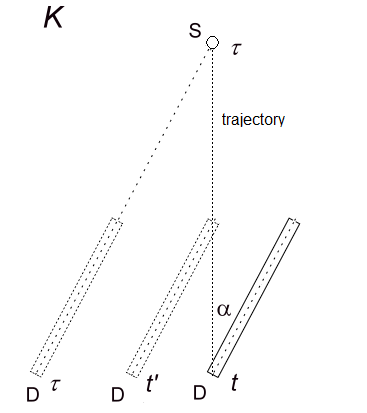}
  \caption{}
  \label{fig:2.2.a}
\end{subfigure}%
\begin{subfigure}{.5\textwidth}
  \centering
  \includegraphics[scale=0.5]{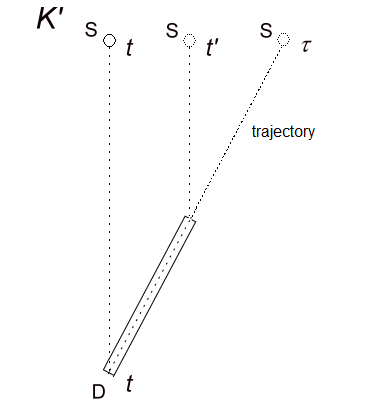}
  \caption{}
  \label{fig:2.2.b}
\end{subfigure}
\caption{\oldref{Figure 2.2.}(a) In system $K$, the source $S$ is at rest and the detector $D$
moves with constant velocity in the horizontal direction. The trajectory of
the photon that leaves the source at time $\tau$ enters the telescope at 
$t^{\prime}$ and reaches the detector at time $t $, is vertical and makes an 
angle $\alpha$ with the axis of the telescope. (b) In the system $K^{\prime }$ 
the detector $D$ is at rest. The trajectory of the photon that leaves the source
at time $\tau$ enters the telescope at $t^{\prime }$ and reaches the
detector at time $t$, is in the direction of the axis of the telescope.}
\vspace{-8pt}
\label{fig:2.2}
\end{figure}

In the second example we have an omni-directional source (e.g. a star) at
rest in $K$, and a telescope that moves with some constant speed. Again,
being the source at rest, the beam and the trajectory coincide in this
system. Since the telescope moves it should be tilted in order for the
photon to enter the telescope and reach the detector, as shown in figure
\oldref{2.2a}\ref{fig:2.2.a}. However, as seen in $K^{\prime }$ (where the 
telescope is at rest) the telescope is oriented in the direction of the trajectory 
of photons, as shown in figure \oldref{2.2b}\ref{fig:2.2.b}. Note that this 
means that the telescope, in both cases, points in the same direction (relative 
to a fixed direction like the vertical one), i.e., it points towards the position of 
the source at time of emission (this is valid in both systems $K$ and $K^{\prime }$).
Note also that in the cases discussed above the difference between SRT and
VRT is at most second order in $v/c$. The difference comes from the
calculation of $\mathbf{\hat{n}}$ in \oldref{(2.2)}(\ref{2.2}) where the 
emission time, $\tau $, is different depending on the component of the star 
velocity along the trajectory. In conclusion, first order ``stellar aberration'' 
cannot be used to distinguish between SRT and VRT.


%% file: 2014arxiv_v14_2_3.tex


\chapter{Fresnel drag}\label{ch:3}

\fancyhf{}
\fancyhead[LE,RO]{3. Fresnel drag}
\fancyhead[RE,LO]{L. Bilbao}
\fancyfoot[RE,LO]{Vibrating Rays Theory}
\fancyfoot[LE,RO]{\thepage}
 
\renewcommand{\headrulewidth}{1pt}
\renewcommand{\footrulewidth}{1pt}

In 1810, Arago \oldref{[R50]}\cite{arago1853memoire} speculated that 
the image formation of stars would be different while approaching to than 
when receding from the source. The focus of the telescope or the deflection 
of start-light by a prism should be different depending on the relative velocity. 
His measurements showed that there were no differences independently of 
the source-detector relative velocity. Fresnel \oldref{[R51]}\cite{fresnel1818influence} 
working out those cases, found that in order to explain Arago's null result, 
the light in a moving medium should be partially dragged according to his well 
known Fresnel coefficient
\begin{equation}
c_{m}=\frac{c}{n}+v\left( 1-\frac{1}{n^{2}}\right)  \label{3.1}
\end{equation}

Fizeau was able to measure the first order coefficient back in 1851 
\oldref{[R52]}\cite{fizeau1851ether}. Fizeau's experiment was repeated 
using ring laser \oldref{[R53]}\oldref{[R54]}\cite{macek1964measurement,bilger1972fresnel}
 that confirmed with high precision that \oldref{(3.1)}(\ref{3.1}) holds 
in a non-dispersive medium.

Note that if we assume that:

(\textit{H1}) The speed of light in a transparent medium is $c/n$ in the
frame of reference of the medium, and,

(\textit{H2}) The Galilean addition of speed holds,
then, the speed of light in a moving medium should be
\begin{equation}
c_{m}=\frac{c}{n}+v  \label{3.2}
\end{equation}
Since \oldref{(3.2)}(\ref{3.2}) is in contradiction with experimental results
\oldref{(3.1)}(\ref{3.1}), either (\textit{H1}) or (\textit{H2}) is wrong 
(of course, both hypothesis may also be wrong).
Rejecting (\textit{H2}) and keeping (\textit{H1}) lead to SRT. The
relativistic derivation of Fresnel's law is due to Laub \oldref{[R55]}\cite{laub1907optik}
and von Laue \oldref{[R56]}\cite{laue1907die}. According to the principle 
of relativity, the velocity of light relative to the proper frame ($K^{\prime }$) 
of a transparent medium (the Galilean frame in which the medium is at rest) 
depends only on the medium. Assuming full drag of light by the medium, 
that is, that the velocity of light $c_{m}^{\prime }$ relative to the proper 
frame of a transparent medium is
\begin{equation}
c_{m}^{\prime }=\frac{c}{n}  \label{3.3}
\end{equation}
the propagation in a moving medium as seen for an observer at rest is then
\begin{equation}
c_{m}=\frac{\frac{c}{n}+v}{1+\frac{v}{nc}}  \label{3.4}
\end{equation}

Note that the above expression has no definite limit when $n\rightarrow 1$
and $v\rightarrow -c$. Let us first take $n\rightarrow 1$, which is
\[
c_{m}=c \, \frac{1+\frac{v}{c}}{1+\frac{v}{c}}=c
\]
therefore, when $v\rightarrow -c$, $c_{m}=c$ since it is independent of $v$.

However, taking first the limit $v\rightarrow -c$, we have
\[
c_{m}=c \, \frac{\frac{1}{n}-1}{1-\frac{1}{n}}=-c
\]
then, when $n\rightarrow 1$, the limit is $c_{m}=-c$ since it is independent
of $n$.

This seems unreasonable since the same limit, first made in a variable and
then in the other, and vice versa, gives very different results: in one case
the light propagates to the right and in the other to the left. As it is
shown below, VRT formula has a definite limit when $n\rightarrow 1$ and $
v\rightarrow -c$.

To the second order in $v$, \oldref{(3.4)}(\ref{3.4}) gives
\begin{equation}
c_{m}=\frac{c}{n}+v\left( 1-\frac{1}{n^{2}}\right) -\frac{v^{2}}{nc}\left( 1-
\frac{1}{n^{2}}\right) +O\left( v^{3}\right)   \label{3.5}
\end{equation}
that coincides with \oldref{(3.1)}(\ref{3.1}) to the first order.

Actually there is another possibility, which is to keep (\textit{H2}) and to
reject (\textit{H1}). As it was pointed out by Clement \oldref{[R57]}\cite{clement1980does} 
and more recently by Drezet \oldref{[R58]}\cite{drezet2005physical} 
the physical origin of the Fresnel drag of light by a moving dielectric 
medium is, in its essential part, independent of the theory of relativity.

Let us suppose that light sources emit waves at a fixed velocity $c$ (the
speed of light in vacuum) in their proper, non-rotating frame of reference.
A transparent moving medium (at speed $v$ relative to the source) can be
modeled using the idea of Clement. That is, the interaction of light with
matter can be divided into two parts: free travel in the vacuum between
atoms, and microscopic interaction of the wave with the atoms (scatter).
Clement used that the speed of light in vacuum is $c$ irrespective of the
source movement. On the contrary, we assume here that the propagation speed
in vacuum is $c$ relative to the source. In order to calculate the speed of
light in a medium the calculation should be performed in the proper system
of the source. Using Clement ideas, but replacing the speed of light in
vacuum $c$ with $c-v$, the mean velocity of light in a moving medium
(relative to the source) is given by
\begin{equation}
c_{m}=\frac{c-v}{1+\left( n-1\right) \frac{c-v}{c}}+v  \label{3.6}
\end{equation}

Note that in contrast with SRT, the above expression has a definite limit
when $n\rightarrow 1$ and $v\rightarrow -c$, that is
\[
c_{m}=c
\]
as expected.

To the second order in $v$, \oldref{(3.6)}(\ref{3.6}) becomes
\begin{equation}
c_{m}=\frac{c}{n}+v\left( 1-\frac{1}{n^{2}}\right) -\frac{v^{2}}{n^{2}c}
\left( 1-\frac{1}{n}\right) +O\left( v^{3}\right)   \label{3.7}
\end{equation}
very similar to the relativistic expression \oldref{(3.5)}(\ref{3.5}).
Due the dispersion of the transparent media, the difference between
(\ref{3.5}) and (\ref{3.7}) is probably hard to detect even for a
medium with low optical dispersion.

Note that the velocity in the proper frame of the medium, $c_{m}^{\prime }$,
is therefore,
\begin{equation}
c_{m}^{\prime }=\frac{c+u}{1+\left( n-1\right) \frac{c+u}{c}}  \label{3.8}
\end{equation}
where $u=-v$ is the speed of the source as seen from the medium. To the
second order in $u$, we get
\begin{equation}
c_{m}=\frac{c}{n}\left( 1+\frac{u}{nc}-\frac{n-1}{n^{2}}\ \frac{u^{2}}{c^{2}}
\right) +O\left( u^{3}\right)  \label{3.9}
\end{equation}

This means that under VRT the speed of light in a medium is not given by 
\oldref{(3.3)}(\ref{3.3}) because it depends also on the speed of the source. 
In other words the transparent medium does not extinguish the speed of the source.

In conclusion, it is possible to explain the Fresnel drag under Galilean
transformation if we assume that the effect of a transparent medium on the
propagation of light is a retardation effect where the mean speed is given
by \oldref{(3.6)}(\ref{3.6}).


%% file: 2014arxiv_v14_2_4.tex


\chapter{Doppler effect and the twin paradox}\label{ch:4}

\fancyhf{}
\fancyhead[LE,RO]{4. Doppler effect}
\fancyhead[RE,LO]{L. Bilbao}
\fancyfoot[RE,LO]{Vibrating Rays Theory}
\fancyfoot[LE,RO]{\thepage}
 
\renewcommand{\headrulewidth}{1pt}
\renewcommand{\footrulewidth}{1pt}

\section{Doppler effect}\label{sec:4.1}

In the case of optical Doppler effect, there are very few situations in
which there is an independent measurement of the shift in frequency and
velocity of the source. The first attempt to verify the Doppler formula for
light was made by Belopolsky \oldref{[R59]}\cite{belopolsky1901fizeau} 
who used an arrangement of two
counter-rotating discs with mirrors arranged around the edges, the light
undergoing six reflections in all, from either disc in turn. Belopolsky used
the Sun as a source of light, and measured the Doppler shift in the
Fraunhofer lines using a prism spectrometer.

One of the most important applications of the Doppler effect is rocket and
satellite tracking \oldref{[R60]}\cite{gill1965doppler}. A standard method of 
measuring the frequency of an
incoming radio signal is to mix it with the output of a standard oscillator,
and measure the resulting beat frequency, which will usually be quite a low
frequency compared with that of the incoming signal. A signal generated at
Earth is received back by using reflection or ``transponder'' Doppler, in
which the moving vehicle does not generate a frequency of its own but merely
reflects or retransmit a signal from a stationary transmitter. The reflected
signal is then mixed with some of the outgoing signal and the beat frequency
measured. The use of a receiver which amplifies and retransmits the signal,
instead of passive reflection, has some advantages, particularly in regard
to the reduction of transmitter power. Relativistic Doppler formula does not
distinguish between passive and active reflection (although, the use of the
SRT Doppler formula in non-inertial frames should be performed with caution,
see Section \oldref{4.3}\ref{sec:4.3}). On the contrary, VRT gives different 
results for passive and active reflection. Spacecraft anomalies may be 
explained taking account of these differences.

It is often assumed without question that the Doppler displacement would
take the same time to reach a distant point as the light itself\oldref{[R10]}
\cite{dingle1960dopplerb}. SRT and
VRT are based on a postulate of relativity, which says that motion is
essentially a relation between two or more bodies and is a meaningless term
when applied to a single body, but they differ in that SRT postulates that
all measurements of the velocity of light with respect to a body, whatever
the motion of that body, will yield the constant value $c$, while VRT
postulates that the velocity of light is always $c$ with respect to the body
that emits the light, no matter how the body may be moving at the time of
emission or later \oldref{[R10]}\cite{dingle1960dopplerb}. Hence immediate 
transmission of the Doppler effect is necessary in the latter theory.

The so-called ``Pioneer anomaly'' \oldref{[R5]}\cite{anderson1998indication} and 
``flyby anomaly'' \oldref{[R6]}\cite{anderson2008anomalous} are examples
where the measured Doppler differs from the predictions of SRT. The Pioneer
anomaly was found in the microwave signal received from the space probes
Pioneer 10 and Pioneer 11 when they were in the outer regions of the solar
system. For distances between 20 and 70 AU a small drift of the Doppler
frequency of 6 nHz/s was found. Also, two other small oscillatory terms with
periods of one day and one year, respectively, were detected. The flyby
refers to a mismatch between extrapolated and measured post-encounter
Doppler data. These phenomena could be an indication that the Relativistic
Doppler formulas do not describe correctly the problem.

In the following $t_{1}$ is the epoch of transmission from Earth; $t_{2}$,
the epoch of interaction of the signal with the spacecraft; and $t_{3}$ the
epoch of reception back at Earth. All of these times are referred to the
corresponding inertial frame (ex., Barycentric Dynamical Timescale, which is
a coordinate time at the solar barycenter system for the Pioneer). The
vectors $\mathbf{r}_{1}$, $\mathbf{r}_{2}$, and $\mathbf{r}_{3}$ represent
the positions of the corresponding antennas at the corresponding epoch; $
\mathbf{v}_{1}$, $\mathbf{v}_{2}$, and $\mathbf{v}_{3}$ represent the
velocities of the spacecraft; and $\mathbf{u}_{1}$, $\mathbf{u}_{2}$, and $
\mathbf{u}_{3}$ represent the velocities of the Earth antennas.

According to SRT the expected frequency at the receiver at time $t_{3}$ of a
source of frequency $f_{0}$, whether actively or passively reflected, can be
expressed as
\begin{equation}
f_{SRT}=f_{0} \, \frac{\gamma _{u_{3}}}{\gamma _{u_{1}}}\frac{1-\mathbf{\hat{r}}_{12}\cdot \mathbf{v}_{2}/c}{1-\mathbf{\hat{r}}_{12}\cdot \mathbf{u}_{1}/c}
\frac{1-\mathbf{\hat{r}}_{23}\cdot \mathbf{u}_{3}/c}{1-\mathbf{\hat{r}}_{23}\cdot \mathbf{v}_{2}/c}  \label{4.1}
\end{equation}
where
\begin{equation}
\gamma _{u_{1,3}}=\frac{1}{\sqrt{1-\left( u_{1,3}/c\right) ^{2}}}
\label{4.2}
\end{equation}
and the unit vectors difference, $\hat{\mathbf{r}}_{12}$ and $\hat{\mathbf{r}}_{23}$ are defined as
\begin{equation}
\mathbf{\hat{r}}_{12}=\frac{\mathbf{r}_{2}-\mathbf{r}_{1}}{\left\vert
\mathbf{r}_{2}-\mathbf{r}_{1}\right\vert }\text{ , \ \ \ \ \ \ \ \ \ \ \ \ \
\ }\mathbf{\hat{r}}_{23}=\frac{\mathbf{r}_{3}-\mathbf{r}_{2}}{\left\vert
\mathbf{r}_{3}-\mathbf{r}_{2}\right\vert }\text{\ }  \label{4.3}
\end{equation}

According to VRT there are two possibilities: passive or active reflection.
In case of passive reflection the measured frequency is
\begin{equation}
f_{VRT\text{passive}}=f_{0} \, \frac{1-\mathbf{\hat{r}}_{22}\cdot \left( \mathbf{v}_{2}-\mathbf{u}_{2}\right) /c}{1+\mathbf{\hat{r}}_{22}\cdot \left( \mathbf{v}_{2}-\mathbf{u}_{2}\right) /c}  \label{4.4}
\end{equation}
where the unit vector $\mathbf{\hat{r}}_{22}$ points from the transmitting
station at epoch $t_{2}$ to the spacecraft at the \textit{same epoch}. For a
stationary source, \oldref{(4.4)}(\ref{4.4})  coincides with the SRT formula 
\oldref{(4.1)}\ref{4.1}), and for a uniformly moving source, \oldref{(4.4)}(\ref{4.4})
and \oldref{(4.1)}(\ref{4.1}) are equal up to second order in the velocities.

In case of an active reflection VRT predicts
\begin{equation}
f_{VRT\text{active}}=f_{0}\left( 1-\mathbf{\hat{r}}_{22}\cdot \frac{\mathbf{v}_{2}-\mathbf{u}_{2}}{c}\right) \left( 1-\mathbf{\hat{r}}_{33}\cdot \frac{\mathbf{u}_{3}-\mathbf{v}_{3}}{c}\right)  \label{4.5}
\end{equation}
where the unit vector $\hat{\mathbf{r}}_{33}$ points from the spacecraft at
epoch $t_{3}$ to the receiving station at the \textit{same epoch}.

As an example, consider a case where the transmitting station is at rest and
the unit vectors do not change appreciably during the measurement, that is, $
\hat{\mathbf{r}}_{12}\approx \hat{\mathbf{r}}_{22}\approx -\hat{\mathbf{r}}
_{23}\approx -\hat{\mathbf{r}}_{33}\approx \hat{\mathbf{r}}$, then
\begin{equation}
f_{VRT\text{active}}-f_{SRT}=f_{0}\left( 1-\mathbf{\hat{r}}\cdot \frac{\mathbf{v}_{2}}{c}\right) \left( 1-\mathbf{\hat{r}}\cdot \frac{\mathbf{v}_{3}}{c}-\frac{1}{1+\mathbf{\hat{r}}\cdot \mathbf{v}_{2}/c}\right)  \label{4.6}
\end{equation}

To first order in $v/c$ the difference between SRT and VRT active reflection is
\begin{equation}
f_{VRT\text{active}}-f_{SRT}\approx f_{0}\ \mathbf{\hat{r}}\cdot \frac{\mathbf{v}_{2}-\mathbf{v}_{3}}{c}  \label{4.7}
\end{equation}

As it will be shown in Chapter \oldref{6}\ref{ch:6}, the Pioneer anomaly seems to be related to
the above difference, produced during the downlink leg, where the time
interval between emission and reception is large and the difference 
$\mathbf{v}_{2}-\mathbf{v}_{3}$ becomes measurable. Instead, the flyby 
anomaly can be interpreted as second order term differences.

\section{Twin paradox}\label{sec:4.2}

Consider the twin paradox where a twin ($A$) remains on Earth and the other 
($B$) travels to a distance $D$ at speed $v=\beta c$ measured in the Earth
system, and then returns back to Earth at the same speed. Calling $T=D/v$,
the total journey time is $2T$. It is interesting to see what conclusion
regarding asymmetric aging is obtained from the Doppler shift received by
each twin. It is assumed that each twin emits a signal with frequency $f_{0}$
constant and unchanging in her or his own reference system.

According to the interpretation of the twin who remains on Earth, she or he
will receive the Doppler signal with a frequency
\[
f_{-}=f_{0} \, \frac{\sqrt{1-\beta }}{\sqrt{1+\beta }}
\]
until time
\[
\Delta t_{-}=T+\frac{D}{c}=T\left( 1+\beta \right)
\]
From that moment up to $2T$, that is, during an interval
\[
\Delta t_{+}=2T-\left( T+\frac{D}{c}\right) =T\left( 1-\beta \right)
\]
will receive a frequency
\[
f_{+}=f_{0} \, \frac{\sqrt{1+\beta }}{\sqrt{1-\beta }}
\]

The total number of cycles received from $B$ will be
\[
N_{B}=f_{-}\Delta t_{-}+f_{+}\Delta t_{+}=\frac{2Tf_{0}}{\gamma }
\]
being
\[
\gamma =\frac{1}{\sqrt{1-\beta ^{2}}}
\]
Meanwhile the number of cycles emitted by $A$ (and, therefore, received by $
B $) is
\[
N_{A}=2Tf_{0}
\]
As the number of cycles (number of beats) is proportional to the age of each
twin, and
\[
N_{B}<N_{A}
\]
it follows that $B$ is younger than $A$ when reunited. The trip duration was
$2T$ for $A$ and $2Y$ for $B$, being $Y=T/\gamma $.

The analysis from the point of view of $B$ leads to the same results. $B$
receives a receding Doppler over half of his journey and an approaching
Doppler over the other half. Therefore, the total count of cycles received
from $A$ is
\[
N_{A}^{\prime }=Yf_{0}\frac{\sqrt{1-\beta }}{\sqrt{1+\beta }}+Yf_{0}\frac{\sqrt{1+\beta }}{\sqrt{1-\beta }}
\]
that is
\[
N_{A}^{\prime }=2\gamma Yf_{0}
\]
While the number of cycles sent is
\[
N_{B}^{\prime }=2Yf_{0}
\]
therefore
\[
N_{B}^{\prime }<N_{A}^{\prime }
\]
thus, $B$ is younger than $A$ upon reunion. Also note that
\begin{eqnarray*}
N_{A}^{\prime } &=&N_{A} \\
N_{B}^{\prime } &=&N_{B}
\end{eqnarray*}
as expected.

In addition to determining the age, the twins try to get the path of the
other twin from the Doppler signal in a similar way to that used in
satellite tracking. The difference is that the received signal is generated
by the other twin instead of using the method of active reflection.

According to SRT, from the received Doppler, $f(t)$, at time $t$ it is
possible to know the velocity, $v$, at the previous instant of emission $
\tau $, from
\[
v\left( \tau \right) =c \, \frac{1-z^{2}\left( t\right) }{1+z^{2}\left( t\right)
}
\]
being
\[
z\left( t\right) =\frac{f\left( t\right) }{f_{0}}
\]
and
\begin{equation}
\tau =t-\frac{x\left( \tau \right) }{c}  \label{4.8}
\end{equation}
where $x(\tau )$ is the relative position of the other twin at the instant $
\tau $. Integrating the velocity it is possible to obtain the trajectory
\begin{equation}
x\left( \tau \right) -x\left( \tau _{0}\right) =\int_{\tau _{0}}^{\tau}v\left( \tau
^{\prime }\right) d\tau ^{\prime }=\frac{c}{2}\int_{t_{0}}^{t}\left(
1-z^{2}\left( t\right) \right) dt  \label{4.9}
\end{equation}

As stated before, twin $A$ receives from $B$ a constant $z_{-}$ signal
during a time interval $\Delta t_{-}$ and another constant signal signal $
z_{+}$ during a time interval $\Delta t_{+}$. Using the above expression
twin $A$ calculates that $B$ displacement during the first part of the
journey is
\[
\Delta x_{-}=\frac{c}{2}\int_{0}^{\Delta t_{-}}\left( 1-z_{-}^{2}\right)
dt=vT=D
\]
while during the second part is
\[
\Delta x_{+}=\frac{c}{2}\int_{\Delta t_{-}}^{2T}\left( 1-z_{+}^{2}\right)
dt=-vT=-D
\]
As expected, it holds that
\[
\Delta x_{-}+\Delta x_{+}=0
\]
So according to $A$, $B$ is away at a constant speed for a time $T$ to reach
a distance $D$, and from there begins the return to $A$, where she or he
arrives at time $2T$. Although the received Doppler frequency is not
symmetric, it follows that $B$'s trip is symmetric. This is so because,
according to SRT, the propagation of the Doppler signal is not instantaneous.

Viewed from $B$, the other twin $A$ moves away during a given period and
then begins the return. Now, the difference is that $B$ observes a
symmetrical Doppler: half the time, between $0$ and $Y$, observes $z_{-}$
and half the time, between $Y$ and $2Y$, a $z_{+}$ Doppler is measured.
Using \oldref{(4.9)}(\ref{4.9}) the displacement during the first part of the 
journey is
\[
\Delta x_{-}=\frac{c}{2}\int_{0}^{\Upsilon }\left( 1-z_{-}^{2}\right) dt=
\frac{\beta }{1+\beta }cY=\frac{D}{\gamma \left( 1+\beta \right) }
\]
while during the approaching stage it is
\[
\Delta x_{+}=\frac{c}{2}\int_{\Upsilon }^{2\Upsilon }\left(
1-z_{+}^{2}\right) dt=-\frac{\beta }{1-\beta }cY=-\frac{D}{\gamma \left(
1-\beta \right) }
\]
thereby the total displacement is
\[
\Delta x_{-}+\Delta x_{+}\neq 0
\]
that seems a contradiction, since a zero displacement is expected in a round
trip.

Another way to see this effect is that in order to fit the outward and
inward trajectories, as seen from $B$, a jump in space and time (backwards!)
should be included around time $Y$. The spatial jump is
\[
\Delta x=2\gamma \beta D
\]
(away from $B$) while the temporal jump is
\[
\Delta t=-\frac{\Delta x}{c}
\]
which is \textit{backwards in time} (see example below). A similar result is
obtained using the ``parallax distance'' \oldref{[R61]}\cite{unruh1981parallax}, 
since both cases involve
one-way propagation of light. Of course, around $Y$ the reference system
attached to the twin $B$ is non-inertial. However, integration is performed
from the left and to the right of this point, during inertial stages. In
addition, the acceleration and deceleration stages can be made as small as
desired compared to $D/c$ while $\gamma $ can be as large as we want, so the
jump can never be neglected (it may be smoothed out, but it always will
include a backwards temporal variation).

If one thinks that the acceleration experienced by the traveler invalidates
the possibility that she or he can use the Doppler effect, it should be
noted, however, that according to SRT this same Doppler effect predicts the
correct result for the asymmetric aging as seen from both $A$ and $B$. The
question is why $B$ can use the Doppler effect to calculate aging, but can
not use it to calculate the trajectory? Moreover, the fact that the Doppler
is symmetrical to $B$ and not to $A$ is essential for the asymmetric aging
and it is precisely this fact that prevents to correctly obtain the path of $
A$ as calculated by $B$.

The Doppler predicted by VRT does not suffer from this problem because the
speed to be considered in the formulas always corresponds to the present
speed. This also shows that, according to VRT, the Doppler is symmetrical
for both twins, both measuring exactly the same displacement, same time, as
the other twin; thus, neither asymmetric aging nor asymmetric trajectories
are present.

According to VRT, the interpretation of both twins is that she or he will
receive the Doppler signal with a frequency
\[
f_{-}=f_{0}\left( 1-\beta \right)
\]
during a time interval
\[
\Delta t_{-}=T
\]
and a frequency
\[
f_{+}=f_{0}\left( 1+\beta \right)
\]
during an interval
\[
\Delta t_{+}=T
\]
The total number of cycles received from either twin will be
\[
N_{A}=N_{B}=f_{-}\Delta t_{-}+f_{+}\Delta t_{+}=2Tf_{0}
\]

Since the propagation of the Doppler shift is instantaneous, the
displacement is
\[
x\left( t\right) -x\left( t_{0}\right) =\int_{t_{0}}^{t}v\left( t^{\prime
}\right) dt^{\prime }  \label{4.10}
\]
therefore both twins calculate that the other twin displacement during the
first part of the journey is
\[
\Delta x_{-}=vT=D
\]
while during the second part is
\[
\Delta x_{+}=-vT=-D
\]
For both twins, it holds that
\[
\Delta x_{-}+\Delta x_{+}=0
\]

\section{Numerical example}\label{sec:4.3}

Let as assume that the speed of the trip is $\beta =v/c=0.6$, and the
duration of each stage is $T=5$ years. Then, the maximum distance is $D=3$
light-years. At the time of departure, $t=0$, the age of the twins is $E$.
Both twin emit a pulse with frequency $f_{0}=2$ year$^{-1}$ (i.e. 2 signals
per year) in which a photograph is exchanged.

\begin{figure}
\begin{center}
\includegraphics[scale=0.5]{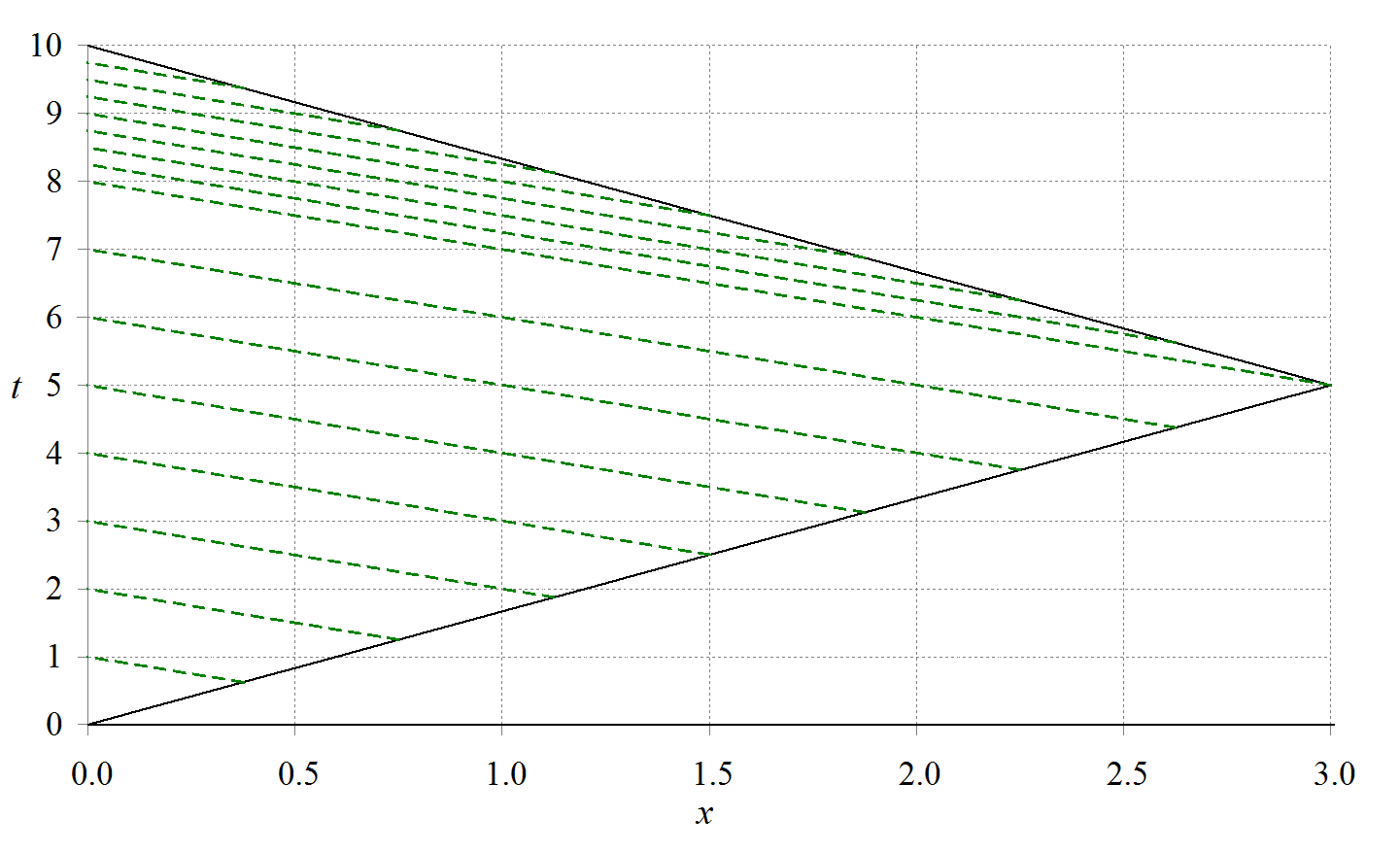}
\caption{Traveler's position (black full lines) calculated at Earth from
the Doppler signals received from the traveler (green dashed lines),
according to SRT using (\ref{4.9}).}
\label{fig:4.1}
\end{center}
\end{figure}

According to SRT the twin who remains on Earth observes the following (see
figure \oldref{4.1}\ref{fig:4.1}). She starts to receive a signal with frequency 
$f_{-}=0.5f_{0}=1$
due to receding Doppler. At $t=1$ receives the first signal, wherein the
photograph shows that the age of the traveler, at the epoch when the pulse
was emitted, was $E+0.5$. Using the speed from the Doppler signal, she
calculates that the pulse departed when $t=0.625$ (in the reference frame of
the Earth), when the spacecraft was at $x=0.375$. Her own age at that time
was $E+0.625$ thus follows that the traveling twin ages more slowly. This
holds up until reception of the eighth pulse at $t=8$, just when she
receives the photograph showing the traveling twin age's as $E+4$, which was
emitted in $t=5$ when the ship was at $x=3$ (maximum distance) and her own
age was $E+5$. It follows that at that time (in the Earth system) the
traveler is one year younger than she. From that moment on, she receives an
approaching Doppler frequency $f_{+}=2f_{0}=4$. For example, at $t=9$ she
receives the photograph of the traveling twin showing an age $E+6$
corresponding to $t=7.5$ when the ship was in its way back in $x=1.5$ and
her own age was $E+7.5$, i.e., the traveler continues to age more slowly
than she does. And so on, until the reunion in $x=0$, which occurs at $t=10$
, when her age is $E+10$ and that of the traveler $E+8$. In that period she
sent 20 photographs but received only 16. This shows that the twins agree
with the asymmetric aging.

\begin{figure}
\begin{center}
\includegraphics[scale=0.5]{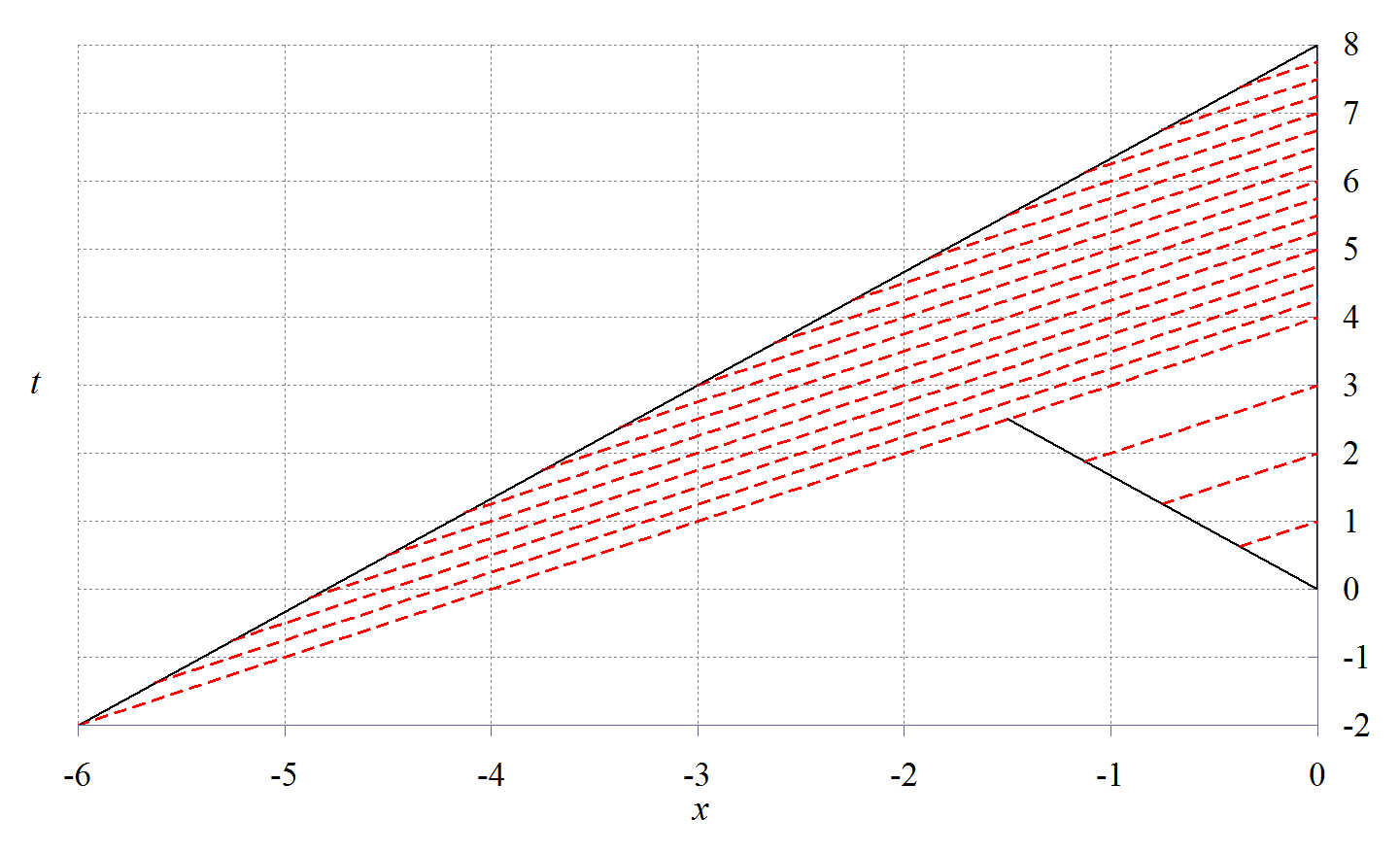}
\caption{Earth position (black full lines) calculated by the traveler
from the Doppler signals  received from the Earth twin (red dashed lines),
according to SRT using (\ref{4.9}).}
\label{fig:4.2}
\end{center}
\end{figure}

The traveling twin observes the following (see figure \oldref{4.2}\ref{fig:4.2}). 
He starts getting a signal with frequency $f_{-}=0.5f_{0}=1$ due to the receding
Doppler. At $t=1$ (measured in his own time) receives the first signal,
wherein the picture shows that the other twin age was $E+0.5$ by the time
the pulse left Earth. From the Doppler shift, he gets the receding speed,
therefore he calculates that the pulse departed at $t=0.625$ (in his
reference system), when the Earth was in $x=-0.375$. His age at that time
was $E+0.625$, thus follows that the Earth twin ages more slowly. This holds
up to the reception of the fourth pulse in $t=4$, just when he receives the
other twin photograph with an age of $E+2$ emitted in $t=2.5$ when the Earth
was in $x=-1.5$ and his own age was $E+2.5$. The traveler interprets that at
that moment the Earth twin was half year younger than himself. From that
moment the return trip begins, therefore, he changes the reference system:
from the system going away from Earth to a system that approaches Earth. He
agrees, with observers in the new system, in calling $t=4$ that instant of
time (this is a mere choice). Also agrees with the new colleagues that the
last received picture shows the Earth twin with age $E+2$. However, on the
new system, that image was emitted in $t=-2$ when the Earth was in $x=-6$,
always according to the simultaneity of the new system. So the traveler must
accept that the image that shows the Earth twin with an age $E+2$, that
according to the old system happened when he was aged $E+2.5$ (i.e. his
Earth twin was younger at that time), corresponds to an older (not younger)
twin on Earth at the same moment! Because, according to the new system, the
picture was emitted when he was $E-2$ (before departure!) instead of the
previously calculated age of $E+2.5$. By only changing the system he must
accept that the Earth twin suddenly aged 4.5 years more. During the return
trip, the Earth twin is aging more slowly but he can not recover the
``sudden'' aging of 4.5 years, so when they
meet again, the traveling twin will be younger than the Earth twin, in
agreement with the calculation performed on Earth.

That is, the traveler must accept that due to the change of system, the same
event on Earth (terrestrial twin aged $E+2$), which in the original system,
when they moved away, was located in $t=2.5$ and $x=-1.5$, now, in the new
system, is located at coordinates $t=-2$ and $x=-6$, which means that in his
own reference system the Earth moved away in space and backwards in time!
(see figure \oldref{4.2}\ref{fig:4.2}). This strange behavior is also observed 
using the ``parallax distance'' \oldref{[R61]}\cite{unruh1981parallax}.

Another way to see this problem is considering that during the first half of
the trip ($t=4$) he receives 4 uniformly distributed signals (receding
Doppler $f_{-}=1$), while during the second half he receives 16 uniformly
distributed signals (approaching Doppler $f_{+}=4$). There is no trajectory
compatible with this ratio between receding and approaching signals and time
of reception, as can be seen in figure \oldref{4.2}\ref{fig:4.2}.

Note that the outgoing and incoming paths of Earth calculated from the
Doppler signal have no intersection: that is, integrating the Doppler
formula forward in time from the beginning of the trip or backward in time
from the end of the trip (both during inertial stages) gives a non
intersecting trajectory. Therefore the traveling twin cannot use the Doppler
signal to calculate the position of the Earth. It is granted that non-local
effect should be taken into consideration when discussing frequency
measurements performed by accelerating observers or sources\oldref{[R62]}
\cite{rothenstein1998frequency}. This means
that while integrating Doppler signals in a non-inertial frame, one should
be aware of the previous history and some corrections (spatial and temporal
jumps) may be applied. Further, the non-local effects remain during a
subsequent inertial stage.

The case of ``radar distance'' \oldref{[R63]}\cite{dolby2001radar} is
different, since a single valued solution exists for the trajectory. A radar
signal is emitted by one twin and after being passively reflected, it is
received back in a later time. From the information of the time of emission
and the time of reception a radar distance is calculated. The fact that the
solution using passive reflection is different from the solution that use
one way propagation (parallax or Doppler) is another demonstration of the
difference, under SRT, between a moving source and a moving image.

Also, note that the asymmetric aging depends on how the simultaneity in the
new system is defined as compared with the simultaneity in the old system.
If simultaneity were arbitrary then relative age upon reunion would depend
on this choice, which is absurd. This implies that simultaneity is not
relative, contrary to some works \oldref{[R64]}\cite{kassner2012ways}.

These results also bring to mind the following. The Earth in its orbit
around the Sun, in a given epoch matches a reference system that approaches,
say, a group of distant stars, but six months later, will match a system
that moves away from the stars, similar to the change of system that the
traveling twin undergoes. This would imply periodical changes in the
parallax distance and in what is defined as simultaneous. Anyway, since the
variation of the velocity of the Earth is $\Delta v_{Earth}/c\approx 10^{-4}$
it may be unimportant in celestial observations.

\begin{figure}
\begin{center}
\includegraphics[scale=0.5]{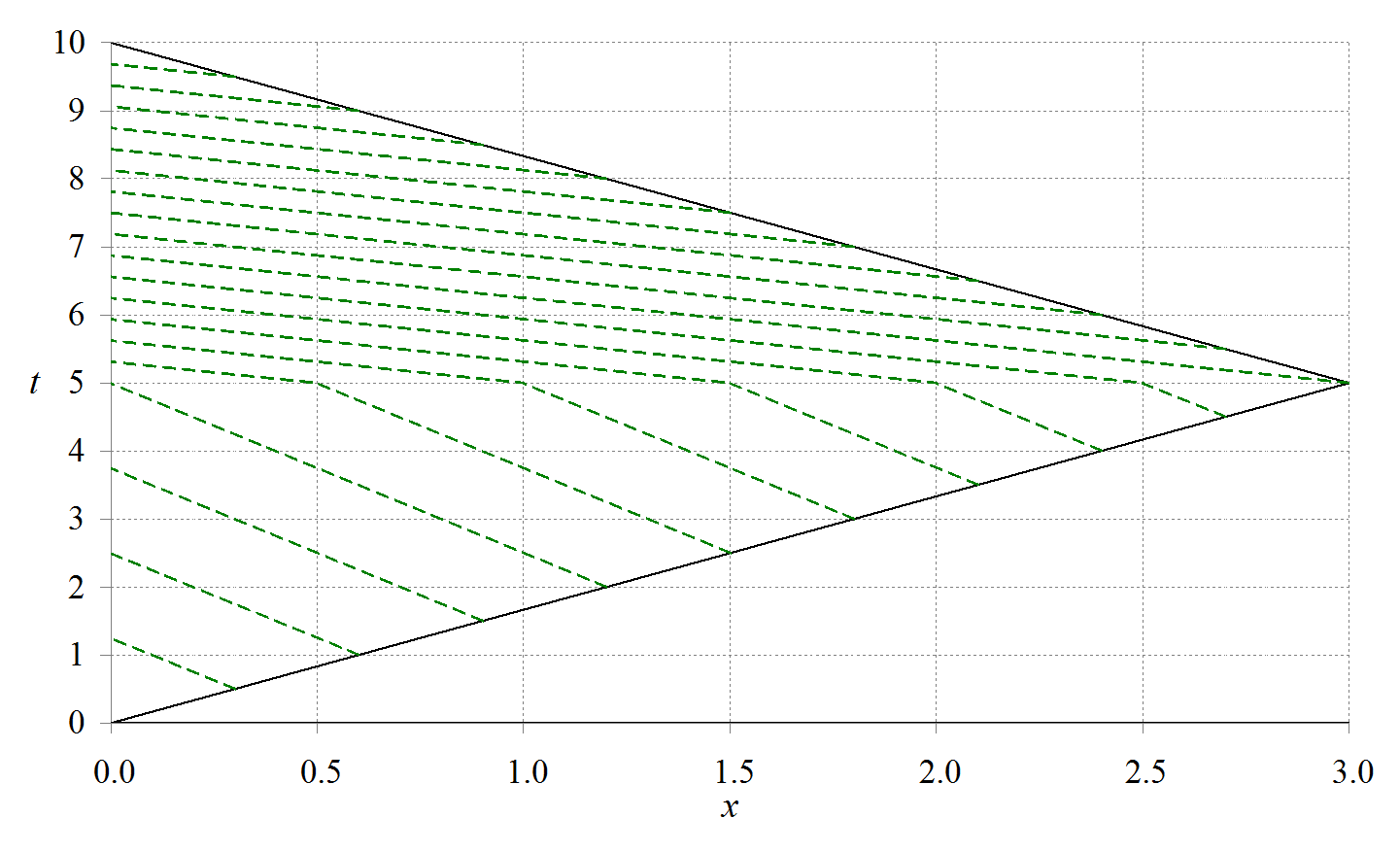}
\caption{Trajectory of a twin (black full lines) and her or his emitted
signals (green dashed lines), according to VRT. From the trajectory of the
signals it is apparent that the speed of the light instantaneously follows
that of the source.}
\label{fig:4.4}
\end{center}
\end{figure}

From VRT point of view no asymmetries are observed. Both twins measure
exactly the same Doppler, same displacement, and same aging. They receive 4
receding Doppler signals ($f_{-}=0.8$) during half of the round trip time
and 16 approaching Doppler signals ($f_{+}=3.2$) during the other half (see
figures \oldref{4.3}\ref{fig:4.4} and \oldref{4.4}\ref{fig:4.4}). 
They exchange a total of 20 signals, pair to pair
emitted and received at the same time. The graphs are identical for both
twins in her or his reference system. The speed of the signals is
instantaneously linked to that of the source, as it is apparent in figure
\oldref{4.4}\ref{fig:4.4}. Anyhow, the speed of the information (the exchanged photography) is not
instantaneous, it moves at $c$ plus the mean speed of the source between
emission and reception. The remarkable aspect of VRT showing the presence of
both, instantaneous and delayed action at a distance will be clarified in
Chapter\oldref{8}\ref{ch:8} where an electrodynamic theory based on VRT is presented.

\begin{figure}
\begin{center}
\includegraphics[scale=0.5]{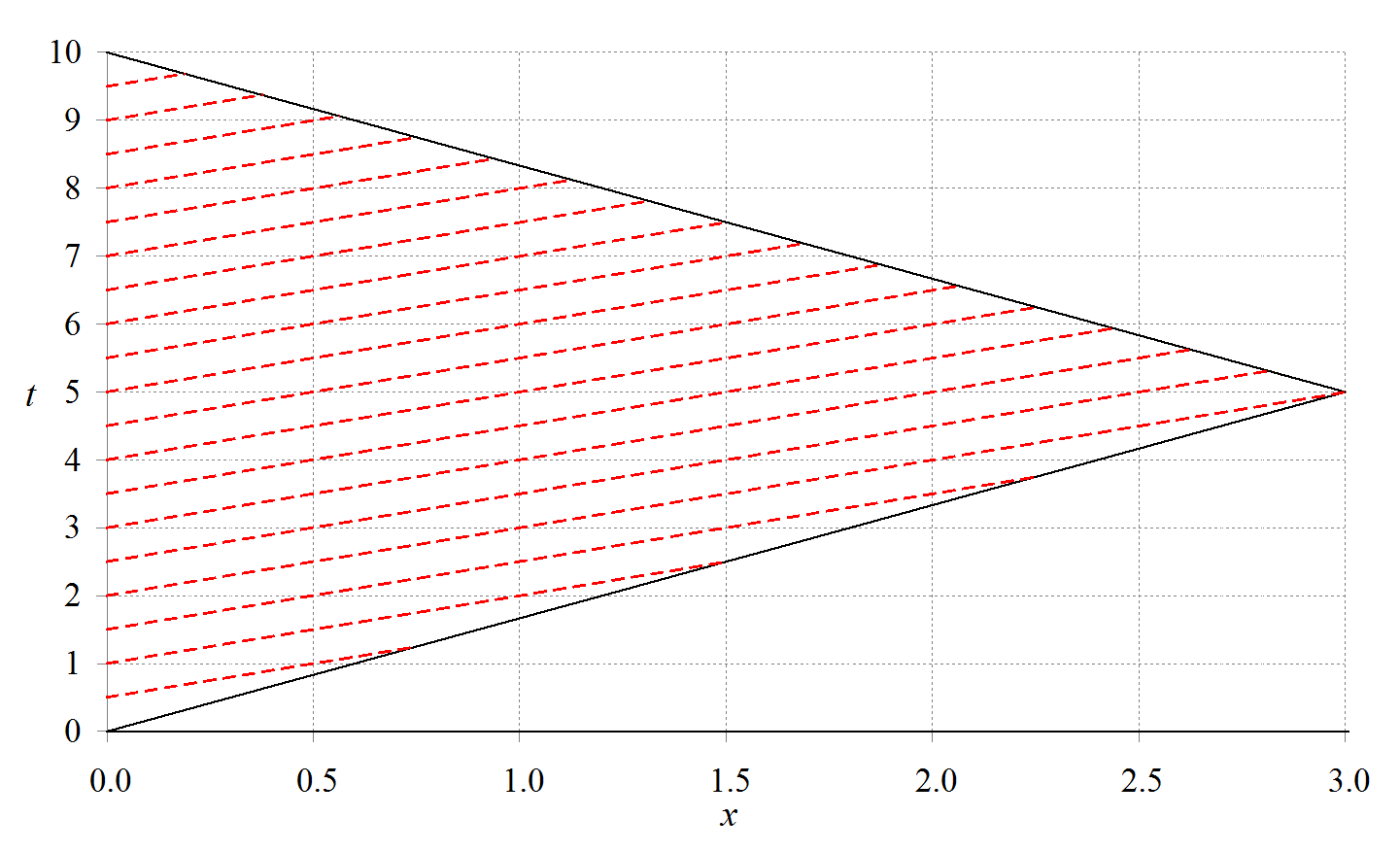}
\caption{Trajectory of a twin (black full lines) calculated by the other twin
from the Doppler signals received from the former (red dashed lines),
according to VRT using (\ref{4.10}). Both twins observe the same.}
\label{fig:4.3}
\end{center}
\end{figure}

\begin{figure}
\begin{center}
\includegraphics[scale=0.5]{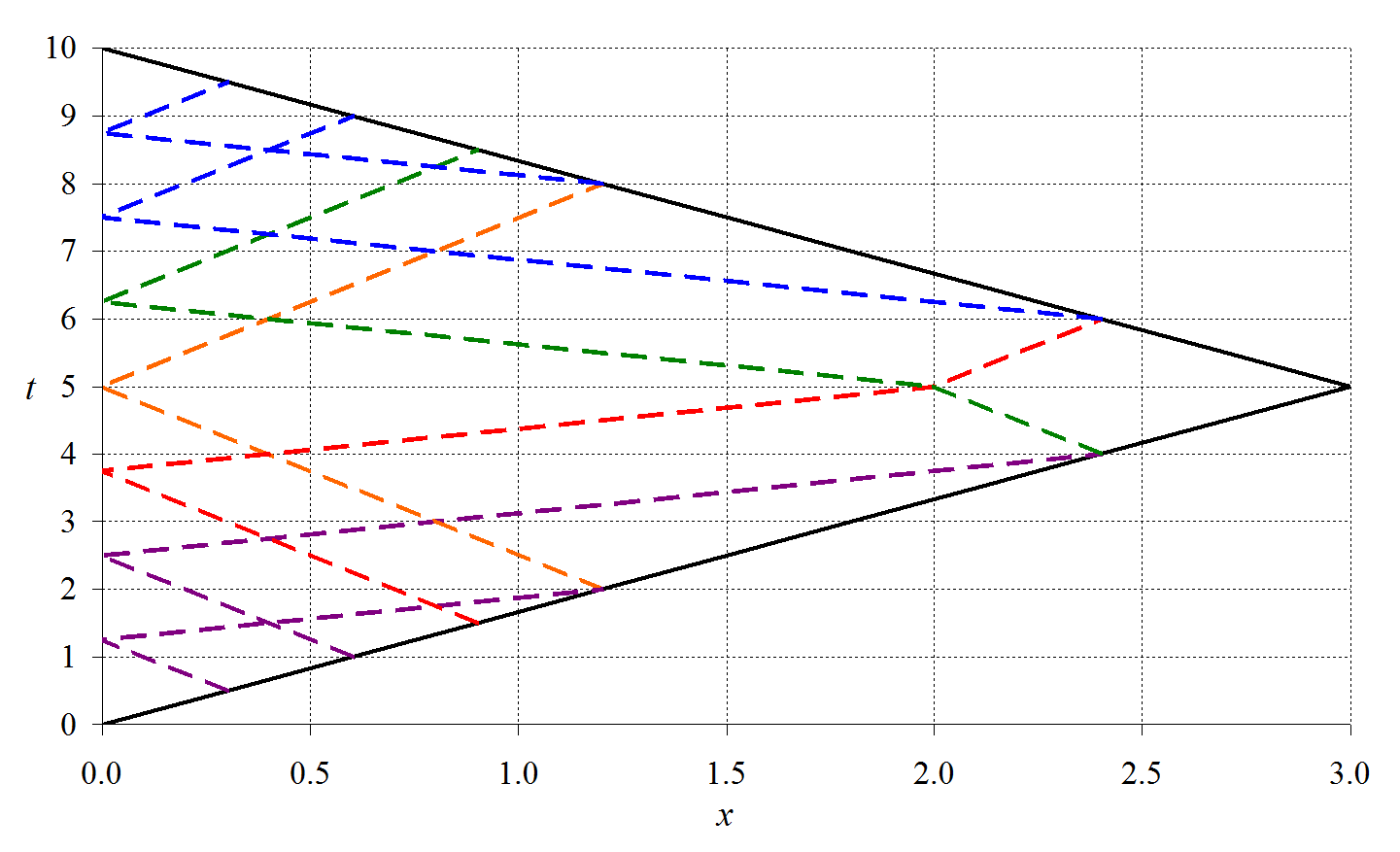}
\caption{Trajectory of a twin (black full lines) and some of her or his
emitted radar signals (dashed lines) used for measuring the distance of the
other twin, according to VRT. From the trajectory of the signals it is
apparent that the speed of the light instantaneously follows that of the
source.}
\label{fig:4.5}
\end{center}
\end{figure}

Finally, in figure \oldref{4.5}\ref{fig:4.5} we show the radar signal emitted by 
a twin (any of them) as seen in the reference system of the other twin. The 
first and second signals, emitted at $t=0.5$ and $t=1$, respectively (in violet),
travels at $c-v$ before reflection and at $c+v$ after reflection because the
source is going away during the full round trip of the signal. The third
signal, emitted at $t=1.5$ (in red), as the previous one travels at $c-v$
before reflection and at $c+v$ after reflection until $t=5$ when the source
starts to return, thereafter the speed becomes $c-v$. The fourth signal (in
orange), emitted at $t=2$, travels at $c-v$ before reflection (because the
source is going away) and also at $c-v$ after reflection (because the source
is approaching), in both cases the direction of propagation is opposite to
that of the source. The subsequent signals are clearly understood by time
reversal. From this it follows that, under VRT, all distances and times
evaluated from either Dopper, parallax or radar coincide, and correspond to
the actual distance and time.

Considering the instantaneous aspect of the Doppler effect in VRT, a
possible test of this theory could be in the evaluation of the dynamics of
supernovae as inferred from visual observations and, simultaneously, from
Doppler measurements. According to VRT bulk velocities determined from
Doppler signals should be advanced in time relative to those visually
observed. Indications of this time differences are present in the
obsevations of supernova SN 1987A \oldref{[R65]}\cite{dewey2008hetg}. 
Precise observations of this kind of
phenomena can shed light on certain aspects of VRT, like a possible upper
limit of the velocity of propagation of the ``instantaneous'' Doppler effect 
from accelerated sources.


%% file: 2014arxiv_v14_2_5.tex


\chapter{Sagnac experiment and time in a rotating frame}\label{ch:5}

\fancyhf{}

\fancyhead[LE,RO]{5. Sagnac experiment}
\fancyhead[RE,LO]{L. Bilbao}
\fancyfoot[RE,LO]{Vibrating Rays Theory}
\fancyfoot[LE,RO]{\thepage}
 
\renewcommand{\headrulewidth}{1pt}
\renewcommand{\footrulewidth}{1pt}

\section{Sagnac experiment}\label{sec:5.1}

In all the Sagnac-type experiments two beams of light travel in opposite
directions about a closed path on a turntable. When the turntable rotates a
fringe shift is observed which is directly proportional to the angular
velocity\oldref{[R66]}\oldref{[R67]}\oldref{[R68]}
\cite{sagnac1913comptes,sagnac1913comptesb,post1967sagnac}. 
The Michelson-Gale experiment \oldref{[R69]}\oldref{[R70]}
\cite{michelson1925effect1,michelson1925effect2} is another demonstration 
of the Sagnac effect.

Actually, the time difference is not the result of acceleration (or of
enclosed area) but only of velocity and length as has been demonstrated with
the fiber optic conveyor Sagnac experiment, WZYL's experiment
\oldref{[R71]}\cite{wang2003modified}. Tartaglia and Ruggiero
\oldref{[R72]}\cite{tartaglia2004sagnac} and Wucknitz
\oldref{[R73]}\cite{wucknitz2004sagnac} come to the conclusion that
acceleration is not the prime reason for the Sagnac effect. According to
Wucknitz an equivalent situation to the conveyor experiment is to replace
the fiber by mirrors, since everything the fiber does, is to guide the light
around the wheels.

WZYL's experiment actually shows that what matters is that light moves along
a closed circuit and that the observer is in motion with respect to the
circuit. The observer needs not rotate for the effect to appear. It is
evident that no connection exists with rotations and with the area enclosed
in the path of light. What matters are the length of the contour and the
speed of the observer.

The observed shift Z of the Sagnac experiment exhibits the following
features:
\begin{itemize}
\item[(a)] obeys formula
\begin{equation}
\Delta Z=\frac{4\mathbf{\Omega }\cdot \mathbf{A}}{c\lambda }=\frac{2}{c\lambda }\int \mathbf{v}\cdot d\mathbf{r}  \label{5.1}
\end{equation}
where $\mathbf{A}$ is the enclosed area, $\mathbf{\Omega }$ the rotation
angular frequency, $\lambda $ the wavelength, $\mathbf{v}$ the local speed
of the circuit and $\mathbf{r}$ a coordinate along the circuit,
\item[(b)] does not depend on the shape of surface area $\mathbf{A}$,
\item[(c)] does not depend on the location of the center of rotation,
\item[(d)] does not depend on the presence of a commoving refracting medium
in the path of the beams.
\end{itemize}

An alternative expression to \oldref{(5.1)}(\ref{5.1}) is given by the 
time lag $\Delta t$
\begin{equation}
\Delta t=\frac{\lambda }{c}\Delta Z  \label{5.2}
\end{equation}

According to VRT it is necessary to solve the problem in a non-rotating
reference system attached to the source. In other words, Faraday's
``vibrating rays'' do not rotate with the source in a similar way that a 
magnetic field lines do not rotate with the rotation of a magnet
\oldref{[R74]}\cite{barnett1912electromagnetic}. 
The following example shows how to use this principle.

\begin{figure}[h]
\centering
\begin{subfigure}{.5\textwidth}
  \centering
  \includegraphics[width=.8\textwidth]{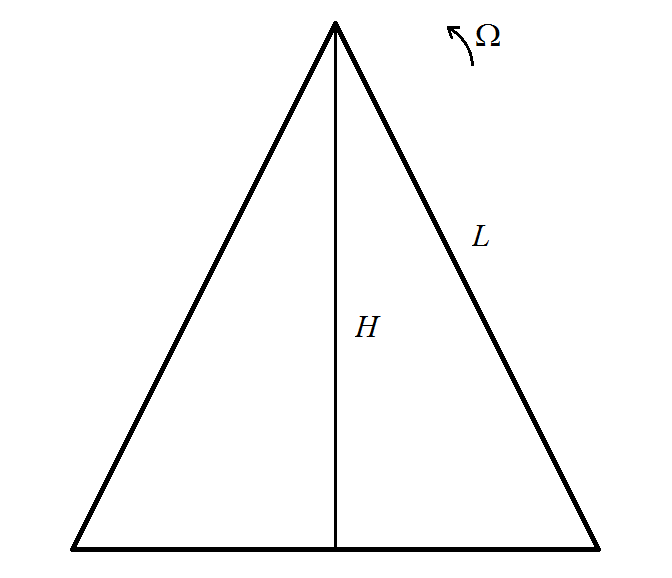}
  \caption{}
  \label{fig:5.1.a}
\end{subfigure}%
\begin{subfigure}{.5\textwidth}
  \centering
  \includegraphics[width=.8\textwidth]{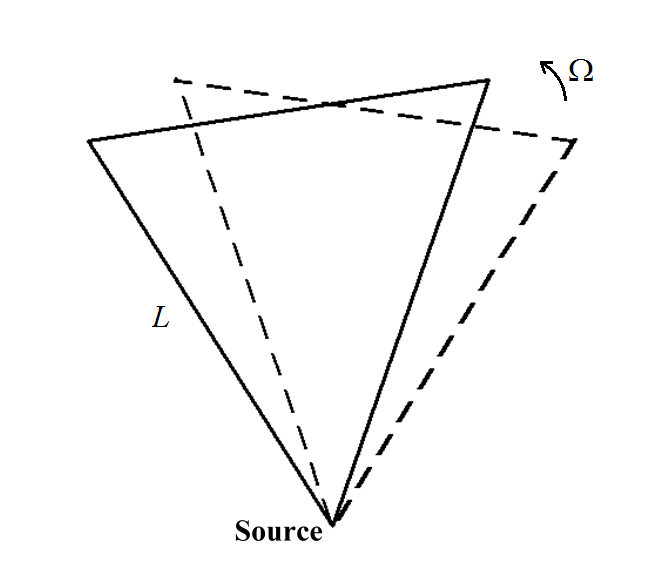}
  \caption{}
  \label{fig:5.1.b}
\end{subfigure}
\caption{(a) Rotating equilateral triangle, (b) the triangle at two
different epochs (dashed and full line) as seen from a non-rotating
reference system attached to the source.}
\label{fig:5.1}
\end{figure}

Consider a triangular interferometer, formed by an equilateral triangle of
side $L$, area $A$ (figure \oldref{5.1a}\ref{fig:5.1.a}). Suppose that 
the source is in one of the vertices and the interferometer rotates with 
angular velocity $\Omega $ counterclockwise. It is assumed that light 
travels in vacuum. As seen from the non rotating system fixed to the 
source, the rest of the interferometer rotates around the source 
(figure \oldref{5.1b}\ref{fig:5.1.b}). Later we will see the case in 
which there is a transparent medium in the interferometer.

The vertices move with speed $\Omega L$ perpendicular to the line that
joints the source and the vertices. The speed is considered constant during
the time it takes the light to perform a turn around the interferometer. For
a signal that travels counterclockwise, the total transit time is
\begin{equation}
t_{1}=\frac{2L}{c}+\frac{L}{c-\Omega H}  \label{5.3}
\end{equation}
while for the clockwise signal it is
\begin{equation}
t_{2}=\frac{2L}{c}+\frac{L}{c+\Omega H}  \label{5.4}
\end{equation}

\begin{SCfigure}[0.7][h]
\caption{Counter-clockwise (red full line) and clockwise (blue full line)
signals, as seen from a non-rotating reference system attached to the
source. The time lag is simply due to the geometrical path difference
between red and blue paths.}
\includegraphics[scale=0.4]{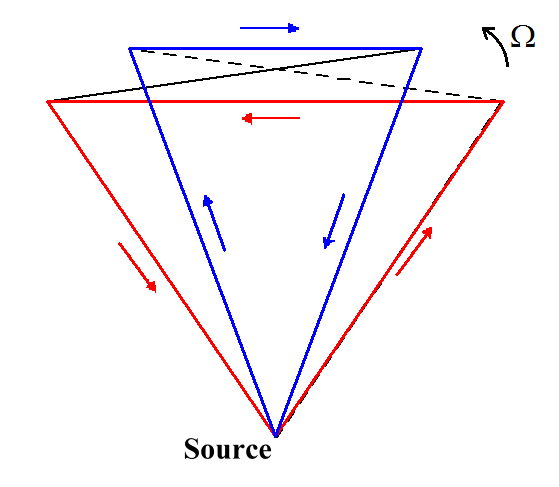}
\label{fig:5.2}
\end{SCfigure}

Therefore, to the first order in $\Omega H/c$ the total time lag is
\begin{equation}
\Delta t=\frac{2L\Omega H}{c^{2}}=\frac{4A\Omega }{c^{2}}  \label{5.5}
\end{equation}
as expected. Under VRT the Sagnac time lag is simply interpreted as the
difference in path between the counter-clockwise and clockwise pulses, as
depicted in figure \oldref{5.2}\ref{fig:5.2}.

More detailed Sagnac-type interferometers include the cases where the light
travels through a refracting, non dispersive medium \oldref{[R68]}\cite{post1967sagnac} 
for which there are different possibilities, those are:
Medium and interferometer rotating (as mentioned above)
\begin{equation}
\Delta Z=\frac{2}{c\lambda }\int \mathbf{v}\cdot d\mathbf{r}  \label{5.6}
\end{equation}
Medium at rest, interferometer rotating
\begin{equation}
\Delta Z=\frac{2}{c\lambda }\int n^{2}\mathbf{v}\cdot d\mathbf{r}
\label{5.7}
\end{equation}
Medium rotating, interferometer at rest
\begin{equation}
\Delta Z=\frac{2}{c\lambda }\int \left( n^{2}-1\right) \mathbf{v}\cdot d
\mathbf{r}  \label{5.8}
\end{equation}
where $n$ is the index of refraction of the medium.

Accordingly VRT predicts the same results. Using the example of the triangle
shown above, for a commoving transparent medium one gets
\begin{equation}
t_{1}=\frac{2nL}{c}+\frac{L}{\frac{c}{n}-\frac{\Omega H}{n^{2}}}  \label{5.9}
\end{equation}
and
\begin{equation}
t_{2}=\frac{2nL}{c}+\frac{L}{\frac{c}{n}+\frac{\Omega H}{n^{2}}}
\label{5.10}
\end{equation}
From these equations the time lag results
\begin{equation}
\Delta t=\frac{2L\Omega H}{c^{2}}=\frac{4A\Omega }{c^{2}}  \label{5.11}
\end{equation}
in agreement with \oldref{(5.6)}(\ref{5.6}).

For the case medium at rest, interferometer rotating we get the following.
Counterclockwise pulse
\begin{equation}
t_{1}=\frac{3nL}{c}+\frac{n^{2}L\Omega H}{c^{2}}+O\left( \Omega ^{2}\right)
\label{5.12}
\end{equation}
Clockwise pulse
\begin{equation}
t_{1}=\frac{3nL}{c}-\frac{n^{2}L\Omega H}{c^{2}}+O\left( \Omega ^{2}\right)
\label{5.13}
\end{equation}
Time lag
\begin{equation}
\Delta t=\frac{2n^{2}L\Omega H}{c^{2}}=\frac{4n^{2}A\Omega }{c^{2}}
\label{5.14}
\end{equation}
identical to \oldref{(5.7)}(\ref{5.7}).

Finally, medium rotating, interferometer at rest, counterclockwise pulse
\begin{equation}
t_{1}=\frac{3L}{\frac{c}{n}+\frac{\Omega H}{3}\left( 1-\frac{1}{n^{2}}\right) }  \label{5.15}
\end{equation}
clockwise pulse
\begin{equation}
t_{2}=\frac{3L}{\frac{c}{n}-\frac{\Omega H}{3}\left( 1-\frac{1}{n^{2}}\right) }  \label{5.16}
\end{equation}
and the time lag
\begin{equation}
\Delta t=\frac{2\left( n^{2}-1\right) L\Omega H}{c^{2}}=\frac{4\left(
n^{2}-1\right) A\Omega }{c^{2}}  \label{5.17}
\end{equation}
in agreement with \oldref{(5.8)}(\ref{5.8}).

\section{Time in rotating frames}\label{sec:5.2}

According to VRT in a Sagnac experiment the time lag is not uniformly
distributed around the closed path. No time lag occurs in the two arms in
which one of the vertices is the source. The largest time lag occurs in the
arm furthest from the source (a similar conclusion is found in
\oldref{[R75]}\cite{moon1991sagnac}). 
Other branches have an intermediate time lag. On the contrary, according to SRT
the time lag only appears when considering the global circuit. In each arm
no delay will be measured for the speed of light is locally isotropic (see,
for example, \oldref{[R73]}\cite{wucknitz2004sagnac}).

\begin{SCfigure}[0.8][h]
\caption{Sector $OBC$ on the rotating disk.}
\includegraphics[scale=0.4]{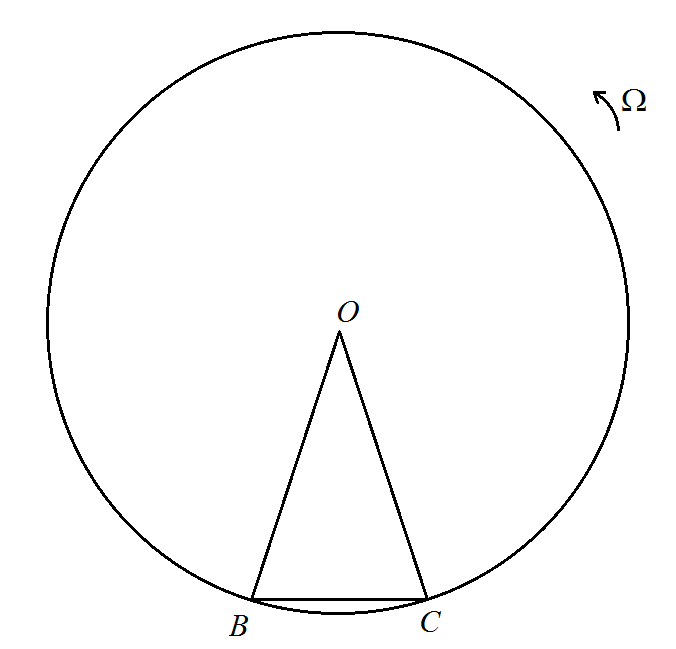}
\label{fig:5.3}
\end{SCfigure}

Consider the case when light propagating on a rotating disk does not travel
around the whole circumference, but rather only around a part of it. To
simplify, we consider this part to be a sector $OBC$ (figure \oldref{5.3}\ref{fig:5.3}).

Two light pulses start simultaneously from $O$, one directed to $C$, then to
$B$ and back to $O$, while the other headed to $B$, then to $C$ and finally
toward the center $O$. It is clear that both pulses will arrive at $O$ at
different instants. There is a physical time lag, it does not depend on
clock synchronization. The question is where the time lag is \textit{produced}. 
Note that this is different from asking where the time lag is \textit{measured}, 
since measurement instruments may add extra time lag (i.e. through the 
measurement cables), therefore hiding the location of the time lag.

In order to find out where the time lag is produced one should compare both
paths stage by stage. Note that, according to SRT, the time lag will be the
same regardless of the source location, whether at $O$, $B$ or $C$ the
results will be identical. Each pulse includes a radial motion towards the
border, then a circumferential motion, and, finally a radial motion toward
the center. Any time lag produced during the radial motion (i.e. different
time between the movement outward and toward the center) must be equal for
both pulses since all radial directions are physically identical. Therefore,
it follows that the time lag is produced during the circumferential motion.
At least, the observation of the process by a non-rotating observer shows
that the time lag emerges in the motion along the peripheral path. But, in a
system fixed to the disk the local speed of light is the same in all
directions (experiments on Earth shows the isotropy of space in a rotating
system \oldref{[R76]}\cite{brillet1979improved}), then the circumferential motion 
of the two pulses from $B$ to $
C$ and from $C$ to $B$, respectively, cannot produce a time lag, in
contradiction with the previous statement. Thus, according to SRT, and
taking into account the Sagnac time lag, and the local isotropy of the speed
of light a contradiction is apparent (Selleri's paradox\oldref{[R77]}\oldref{[R78]}
\cite{selleri1997noninvariant,selleri2004sagnac} is another way to 
show the contradiction).

A colleague of mine gave me the following explanation on the above
contradiction: ``\textit{The two pulses departing from }$O$\textit{\ will
arrive at different moments back to }$O$\textit{. And neither of the stages
(circumferential or radial) when carried out separately, has been found
responsible for this! On the face of it, it seems that the circumferential
motion must be the culprit. At least, the observation of the process by the
Lab observer shows that the time delay emerges in the motion along the
peripheral path. But, there will be no delay by the synchronized clocks of
the disk. With all that, however, two different radial lines, as was shown
above, do not contribute to time delay of the signals converging to (or
diverging from) the center. Thus, there is no time discrepancy along any
path that does not form a closed loop. Only by comparing both arrivals at
the same detector, and only if the paths form a loop enclosing a finite
area, and only when each pulse keeps moving one way along its respective
path, will there emerge a time delay. Thus, an observation of two objects,
starting simultaneously from one place on a spinning disk, moving with equal
speed along two different paths of equal length, and arriving at another
place, reveals one of the most weird results of Relativity: there is no such
thing as one single time for all space in a rotating frame.}''

I can hardly agree with the above explanation (reducing the contradiction to
a problem of time definition) since the time lag is a physical phenomenon,
thus it should be produced somewhere in the light path. Although difficult
to measure, for the measurement apparatus is also rotating (at least in
part) and thus may add some extra time lag, the time lag should be present
in one or more legs of the interferometer that does not form a closed loop.

VRT does not suffer from the above contradiction. Since the speed of
propagation depends on the source movement, radial directions are not all
equivalent, unless the source is in the center of rotation, as in the
previous example. If the source were placed in $B$, then the time lag would
be produced in the radial motion as the difference between $OC$ and $CO$
movements. If the source were placed in $C$ the time lag would be produced
in the radial motion as the difference between $OB$ and $BO$ movements.
Finally, when the source is at $O$ there is no time lag in the radial
stages; the time lag is produced in the circumferential motion.

The final question is why the circumferential time lag is acceptable under
VRT but not under SRT, since local isotropy has been measured. The point
here is that local isotropy in rotating platform has been always measured
using \textit{local} sources, that is, with the source located in the same
(or close enough) arm where the speed of light is under test (see, for
example, \oldref{[R79]}\cite{antonini2005test}). Therefore, no time 
lag is expected in that arm. This fact
suggest a way to test VRT against SRT: the idea is to use two sources, one
at $O$ and the other at $B$ (or $C$). For the light produced at $O$, the
circumferential speed of light will be anisotropic, while for the light
produced at $B$ (or $C$) it will be isotropic, the difference between them
will be independent of the synchronization procedure, for there is no
synchronization procedure that can eliminate simultaneously the measured
time lag for both beams. In Chapter \oldref{9}\ref{ch:9} we will show the 
setup and results of an experiment based on this idea.


%% file: 2014arxiv_v14_2_6.tex


\chapter{Spacecraft anomalies}\label{ch:6}

\fancyhf{}

\fancyhead[LE,RO]{6. Spacecraft anomalies}
\fancyhead[RE,LO]{L. Bilbao}
\fancyfoot[RE,LO]{Vibrating Rays Theory}
\fancyfoot[LE,RO]{\thepage}
 
\renewcommand{\headrulewidth}{1pt}
\renewcommand{\footrulewidth}{1pt}

\section{Pioneer anomaly}\label{sec:6.1}

\subsection{Possible explanation}\label{sub:6.1.1}

The Pioneer anomaly refers to the fact that the received Doppler frequency
differs from the modeled one by a blue shift that varies almost linearly
with time, and whose derivative is
\begin{equation}
\frac{d\left( \Delta f\right) }{dt}\approx -6\times 10^{-9}\ \text{Hz/s}
\label{6.1}
\end{equation}
where $\Delta f$ is the frequency difference between the measured and the
modeled values.

Since the spacecraft flies away from Earth, there is a Doppler shift towards
red. However, the measured redshift value is smaller than the expected one.
This is equivalent to say that the spacecraft has a lower speed (in module)
than it would at that position of the orbit. Precisely because of this
equivalence it can be interpreted that the spacecraft has slowed down, thus
the name anomalous acceleration.

As it was mentioned in Chapter \oldref{1}\ref{ch:1}, the main difference in 
Doppler (to first order) between VRT and SRT is that in the case of a source 
with variable speed, SRT relates to the speed of the source at the time of 
emission, while VRT relates to the speed of the source at the time of reception.

If VRT were valid, it automatically invalidates all calculations and data
analysis of Pioneer which are based on SRT. So, it is not simple to make a
direct comparison between the expected results from SRT and VRT. Further,
there are other sources that may produce an anomaly, for example, thermal
effects. However, we can evaluate the difference mentioned in the preceding
paragraph, at least as an order of magnitude and, in turn, to see whether or
not the main features predicted by VRT are present in the measurements.

The signal from Earth (uplink) is emitted with a frequency $f_{0}\approx 2.3$
GHz. As the ship receives the signal, it sends another signal to Earth
(downlink), which is in phase with the received one. A simplified model is
to assume that the emitter is at rest in the solar barycenter system and the
spacecraft initially moves in the gravitational field of the Sun, i.e., its
velocity decreases with time. We also assume that at large distance the
position unit vectors do not change appreciably.

Calling $t_{2}$ the emission time of the downlink signal from the spacecraft
toward Earth and $t_{3}$ the reception time at Earth, the first order
difference of the Doppler shift between VRT active reflection and SRT is
(see Chapter \oldref{4}\ref{ch:4})
\begin{equation}
f_{VRT\text{active}}-f_{SRT}\approx f_{0}\ \mathbf{\hat{r}}\cdot \frac{\mathbf{v}_{2}-\mathbf{v}_{3}}{c}  \label{6.2}
\end{equation}

That is, the velocity used in SRT formula is that at the time of emission
while according to VRT is the corresponding at the time of reception.

Since the spacecraft slows down as it moves away, then $\hat{\mathbf{r}}
\cdot \left( \mathbf{v}_{2}-\mathbf{v}_{3}\right) >0$, therefore the
difference corresponds to a small blue shift mounted over the large red
shift as it has been observed in the Pioneer anomaly. It should be noted
that this difference appears because of the active reflection produced by
the onboard transmitter. In case of a passive reflection (for example, by
means of a mirror) the above difference vanishes (see Chapter \oldref{4}\ref{ch:4}).

\subsection{Main term}\label{sub:6.1.2}

Using that the variation of the velocity of the spacecraft between the time
of emission and reception is approximately
\begin{equation}
\mathbf{v}_{2}-\mathbf{v}_{3}\approx \mathbf{a}\left( t_{2}-t_{3}\right)
\label{6.3}
\end{equation}
where $\mathbf{a}$ is a mean acceleration during the downlink interval, we
get
\begin{equation}
\Delta f\approx -f_{0}\,\frac{\mathbf{\hat{r}}\cdot \mathbf{a}}{c^{2}}\left(
t_{3}-t_{2}\right)  \label{6.4}
\end{equation}

An estimate for the duration of the downlink is simply
\begin{equation}
t_{3}-t_{2}\approx \frac{r}{c}  \label{6.5}
\end{equation}
where $r$ is a mean position of the spacecraft between $t_{2}$ and $t_{3}$,
therefore
\begin{equation}
\Delta f\approx -f_{0}\,\frac{\mathbf{r}\cdot \mathbf{a}}{c^{2}}  \label{6.6}
\end{equation}

Since
\begin{equation}
\mathbf{a=-}\frac{GM}{r^{2}}\,\hat{\mathbf{r}}  \label{6.7}
\end{equation}
then, the time derivative becomes
\begin{equation}
\frac{d\left( \Delta f\right) }{dt}\approx f_{0}\frac{\mathbf{v}\cdot
\mathbf{a}}{c^{2}}  \label{6.8}
\end{equation}
where $G$ is the gravitational constant, $M$ the mass of the Sun.

If the difference (\ref{6.8}) (6.8) is interpreted as an anomalous acceleration we get
\begin{equation}
a_{a}\approx \frac{v}{c}\,a  \label{6.9}
\end{equation}

That is, the so-called anomalous acceleration is $v/c$ times the actual
acceleration of the spacecraft.

As a numerical example of a characteristic value consider the anomalous
acceleration detected at the shortest distance of the Cassini spacecraft
during solar conjunction in June, 2002. The spacecraft was at a distance of
7.42 AU moving at a speed of 5.76 km/s. The anomalous acceleration given by
\oldref{(6.9)}(\ref{6.9}) is $a_{a}\approx 2\times 10^{-9}$ m/s$^{2}$ of 
the same order of the measured one ($a_{a}\approx 2.7\times 10^{-9}$ m/s$^{2}$). 
The closest distance at which the Pioneer anomaly has been detected was 
about 20 AU. Using data from HORIZONS Web-Interface \oldref{[R81]}\cite{nasa2014jpl}
the anomalous acceleration predicted by \oldref{(6.9)}(\ref{6.9}) at that distance is 
$a_{a}\approx 7.3\times 10^{-9}$ m/s$^{2}$ of the same order as the measured one.

Note that \oldref{(6.9)}(\ref{6.9}) predicts a decreasing anomaly in contrast
with the measured values where an almost constant value has been obtained. 
As it was mentioned above, there are many reasons for this discrepancy to appear. 
Of course, VRT may be wrong, but assuming its validity the whole modeling of
the spacecraft should be modified. Corrections like \oldref{(6.10)}(\ref{6.10}) 
should also be included in VRT. Also, there are other partial explanations on the
phenomenon like thermal effects \oldref{[R82]}\cite{turyshev2010pioneer} 
which can be as large as one-third of the total effect, or even more.

According to Markwardt \oldref{[R80]}\cite{markwardt2002independent} 
the expected frequency at the receiver includes an additional Doppler effect 
caused by small effective path length changes, given by
\begin{equation}
\Delta f_{path}=-\frac{2f_{0}}{c}\frac{dl}{dt}  \label{6.10}
\end{equation}
where $dl/dt$ is the rate of change of effective photon trajectory path
length along the line of sight. This is a first order effect that can
partially hide the difference between SRT and VRT. Therefore, a more careful
analysis should take into account the additional contribution of \oldref{(6.10)}(\ref{6.10}) in
\oldref{(6.9)}(\ref{6.9}).

Further, other first order effects may appear, for example, by a slight
rotation of the orbital plane. Due to spacecraft maneuvers or random
perturbations the orbital parameters are obtained by periodically fitting
the measurements with theoretical orbits. Therefore there is no
straightforward way to weight the importance of these fittings in
\oldref{(6.9)}(\ref{6.9}). In other words, data acquisition and analysis 
may hide part of the Vibrating Rays Theory signature.

Although the calculated values using VRT with SRT orbits are smaller than
those measured, we cannot forget that it is impossible to make a direct
comparison because the entire analysis of the Pioneer anomaly is done from
the relativistic point of view. Observations in Earth are converted to the
solar system barycenter. This is done in two stages. In the first one,
measurements of clocks on Earth are converted to measurements of a
hypothetical clock in a perfectly circular orbit and in a uniform
gravitational field around the Sun. The time measured by this hypothetical
clock is called TDB (barycentric Dynamical Time). In the second stage, these
data are limited to measurements in the barycentric system of the Sun. The
results after these reductions correspond to having a transmitter and a
receiver in the same frame of reference, but calculated according to SRT
transformations. Therefore, there is no straightforward way to fit the whole
data according to VRT. Probably, the whole data acquisition and analysis
should be reformulated.

\subsection{Annual term}\label{sub:6.1.3}

Apart from the residual referred to in the preceding paragraph there is also
an annual term. According to Anderson et al. \oldref{[R83]}\cite{anderson2002study} 
the problem is due to modeling errors of the parameters that determine 
the spacecraft orientation with respect to the reference system. Anyway, 
Levy et al. \oldref{[R84]}\cite{levy2008pioneer} claims that
errors such as errors in the Earth's ephemeris, the orientation of the
Earth's spin axis or the station's coordinates are strongly constrained by
other observational methods and it seems difficult to change them enough to
explain the periodic anomaly.

The advantage of studying the annual term over the main term, is that the
former is less sensitive to corrections like the thermal effect or small
path length changes included in (\ref{6.10}) (6.10). None of these terms 
may be caused by the Earth orbital position.

Under VRT model, the annual term is explained with the difference between
the velocity at time of emission and the moment of detection, that depends
on whether the spacecraft is in opposition or in conjunction relative to the
Sun. When the spacecraft is in conjunction light takes longer to get back to
Earth than in opposition. The time difference between emission and reception
will be increased by the time the light takes in crossing the Earth orbit.
Therefore, according to \oldref{(6.4)}(\ref{6.4}) the anomaly will be larger. 
Specifically, taking into account the delay due to the position of Earth in 
its orbit, in opposition \oldref{(6.5)}(\ref{6.5}) should be written as
\begin{equation}
t_{3}-t_{2}\approx \frac{r+R_{orb}}{c}  \label{6.11}
\end{equation}
while in conjunction it would be
\begin{equation}
t_{3}-t_{2}\approx \frac{r-R_{orb}}{c}  \label{6.12}
\end{equation}
where $R_{orb}$ is the mean orbital radius of Earth.

Therefore, an estimate of the magnitude of the amplitude of the 
annual term is
\begin{equation}
\Delta f\approx f_{0} \, \frac{aR_{orb}}{c^{2}}  \label{6.13}
\end{equation}

For the case of Pioneer 10 at 40 AU we get
\begin{equation}
\Delta f\approx 14\text{ mHz}  \label{6.14}
\end{equation}
and at 69 AU
\begin{equation}
\Delta f\approx 4.8\text{ mHz}  \label{6.15}
\end{equation}

\begin{figure}[h]
\begin{center}
\includegraphics[scale=0.5]{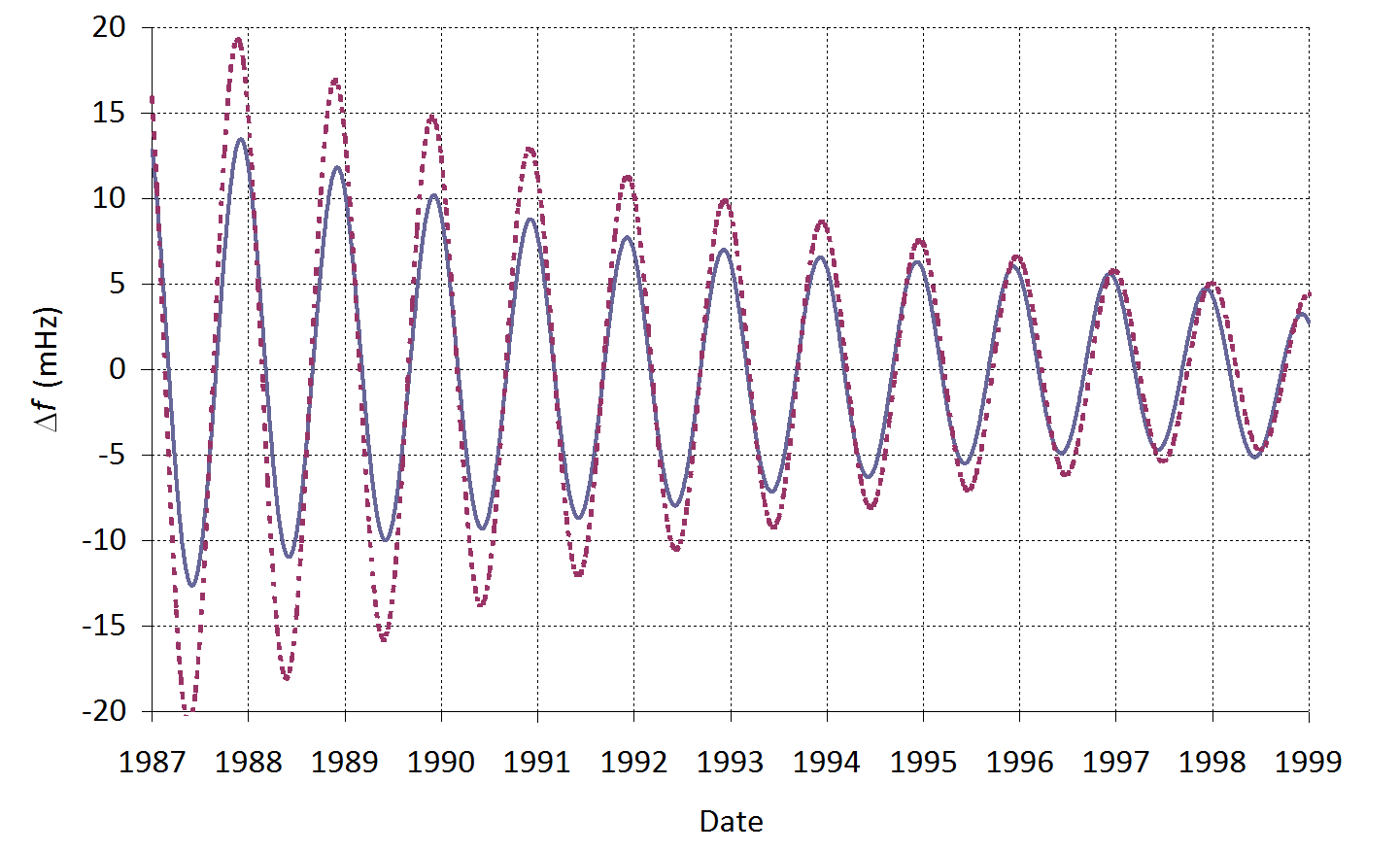}
\caption{Annual variation of the frequency difference between VRT and SRT
(full line) and anomalous dumped sine best fit of the 50 days average
measured by Turyshev et al. \oldref{[R85]}\cite{turyshev1999apparent} 
(dashed line), for Pioneer 10 from January 1987 to January 1999.}
\label{fig:6.1}
\end{center}
\end{figure}

Using data from HORIZONS Web-Interface \oldref{[R81]}\cite{nasa2014jpl} 
the corresponding $\Delta f$ predicted by VRT is plotted in figure
\oldref{6.1}\ref{fig:6.1}. Also the dumped sine best fit of
the 50 days average measured by Turyshev et al.\oldref{[R85]}
\cite{turyshev1999apparent} is plotted showing a good agreement 
between measurements and VRT prediction. The agreement is statistically 
significant, $p<10^{-4}$, $R^{2}=0.879$. The negative peaks
(i.e., maximum anomalous acceleration) occur during conjunction when the
Earth is further apart from the spacecraft, and positive peaks during
opposition. Also, the amplitude is larger at the beginning of the plotted
interval and decreases with time, in good agreement with the observed
behavior \oldref{[R83]}\oldref{[R85]}\cite{anderson2002study,turyshev1999apparent}.

\subsection{Saturn encounter}\label{sub:6.1.4}

At the encounter of Pioneer 11 with Saturn an alteration of the anomaly was
observed. It occurred in a very short time scale of the order of 1 day,
during the time the spacecraft was within the influence of the gravity of
Saturn. Under VRT this is expected since in \oldref{(6.9)}(\ref{6.9})
the actual acceleration must be used. During a planetary encounter the 
acceleration is given by the planet gravity, therefore both the intensity 
and direction change in a short time scale. No data is available in order
to make a detailed numerical evaluation.

\section{Flyby anomaly}\label{sec:6.2}

\subsection{General considerations}\label{sub:6.2.1}

Like the Pioneer anomaly, the Earth flyby anomaly is a real effect inherent
to the tracking of spacecraft. The anomalies are associated to a modeling
problem, in the sense that relativistic Doppler includes terms that are
absent in the measured signals. Therefore, the flyby anomaly should be
present in both Doppler data and range data. The empirical equation of the
flyby anomaly is given by Anderson et al. \oldref{[R6]}\cite{anderson2008anomalous}.
As an example of the possible explanation of the flyby anomaly we will
compare the prediction of Doppler according to SRT and VRT for the case of a
moving source and a moving spacecraft that approach each other up to a close
encounter and then start to separate.

\subsection{Simplified case}\label{sec:6.2.2}

Consider an antenna that moves with a constant velocity $u$ along the axis $
x $, and a spacecraft approaching at speed $v_{a}$ making an angle $\alpha
_{a} $ with the horizontal axis, and flying away with a speed $v_{r}$ making
an angle $\alpha _{r}$ with the horizontal axis. Suppose that the antenna
emits at a frequency $f_{0}$ in its own system. The signal is emitted
towards the spacecraft (uplink) and returns to the antenna after being
actively reflected by the spacecraft circuit (downlink). If $u<<v<<c$,
according to SRT, \oldref{(4.1)}(\ref{4.1}), the frequency received back 
by the antenna during approach is
\begin{equation}
\left. \frac{f_{a}}{f_{0}}\right\vert _{SRT}=1+\frac{2v_{a}}{c}-\frac{2u\cos
\alpha _{a}}{c}+\frac{2v_{a}^{2}}{c^{2}}-\frac{4v_{a}u\cos \alpha _{a}}{c^{2}}+\ldots  \label{6.16}
\end{equation}
while during receding
\begin{equation}
\left. \frac{f_{a}}{f_{0}}\right\vert _{SRT}=1-\frac{2v_{r}}{c}+\frac{2u\cos
\alpha _{r}}{c}+\frac{2v_{r}^{2}}{c^{2}}-\frac{4v_{r}u\cos \alpha _{r}}{c^{2}}+\ldots  \label{6.17}
\end{equation}

Therefore, the expected frequency shift difference between the outgoing and
incoming spacecraft is
\begin{align}
\left.  \frac{\Delta f}{f_{0}}\right\vert _{SRT}  & =1-\frac{2\left(
v_{r}+v_{a}\right)  }{c}+\frac{2u\left(  \cos\alpha_{r}+\cos\alpha_{a}\right)
}{c} \label{6.18}\\
& +\frac{2\left(  v_{r}^{2}-v_{a}^{2}\right)  }{c^{2}}-\frac{4u\left(
v_{r}\cos\alpha_{r}-v_{a}\cos\alpha_{a}\right)  }{c^{2}}+\ldots \nonumber
\label{6.18}
\end{align}

According to VRT, there are two possibilities: passive or active reflection.
In case of passive reflection, from \oldref{(4.4)}(\ref{4.4}) the same expression 
as SRT is obtained, therefore
\begin{equation}
\left. \frac{\Delta f}{f_{0}}\right\vert _{VRT\text{passive}}-\left. \frac{\Delta f}{f_{0}}\right\vert _{SRT}=0+O\left( u^{3},u^{2}v,uv^{2},v^{3}\right)
\label{6.19}
\end{equation}

For active reflection, from \oldref{(4.5)}(\ref{4.5})) both legs (uplink and downlink) 
contribute with identical expressions, this means that the Doppler shift is 
symmetrical in the sense that two moving bodies detect the same shift relative 
to each other. This is the main difference between an active and a passive
reflection. Thus considering the uplink and downlink frequency during
approach we get
\begin{equation}
\left. \frac{f}{f_{0}}\right\vert _{VRT\text{active}}=\left( \frac{c+v_{a}-u\cos \alpha _{a}}{c}\right) ^{2}  \label{6.20}
\end{equation}
and while receding
\begin{equation}
\left. \frac{f}{f_{0}}\right\vert _{VRT\text{active}}=\left( \frac{c-v_{r}+u\cos \alpha _{r}}{c}\right) ^{2}  \label{6.21}
\end{equation}

In the limit $u<<v$ and $v<<c$, the former reduce to
\begin{equation}
\left. \frac{f}{f_{0}}\right\vert _{VRT\text{active}}=1+\frac{2v_{a}}{c}-
\frac{2u\cos \alpha _{a}}{c}+\frac{v_{a}^{2}}{c^{2}}-\frac{2v_{a}u\cos
\alpha _{a}}{c^{2}}  \label{6.22}
\end{equation}
and the latter
\begin{equation}
\left. \frac{f}{f_{0}}\right\vert _{VRT\text{active}}=1-\frac{2v_{r}}{c}+
\frac{2u\cos \alpha _{r}}{c}+\frac{v_{r}^{2}}{c^{2}}-\frac{2v_{r}u\cos
\alpha _{r}}{c^{2}}  \label{6.23}
\end{equation}

Therefore, the expected frequency shift difference between the outgoing and
ingoing signals is
\begin{align}
\left.  \frac{\Delta f}{f_{0}}\right\vert _{VRT\text{active}}  &
=1-\frac{2\left(  v_{r}+v_{a}\right)  }{c}+\frac{2u\left(  \cos\alpha_{r}%
+\cos\alpha_{a}\right)  }{c}\label{6.24}\\
& +\frac{v_{r}^{2}-v_{a}^{2}}{c^{2}}-\frac{2u\left(  v_{r}\cos\alpha_{r}%
-v_{a}\cos\alpha_{a}\right)  }{c^{2}} \nonumber%
\end{align}

Assuming that the ``measured'' value corresponds to VRT (active reflection) 
and the ``modeled'' shift is given by SRT, then the ``anomalous'' jump 
around the point of maximum approach is
\begin{equation}
\left. \frac{\Delta f}{f_{0}}\right\vert _{\text{anomalus}}=\left. \frac{\Delta f}{f_{0}}\right\vert _{VRT\text{active}}-\left. \frac{\Delta f}{f_{0}}
\right\vert _{SRT}=-\frac{v_{r}^{2}-v_{a}^{2}}{c^{2}}+\frac{2u\left(
v_{r}\cos \alpha _{r}-v_{a}\cos \alpha _{a}\right) }{c^{2}}  \label{6.25}
\end{equation}

When the incoming and outgoing speed are equal ($v_{a}=v_{r}=v$) the above
becomes
\begin{equation}
\left. \frac{\Delta f}{f_{0}}\right\vert _{\text{anomalus}}=\left. \frac{\Delta f}{f_{0}}\right\vert _{VRT\text{active}}-\left. \frac{\Delta f}{f_{0}}
\right\vert _{SRT}=\frac{2uv\left( \cos \alpha _{r}-\cos \alpha _{a}\right)
}{c^{2}}  \label{6.26}
\end{equation}

Associating the speed of the antenna with the tangential speed of Earth 
($u=\Omega _{E}R_{E}$) and the incoming (or outgoing) angle with the
declination of the orbit, the empirical formula by Anderson et al.\oldref{[R6]} 
\cite{anderson2008anomalous} is recovered. Note that the ``anomaly''
corresponds to the first non null term in the difference between SRT and
VRT. Curiously, this is the same numerical term as reported by Mbelek\oldref{[R17]} 
\cite{mbelek2008special} starting from the wrong idea that second order 
terms are absent in Anderson et al. \oldref{[R6]}\cite{anderson2008anomalous}.

Note that if passive reflection were used, from \oldref{(6.19)}(\ref{6.19})
follows that no anomaly would be detected.

\subsection{Detailed evaluation of NEAR case}\label{sub:6.2.3}

Consider the case of NEAR. Taking the data from Anderson et al.
\oldref{[R6]}\cite{anderson2008anomalous} and
assuming an ideal hyperbolic orbit it is possible to simulate the Doppler
residual using values from VRT as the measured ones.

\begin{SCfigure}[0.9][h]
\caption{Aerial view of the DSS-63, 210 foot antenna located near Madrid,
Spain, in August 28, 1980 (photo taken by the author).}
\includegraphics[scale=4]{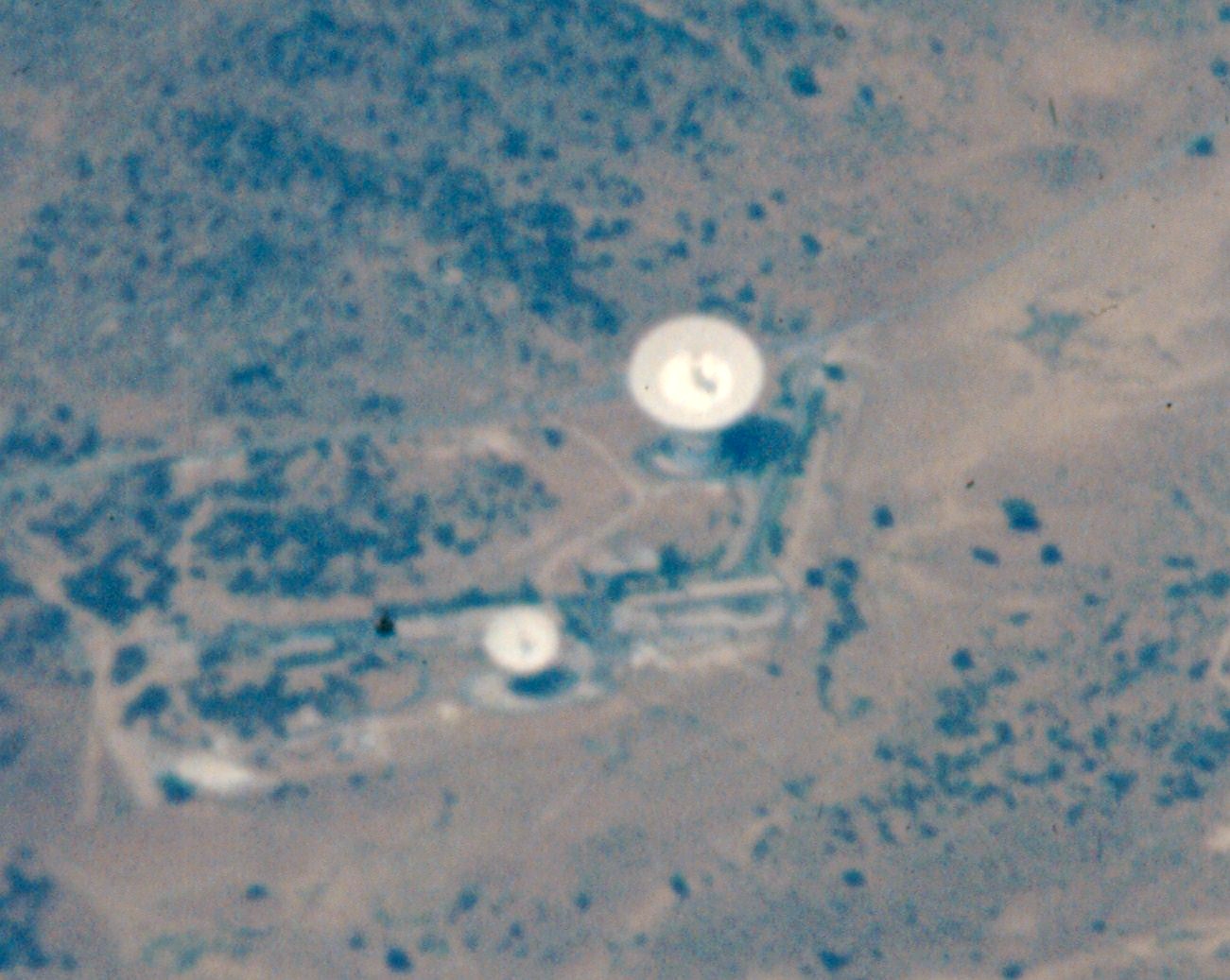}
\label{fig:6.2}
\end{SCfigure}

In order to compare the calculated values using VRT to those modeled from
SRT, a similar procedure to an actual measurement was performed. We have
used the fact that there are 3 antennas located in USA (35${}^{\circ }$ 26'
N, 116${}^{\circ }$ 53' W), Spain (40${}^{\circ }$ 25' N, 4${}^{\circ }$ 15'
W, figure \oldref{6.2}\ref{fig:6.2}) and Australia (35${}^{\circ }$ 24' S, 
148${}^{\circ }$ 59' E), respectively. A full description of the tracking system 
is found in a series of monographs of the Jet Propulsion Laboratory\oldref{[R86]}
\cite{descanso2014dsn}. 

\begin{figure}[h]
\begin{center}
\includegraphics[scale=0.5]{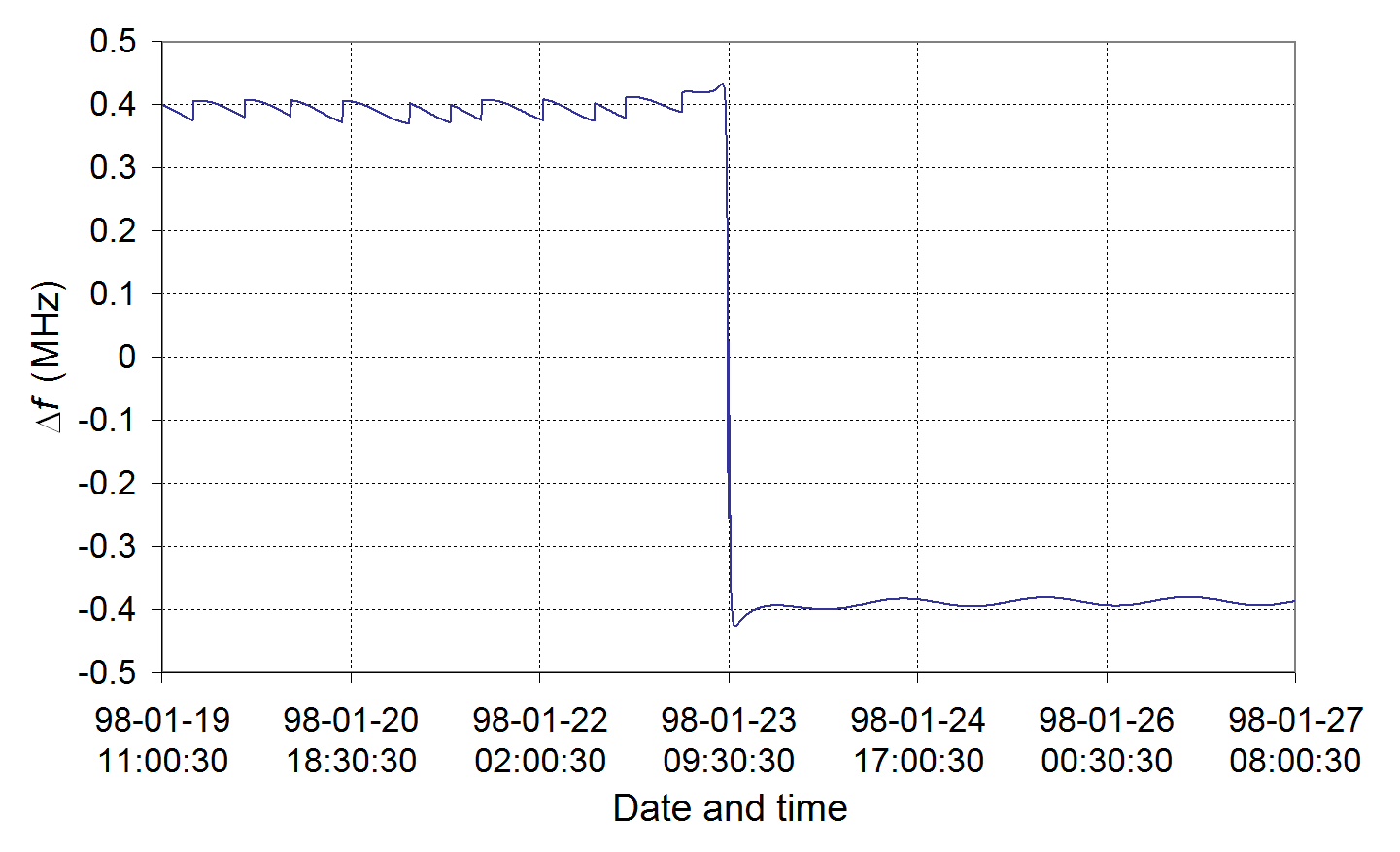}
\caption{Simulated X-band (9.44 GHz) Doppler signal of NEAR spacecraft
under an ideal hyperbolic orbit using VRT. The Earth rotation signature as
well as the switch between antennas are clearly visible.}
\label{fig:6.3}
\end{center}
\end{figure}

The receiving
antenna was chosen as that having a minimum angle between the spacecraft and
the local zenith. The simulated data using VRT are plotted in figure \oldref{6.3}\ref{fig:6.3}.
The switch between receiving antennas is clearly visible during
pre-encounter. After maximum approach the spacecraft remains within the
Canberra cover area, and thus the receiving antenna does not change further.

The simulated residual is obtained by subtracting the SRT Doppler, \oldref{(4.1)}(\ref{4.1}),
from the VRT calculation. We observed, however, that the term that contains
the velocity of the antennas, that is (same notation as in Chapter\oldref{4}\ref{ch:4} is used)
\begin{equation}
d=\frac{\gamma _{u_{3}}}{\gamma _{u_{1}}}\frac{1-\mathbf{\hat{r}}_{23}\cdot
\mathbf{u}_{3}/c}{1-\mathbf{\hat{r}}_{12}\cdot \mathbf{u}_{1}/c}
\label{6.27}
\end{equation}
is not enough to completely remove the \textit{first order} (in $u/c$) Earth
signature. This is so because the velocity of the antennas is not uniform
and the evaluation of the emission time is different for VRT and SRT. Then,
a small, first order related term remains. Anyway, since orbital parameters 
are obtained by periodically fitting the measurements to theoretical orbits, 
thus a similar procedure is needed for VRT. Curiously, by doing so, the first 
order term is removed. The only difference between orbits adjusted by SRT 
and VRT is a slight rotation of the orbit plane, as mentioned above.
Note that in the case of range disagreement (discussed below) two different 
orbital adjustment would be needed by the DSN and the SSN due to the
different propagation speed. In consequence, it will be impossible to fit a 
simultaneous measurement, as it seems to happen with the range disagreement.

After applying the above correction, and subtracting the SRT Doppler, no
first order Earth signature remains in the signal, but second order residual
remains. Each antenna produces a sinusoidal residual with a phase shift at
the moment of maximum approach. Therefore, if we fit the data with the
pre-encounter sinusoid a post-encounter residual remains and vice versa. It
is impossible to reduce the residual to zero by only one fit.

\begin{figure}[h]
\begin{center}
\includegraphics[width=\linewidth]{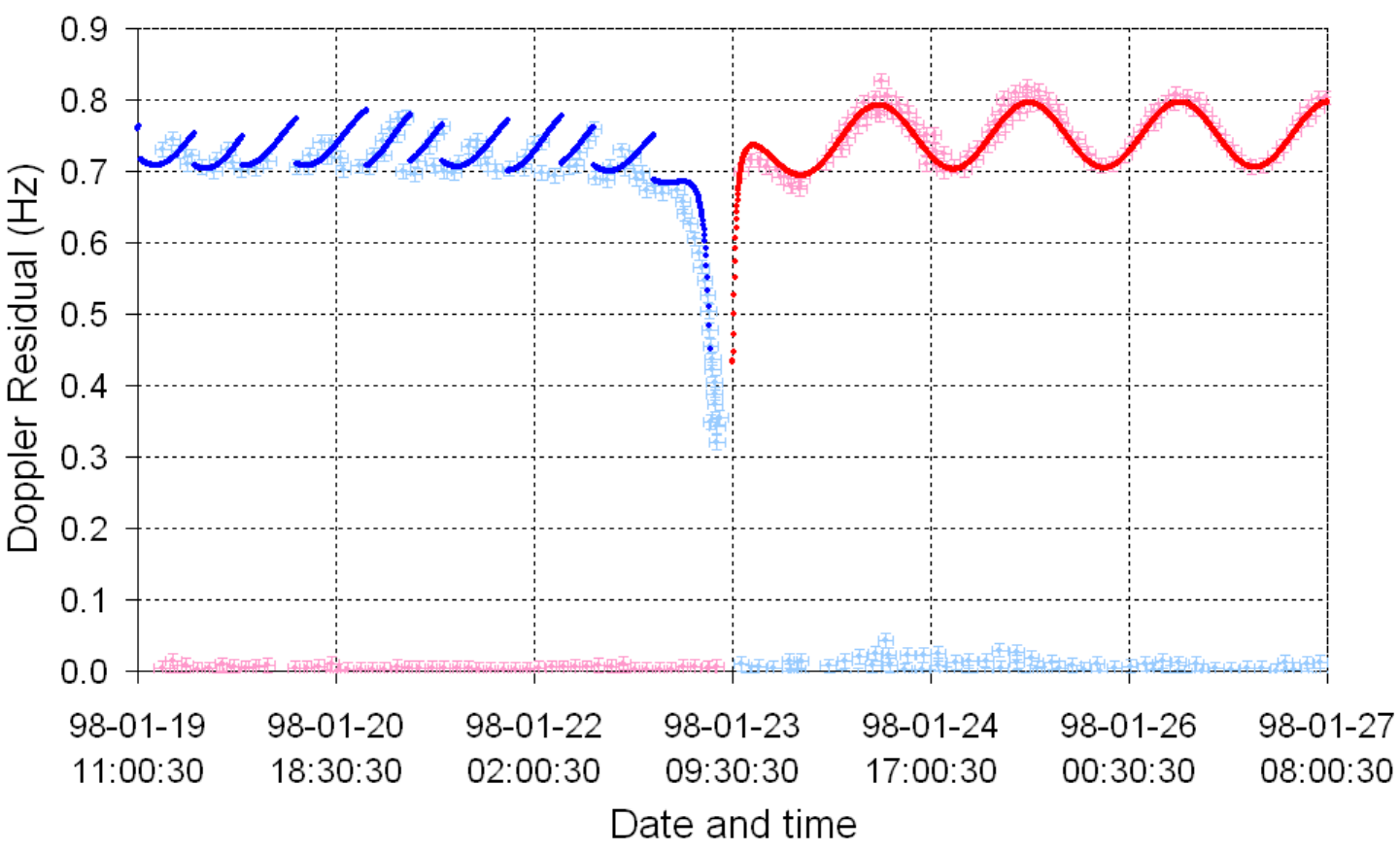}
\caption{Fitting the pre- (right half, in red) and post-encounter (left
half, in blue) simulated X-band Doppler data residual, for the NEAR flyby
under an ideal hyperbolic orbit.}
\label{fig:6.4}
\end{center}
\end{figure}

In figure \oldref{6.4}\ref{fig:6.4} we plot simultaneously the result of 
fitting the residual by pre-encounter data (right half in red, corresponding 
to figure 2a of \oldref{[R6]}\cite{anderson2008anomalous})
and by post-encounter data (left half in blue, corresponding to figure 2b
of \oldref{[R6]}\cite{anderson2008anomalous}).
The adjustment of the experimental data with the VRT model is statistically 
significant, $p<10^{-4}$, $R^{2}=0.896$ for the pre-encounter fitting
and, $p<10^{-4}$, $R^{2}=0.763$ for the post-encounter fitting.

The amplitude and phase (i.e., minima and maxima) of the corresponding 
antenna agree quite well. The post-encounter fitting  (blue) can 
be improved by appropriately setting the exact switching times of the 
antennas (which are unknown to the author). Anyway, the flyby Doppler 
residual exhibits a clean signature of the VRT theory.

Although a small rotation of the orbital plane was used 
to eliminate \textit{first order} effects, the conclusion is that the flyby 
anomaly is mainly related to the \textit{second order} term given by 
\oldref{(6.26)}(\ref{6.26}), The \textit{second order} term that appears 
with different weighting factors in both SRT and VRT formulas seems to 
be overestimated by SRT.

Finally, if the measurement were performed using passive reflection, then
according to VRT, no anomaly should be present. This is also valid for other
measurements involving passive reflection like lunar ranging\oldref{[R87]}
\cite{murphy2008apache}. Therefore,
a possibility for testing VRT is to perform a simultaneous measurement of a
flyby Doppler residual using passive and active reflections. According to
SRT no difference should be measured, while according to VRT passive and
active reflections will produce different results.

\section{Range disagreement.}\label{sec:6.3}

A short track of both Galileo and NEAR flybys was performed by the Space
Surveillance Network (SSN) \oldref{[R7]}\cite{antreasian1998aiaa}, using Millstone 
(42.6$^{\circ }{}$ N, 71.43${}^{\circ }$ W) and Altair (9.18$^{\circ }$ N, 
167.42${}^{\circ }$ E) tracking stations. The NEAR SSN trajectory (passive reflection) 
was found in disagreement with DSN data (active reflection).

We study here with some detail the difference between trajectory
measurements using VRT and SRT theories. For this we consider an emitting or
receiving antenna whose position, relative to the origin of an inertial
coordinate system, is given by $\mathbf{X}_{a}\left( t\right) $, and a space
probe with position determined by $\mathbf{X}_{s}\left( t\right) $. In the
following we denote with a prime the times corresponding to VRT theory, and
without primes those given by SRT. In a two-way measurement $t_{1}$ and $
t_{1}^{\prime }$ refer to the times of emission from the antenna, $t_{2}$
and $t_{2}^{\prime }$ to the times of re-emission (or reflection) from the
probe, and $t_{3}$ to the time of reception by the antenna. The latter is
taken as the reference time, the same for both theories.

According to SRT, times and positions are related by the expressions
\begin{equation}
c\left( t_{3}-t_{2}\right) =\left\vert \mathbf{X}_{a}\left( t_{3}\right) -
\mathbf{X}_{s}\left( t_{2}\right) \right\vert \equiv \left\vert \mathbf{R}
_{23}\right\vert =R_{23}  \label{6.28}
\end{equation}
\begin{equation}
c\left( t_{2}-t_{1}\right) =\left\vert \mathbf{X}_{s}\left( t_{2}\right) -
\mathbf{X}_{a}\left( t_{1}\right) \right\vert \equiv \left\vert \mathbf{R}
_{12}\right\vert =R_{12}  \label{6.29}
\end{equation}

For VRT the corresponding expressions for \textit{active} probe re-emission
are
\begin{equation}
c\left( t_{3}-t_{2a}^{\prime }\right) =\left\vert \mathbf{X}_{a}\left(
t_{3}\right) -\mathbf{X}_{s}\left( t_{3}\right) \right\vert   \label{6.30}
\end{equation}
\begin{equation}
c\left( t_{2a}^{\prime }-t_{1a}^{\prime }\right) =\left\vert \mathbf{X}
_{s}\left( t_{2a}^{\prime }\right) -\mathbf{X}_{a}\left( t_{2a}^{\prime
}\right) \right\vert   \label{6.31}
\end{equation}

Denoting the time difference between theories by $\delta
t_{1a}=t_{1a}^{\prime }-t_{1}$ and $\delta t_{2a}=t_{2a}^{\prime }-t_{2}$,
we can write \oldref{(6.30)}(\ref{6.30}) as
\[
c\left( t_{3}-t_{2}-\delta t_{2a}\right) =\left\vert \mathbf{X}_{a}\left(
t_{3}\right) -\mathbf{X}_{s}\left( t_{2}\right) +\mathbf{X}_{s}\left(
t_{2}\right) -\mathbf{X}_{s}\left( t_{3}\right) \right\vert
\]
which is rewritten, using \oldref{(6.28)}(\ref{6.28}), as
\begin{equation}
R_{23}-c\delta t_{2a}=\left\vert \mathbf{R}_{23}-\mathbf{v}
_{23}R_{23}/c\right\vert   \label{6.32}
\end{equation}
where $\mathbf{v}_{23}$ is the mean velocity of the probe between times $
t_{2}$ and $t_{3}$.

Analogously, for \oldref{(6.31)}(\ref{6.31}),
\[
R_{12}+c\left( \delta t_{2a}-\delta t_{1a}\right) =\left\vert \mathbf{R}
_{12}-\mathbf{u}_{12}R_{12}/c+\mathbf{X}_{s}\left( t_{2a}^{\prime }\right) -
\mathbf{X}_{s}\left( t_{2}\right) -\mathbf{X}_{a}\left( t_{2a}^{\prime
}\right) +\mathbf{X}_{a}\left( t_{2}\right) \right\vert
\]
where $\mathbf{u}_{12}$ is the mean velocity of the antenna between $t_{1}$
and $t_{2}$.

Assuming that during $\delta t_{2}$ the velocity of the probe, $\mathbf{v}
_{2}$, and that of the antenna, $\mathbf{u}_{2}$, have a negligible change,
the last expression can be written as
\begin{equation}
R_{12}+c\left( \delta t_{2a}-\delta t_{1a}\right) =\left\vert \mathbf{R}
_{12}-\mathbf{u}_{12}R_{12}/c+\left( \mathbf{v}_{2}-\mathbf{u}_{2}\right)
\delta t_{2a}\right\vert   \label{6.33}
\end{equation}

Expressions \oldref{(6.32)}(\ref{6.32}) and \oldref{(6.33)}(\ref{6.33})
allow to determine the time differences between theories.

By developing at first order in $v/c$ the r.h.s. of \oldref{(6.32)}(\ref{6.32}) and
\oldref{(6.33)}(\ref{6.33}, the time differences are explicitly written at that order as
\begin{eqnarray}
c^{2}\delta t_{2a} &=&\mathbf{v}_{23}\cdot \mathbf{R}_{23}  \label{6.34} \\
c^{2}\delta t_{1a} &=&\mathbf{v}_{23}\cdot \mathbf{R}_{23}+\mathbf{u}
_{12}\cdot \mathbf{R}_{12}  \nonumber
\end{eqnarray}

We note that at this order of approximation the assumption of constant
velocities of probe and antenna during $\delta t_{2a}$ is not even necessary
(it is anyway useful as an estimation of order of magnitudes).

If at $t_{2}^{\prime }$ the reflection by the probe were \textit{passive}
then, since according to VRT the velocity of light is linked in this case to
that of the antenna, the corresponding $t_{2}^{\prime }$ would be given by
\[
c\left( t_{3}-t_{2p}^{\prime }\right) =\left\vert \mathbf{X}_{a}\left(
t_{2p}^{\prime }\right) -\mathbf{X}_{s}\left( t_{2p}^{\prime }\right)
\right\vert
\]
which indicates that in the system of the antenna, in order to reach it
back, light moving at $c$ needs to travel a distance equal to that existing
between probe and antenna at the time of reflection. Of course, the uplink
relation, \oldref{(6.31)}(\ref{6.31}), remains the same for both cases. 
Proceeding as before we write
\[
R_{23}-c\delta t_{2p}=\left\vert \mathbf{R}_{23}-\mathbf{u}
_{23}R_{23}/c+\left( \mathbf{u}_{2}-\mathbf{v}_{2}\right) \delta
t_{2p}\right\vert
\]

Developing at order $v/c$ we thus obtain for the case of passive reflection
\begin{eqnarray}
c^{2}\delta t_{2p} &=&\mathbf{u}_{23}\cdot \mathbf{R}_{23}  \label{6.35} \\
c^{2}\delta t_{1p} &=&\mathbf{u}_{23}\cdot \mathbf{R}_{23}+\mathbf{u}
_{12}\cdot \mathbf{R}_{12}  \nonumber
\end{eqnarray}

Noting that corrections of proper time for the system of the antenna are of
order $(u/c)^{2}$, differences in trajectory data are thus related at order 
$v/c$ to differences in the times between emission and reception by the
antenna in the reference frame employed.

\begin{figure}[h]
\begin{center}
\includegraphics[scale=0.5]{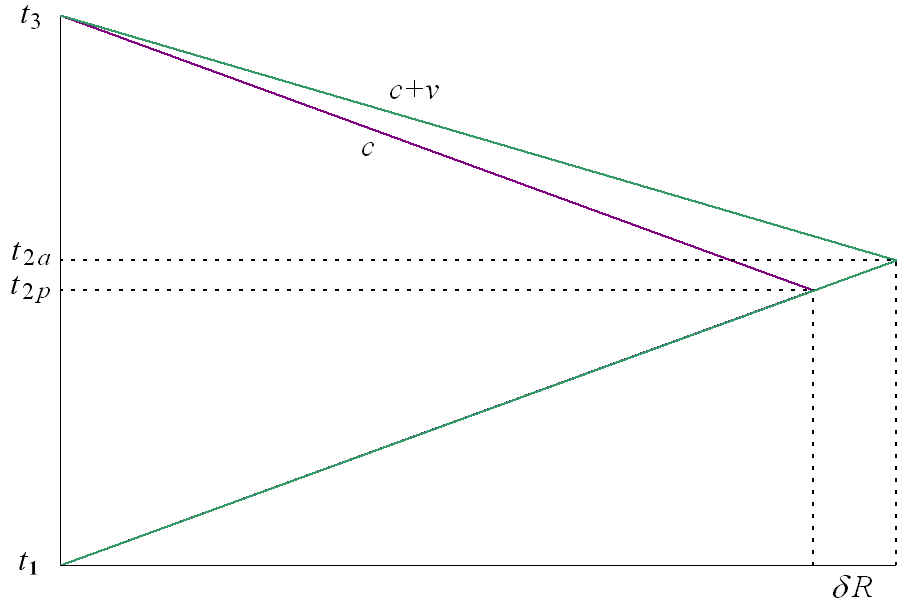}
\caption{Trajectory difference, $\delta R$, at a given epoch ($t_{2p}$)
as interpreted from the time difference between active (green full line) and
passive (violet full line) reflection from a spacecraft (see text).}
\label{fig:6.5}
\end{center}
\end{figure}

The downlink time difference between passive and active reflection (due to
the different propagation speed of the downlink signals) is interpreted as a
trajectory discrepancy rather than a different propagation speed, see figure
\oldref{6.5}\ref{fig:6.5}. In this way, one can write for the difference of trajectory
\[
\delta R=c\left( t_{2p}-t_{2a}\right)
\]
Using
\[
t_{2p}-t_{2a}=\left( t_{2p}-t_{2}\right) -\left( t_{2a}-t_{2}\right) =\delta
t_{2p}-\delta t_{2a}
\]
so that, from \oldref{(6.34)}(\ref{6.34}) and \oldref{(6.35)}(\ref{6.35}), we get
\begin{equation}
\delta R=\frac{\left( \mathbf{u}_{23}-\mathbf{v}_{23}\right) \cdot \mathbf{R}_{23}}{c}  \label{6.36}
\end{equation}
Notice that $\mathbf{v}_{23}-\mathbf{u}_{23}$ is the mean speed 
of the spacecraft relative to the radar between emission and reception,
thus coincides with the ad-hoc fitting of \oldref{(I.1)}(\ref{I.1}).

\begin{figure}[h]
\begin{center}
\includegraphics[width=\linewidth]{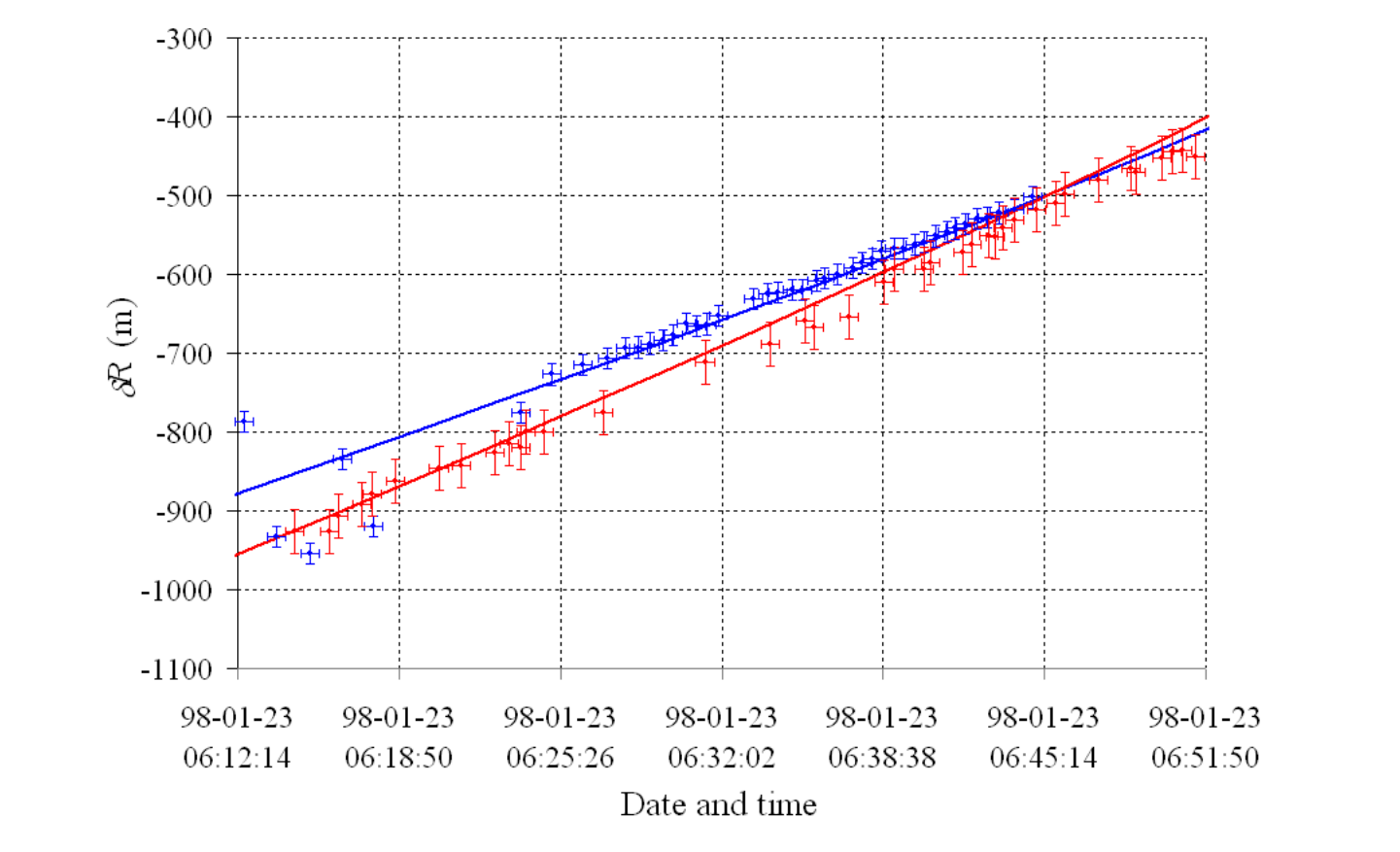}
\caption{Trajectory difference interpreted from the time difference
between VRT passive and active reflection (full lines, Millstone in blue and
Altair in red) obtained from \oldref{(6.36)}(\ref{6.36}), and the 
corresponding measured trajectory disagreement between SSN 
(passive reflection) and DSN (active reflection), for NEAR flyby 
(blue points for Millstone and red points for Altair tracking stations). 
For Millstone, the error bars refer to the uncertainties in the extraction 
of the data from figure 10 of reference \oldref{[R7]}\cite{antreasian1998aiaa}, 
rather than to its tracking error (5 m), while for Altair, the accuracy is 25 m.}
\label{fig:6.6}
\end{center}
\end{figure}

As it was pointed out above, the NEAR SSN trajectory data (passive
reflection) was found in disagreement with DSN data (active reflection).
Further, the disagreement exhibits different slopes for different tracking
stations. In figure \oldref{6.6}\ref{fig:6.6} we plot the trajectory 
difference between VRT passive and active reflection (full line, Millstone 
in blue and Altair in red) for the NEAR flyby obtained from \oldref{(6.36)}(\ref{6.36}).
As a comparison the measured trajectory difference (SSN data based on 
the DSN trajectory), is also plotted (circles with error bars, Millstone in 
blue and Altair in red), showing an excellent agreement. The explanation 
of the different ``intriguing'' \oldref{[R7]}\cite{antreasian1998aiaa} 
slopes from both stations is evident from \oldref{(6.36)}(\ref{6.36}). 
Although the exact location of the radar stations are
unknown to the author, using approximate coordinates, a statistically 
significant fit is obtained for both radar stations ($p<10^{-4}$, 
$R^{2}=0.987$ for Altair, and $p<10^{-4}$, $R^{2}=0.919$ for 
Millstone including the first outliers points). The high statistical 
significance of the fit suggest that these results are probably the 
strongest support to VRT.

%% file: 2014arxiv_v14_3_7.tex


\chapter{VRT and satellite positioning systems}\label{ch:7}

\fancyhf{}

\fancyhead[LE,RO]{7. Satellite positioning systems}
\fancyhead[RE,LO]{F. Minotti, L. Bilbao}
\fancyfoot[RE,LO]{Vibrating Rays Theory}
\fancyfoot[LE,RO]{\thepage}
 
\renewcommand{\headrulewidth}{1pt}
\renewcommand{\footrulewidth}{1pt}

It is natural to inquire into the possibility of detecting effects predicted
by VRT in satellite positioning systems (SPS) such as GPS, GLONASS, GALILEO,
and BeiDou. Indeed, since in VRT the speed of the signal emitted by a
satellite, relative to the GPS detector, depends on the satellite motion,
one expects that positions obtained from the assumption that the signal
speed is independent of the emitting source motion (as done in actual
working detectors) should differ from the true positions. The fact that the
GPS is such a precise tool, seems so to contradict one of the basis of VRT.

On the other hand, the actual determination of position using a SPS requires
to process data from at least four satellites (usually more), which can be
in many different and varying positions and states of motion, so that a more
definite answer can be obtained only by a simulation of a positioning
process, using realistic satellite ephemeris, employing the equations used
in actual detectors, and comparing differences when the broadcast
satellite position and time of emission are received at the detector by a
signal that behaves according to either VRT or SRT.

At its simplest, each satellite continuously broadcasts a distinctive signal that, 
in the case of GPS, is phase-modulated at different frequencies. The highest
phase-modulation-frequency component repeats periodically at predetermined 
times (typically every about 1 ms). When compared with a similar periodic signal 
generated locally in the receiver at the same times (according to the receiver clock), 
a time difference between emission and reception can be computed, which, when
multiplied by the speed of light, gives a value of distance denoted as pseudorange 
(due to its being determined by two different clocks). Additionally, positions of the 
satellites, relative to an Earth fixed coordinate system (ECEF), can be computed 
by the receiver at any time using a periodically updated ephemeris (included as 
part of the satellites signal in the lower phase-modulation-frequency components). 
Using this information, and the assumption that in an inertial system a global time 
definition is possible so that signals travel at the speed of light, the receiver can 
determine its position given data from three satellites. The inertial system employed 
is an earth centered inertial system (ECIS) that accompanies the earth in its orbital 
motion, but does not rotate, while the time employed corresponds to (a slightly shifted) 
Coordinated Universal Time, UTC (USNO) \oldref{[R88]}\cite{usno2016gps}. Since
most receivers clocks are not accurate enough, a correction to the receiver time is 
also calculated, which requires data from at least one more satellite, taking a minimum 
of four satellites to obtain a position fix.

If data from four or more satellites are available the fix can be determined
by minimizing, with respect to the receptor position $\mathbf{x}_{r}$ and
the time of reception $t_{r}$, the function
\begin{equation}
R=\sum_{i=1}^{n}\left[ c\left( t_{r}-t_{i}\right) -\left\vert \mathbf{x}_{r}-
\mathbf{x}_{i}\right\vert \right] ^{2}  \label{7.1}
\end{equation}
where $n$ is the number of satellites employed, each of which has position $
\mathbf{x}_{i}$ at the corresponding time of emission $t_{i}$. It is assumed
in \oldref{(7.1)}(\ref{7.1}) that positions of satellites and receiver are given relative 
to an ECIS, and that the data from all satellites are determined 
from the multiple satellite messages received at a single moment.

To determine the receiver position using VRT would require the minimization
of an expression like \oldref{(7.1)}(\ref{7.1}), but with the satellite positions given 
at the reception time $t_{r}$. The corresponding function could be approximated 
using the broadcast satellite information as
\begin{equation}
R_{VRT}=\sum_{i=1}^{n}\left[ c\left( t_{r}-t_{i}\right) -\left\vert \mathbf{x}_{r}-\mathbf{x}_{i}-\mathbf{\dot{x}}_{i}\left( t_{r}-t_{i}\right)
\right\vert \right] ^{2}  \label{7.2}
\end{equation}
where all satellite data correspond to the emission times $t_{i}$, and where
$\mathbf{\dot{x}}_{i}$ is the corresponding satellite velocity. Note that
expressions \oldref{(7.1)}(\ref{7.1}) and \oldref{(7.2)}(\ref{7.2}) differ 
by terms of order $v_{s}/c$, where $v_{s}$
is the component of satellite velocity on the line between satellite and
receiver. Since typical satellite speeds in the GPS are about 4 km/s one
expects measurable effects if signals from actual satellites would propagate
according to VRT instead of SRT, an effect similar to the range disagreement
discussed in Chapter \oldref{6}\ref{ch:6}.

The purpose of the present notes is to simulate what the position fix would
be if the usual expression \oldref{(7.1)}(\ref{7.1}) is employed, but the 
broadcast satellite data propagate as prescribed by VRT. In such a case, 
the data from satellite ``$i$'' received at time $t_{r}$ would have been 
emitted at the time $t_{i}^{\prime }$ given by
\begin{equation}
t_{i}^{\prime }=t_{r}-\frac{\left\vert \mathbf{x}_{r}-\mathbf{\tilde{x}}_{i}\right\vert }{c}  \label{7.3}
\end{equation}
where $\mathbf{\tilde{x}}_{i}$ is the satellite position at $t_{r}$.

In this way, given the satellite orbits of a given constellation one can
take a receiver position $\mathbf{x}_{r}$ at time $t_{r}$ and compute the
visible satellite positions $\mathbf{\tilde{x}}_{i}$ at that moment, and the
corresponding times of emission $t_{i}^{\prime }$. The corresponding
satellite positions $\mathbf{x}_{i}^{\prime }$ at the times of emission $
t_{i}^{\prime }$ can then be evaluated and used in \oldref{(7.1)}(\ref{7.1})
to determine putative values of receptor position $\mathbf{x}_{r}^{\prime }$ 
and time of reception $t_{r}^{\prime }$. The position ``error'' is thus 
determined as: $\Delta \mathbf{x}_{r}=\mathbf{x}_{r}^{\prime }-\mathbf{x}_{r}$.

Note that the time error, $\Delta t_{r}=t_{r}^{\prime }-t_{r}$, can be
interpreted as the correction to the receiver time. On the other hand, the error 
in position could in principle give a direct hint on possible effects due to signal velocity 
dependence on satellite motion, as predicted by VRT.

We have simulated SPS fixes from modeled GPS and GLONASS. As both give
similar results only the GPS modeling will be presented.

For the GPS we have modeled a constellation of 24 satellites with circular
Keplerian orbits of 26,560 km radius in an ECIS. The satellites are distributed
into six orbital planes, each of which has an inclination of 55 degrees
relative to the Earth equator. The four satellites in each orbital plane are
not equally spaced, but separated by angular differences of 30, 105, and 120
degrees, and with different phases between orbits, as in the actual constellation. 
The orbital planes are equally distributed so that the right ascension of the 
ascending node (RAAN) of the orbits in two consecutive planes differ by 60 degrees.

We have taken in our model typical relative positions of GPS satellites as
shown in \oldref{[R89]}\cite{dana2016gps}. Taking the argument of periapsis as 
90 degrees, the true anomalies of the employed satellites at our initial time 
were (all angles in degrees): for an orbit with RAAN = 17, (80, 110, 215, 335); 
for an orbit with RAAN = 77, (25, 145, 250, 280); for an orbit with 
RAAN = 137, (40, 70,175, 295); for an orbit with RAAN = 197, (105, 210, 
240, 345); for an orbit with RAAN = 257, (0, 30, 135, 255); and for an orbit 
with RAAN = 317, (55,175, 280, 310).

The receiver was considered stationary on the surface of an spherical Earth
of radius 6,371 km, rotating with a period of a sidereal day. All
calculations were made in the ECIS in which satellite and receiver motions
are modeled as described. In this way, given the longitude and latitude of a
receiver, fixes are determined at varying values of the reception time $t_{r}
$, as was described above.

For each $t_{r}$ the calculated position error $\Delta \mathbf{x}_{r}$ is
projected onto the local tangential plane at the position of the receiver to
determine the South-North (SN) and West-East (WE) distance errors, while the
projection of $\Delta \mathbf{x}_{r}$ on the local radial direction gives
the height error.

We first present the results obtained using in the minimization of
expression \oldref{(7.1)}(\ref{7.1}) all satellites that are above 15 degrees 
over the horizon of the receiver, which is the usual practice in open areas.

\begin{figure}[ht]
\begin{center}
\includegraphics[width=\linewidth]{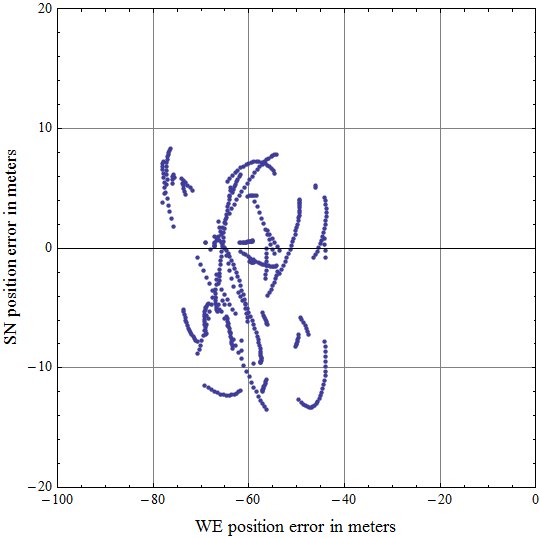}
\caption{WE (horizontal) and SN (vertical) position errors in meters for a
receiver at the equator.}
\label{fig:7.1}
\end{center}
\end{figure}

\begin{figure}[ht]
\begin{center}
\includegraphics[width=\linewidth]{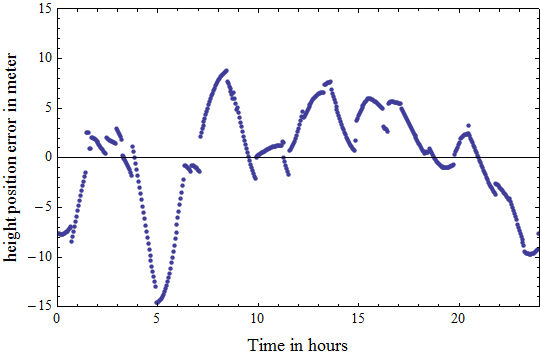}
\caption{ Height error in meters as a function of time in hours for a
receiver at the equator.}
\label{fig:7.2}
\end{center}
\end{figure}

In figures \oldref{7.1}\ref{fig:7.1} and \oldref{7.2}\ref{fig:7.2} we present 
the horizontal (WE and SN) and height errors, respectively, given in meters, 
for a receiver on the equator (the longitude of the receiver is not important, 
due to the Earth rotation), and for fixes done about every three minutes over one day time.

The SN error has a mean of $-1.9$ m with a dispersion of about $\pm 6$ m,
whereas the WE error has a mean of $-61$ m, with a dispersion of about $\pm
8.5$ m. The height error has a mean below 1 cm, and the dispersion is about $
\pm 5$ m.

\begin{figure}[ht]
\begin{center}
\includegraphics[width=\linewidth]{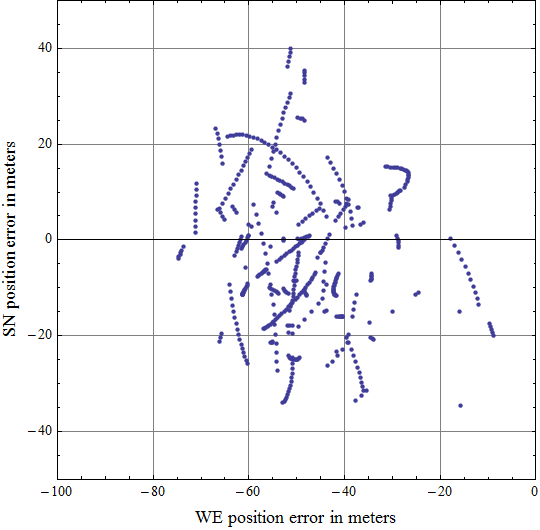}
\caption{WE (horizontal) and SN (vertical) position errors in meters for a
receiver at 45 degree latitude.}
\label{fig:7.3}
\end{center}
\end{figure}

\begin{figure}[ht]
\begin{center}
\includegraphics[width=\linewidth]{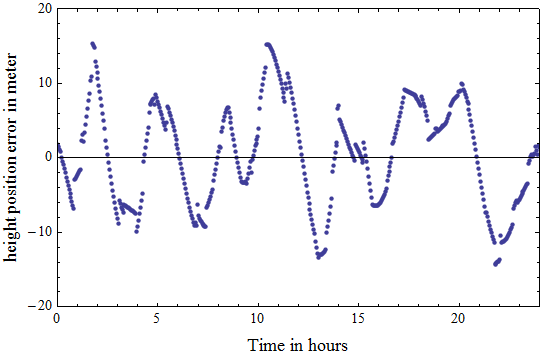}
\caption{Height error in meters as a function of time in hours for a
receiver at 45 degree latitude.}
\label{fig:7.4}
\end{center}
\end{figure}

In figures \oldref{7.3}\ref{fig:7.3} and \oldref{7.4}\ref{fig:7.4} the 
corresponding results for a receiver located at 45 degree latitude are presented. 
In this case the SN error has a mean of $-2.5$ m, with a dispersion of $\pm 15$ m. 
The WE error has a mean of $-48$ m with a dispersion of $\pm 12$ m. The height 
error mean is 0.34 m with a dispersion of $\pm 7$ m.

One can at this point argue that while SN and height positioning are
relatively accurate, the systematic WE errors of mean up to 61 m should easily
show up in the measurements if VRT were valid.

There is however a point to consider in this respect. The satellite
positions are determined from the broadcast message using the orbital
parameters of the satellite orbits which are included in the message. Due to
the non-spherical mass distribution of Earth, tidal forces and other effects, the 
satellite orbits are not perfectly Keplerian, in such a way that, for instance, inclination 
and RAAN are slowly changing. The satellite ephemeris message thus includes values 
of inclination and RAAN at a given time of reference, as well as the rate of change of 
these parameters at that time \oldref{[R90]}\cite{gpsw2016gps}. 
Satellite ephemeris are thus valid for relatively short periods of time, usually 
four hours. As an example, a typical rate of change of the RAAN is about 
$-8\times 10^{-9}$ rad/sec, which is a relatively high value, considering that 
at this rate it would take about 20 minutes for the orbit to change enough to produce 
errors similar to those shown in the example for the WE positioning. 
Not only orbit parameters need periodic corrections, satellite clocks have fluctuations 
and drifts that have to be corrected. Since those corrections imply the assumption of 
signal velocity independent on the motion of the emitter, VRT effects could be hidden, 
for instance, in the way satellite orbits are modeled.   

\begin{figure}[ht]
\begin{center}
\includegraphics[width=\linewidth]{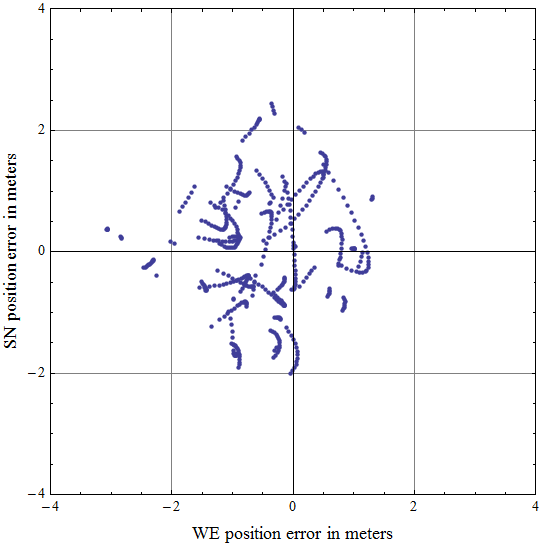}
\caption{EW (horizontal) and SN (vertical) position errors in meters for a
receiver at the equator with calculated satellite positions advanced by 79 milliseconds.}
\label{fig:7.5}
\end{center}
\end{figure}

\begin{figure}[ht]
\begin{center}
\includegraphics[width=\linewidth]{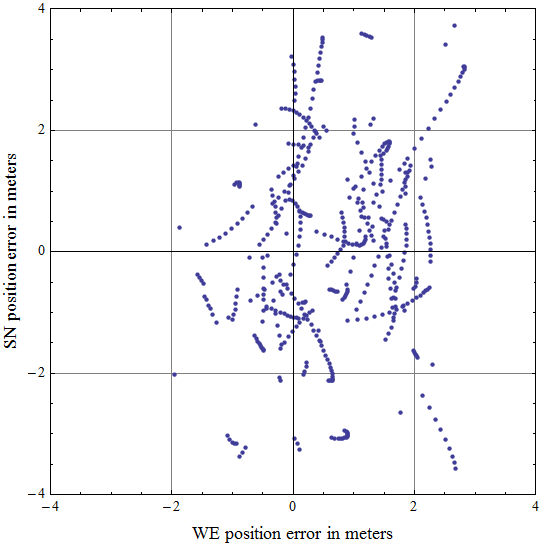}
\caption{Same as figure \oldref{7.5}\ref{fig:7.5}, but for a receiver 
at 45 degree latitude.}
\label{fig:7.6}
\end{center}
\end{figure}

\begin{figure}[ht]
\begin{center}
\includegraphics[width=\linewidth]{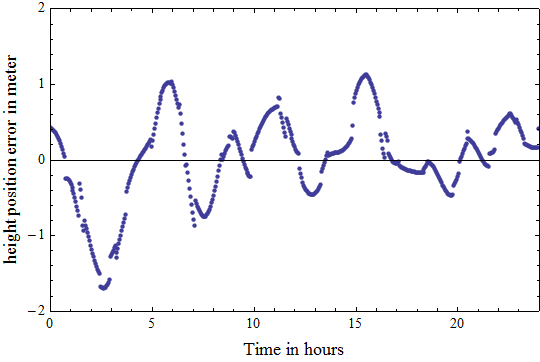}
\caption{Height error in meters as a function of time in hours for a
receiver at the equator with calculated satellite positions advanced by 79 milliseconds.}
\label{fig:7.7}
\end{center}
\end{figure}

\begin{figure}[ht]
\begin{center}
\includegraphics[width=\linewidth]{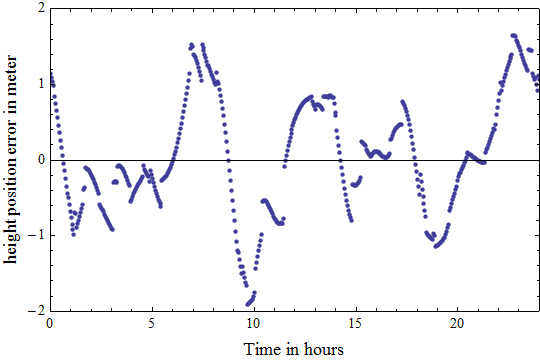}
\caption{Same as figure \oldref{7.7}\ref{fig:7.7}, but for a receiver 
at 45 degree latitude.}
\label{fig:7.8}
\end{center}
\end{figure}

If, for instance, the calculated satellite position were about 79 ms ahead along its 
own orbits from the actual position, the errors and their dispersions, determined 
assuming VRT valid, but interpreting the measurement according to SRT, would be 
very much reduced. This is shown in figures \oldref{7.5}\ref{fig:7.5} and \oldref{7.6}\ref{fig:7.6} 
where the horizontal errors for the receivers at the equator and at 45 degree 
latitude are shown, and in which the minimization of \oldref{(7.1)}(\ref{7.1}) is done 
with the values of $t_{i}^{\prime }$ given by \oldref{(7.3)}(\ref{7.3}) for the correct 
orbits, but along which the assumed satellite positions are advanced by 79 ms. 
Correspondingly, the errors in height determination as a function of time are shown for 
the receivers at the equator and at 45 degree latitude in figures \oldref{7.7}\ref{fig:7.7} 
and \oldref{7.8}\ref{fig:7.8} 

For the receiver at the equator the mean error of position in direction EW is about -0.5 m, 
whereas those in direction SN and in height are below 1 cm, with dispersions all below 1 m.
For the receiver at 45 degree latitude the mean error of position in direction EW is about -0.7 m, 
whereas those in direction SN and in height are below 10 cm, with dispersions all below 1.6 m.

It is thus argued that if VRT is valid, a shift of about 80 ms along the orbit of the satellite 
position could be automatically included, because the satellite orbits themselves are evaluated 
from the very GPS satellite broadcast data, received at a network of stations with precisely 
determined positions on Earth (the so called ``Precise Orbit Determination''), assuming SRT
 as the valid model.

We mention in passing that a small advance in the RAAN of all satellite orbits of about 
$10^{-5}$ rad, which corresponds to only about 20 minutes at the natural rate of change 
of the RAAN, can also result in correct positions (although with larger dispersions), without 
the need for satellite position shift along its orbit (see previous versions of this manuscript). 
It is thus possible to have a combination of position shifting along the orbit and small RAAN 
corrections that result in precise fixes. 

In conclusion, if VRT were valid, the standard procedure used in SPS could result in non-
measurable differences in position determination relative to SRT.

%% file: 2014arxiv_v14_3_8.tex


\chapter{An electrodynamic theory based on the Vibrating Rays Theory
(VRT) of electromagnetism}\label{ch:8}

\fancyhf{}

\fancyhead[LE,RO]{8. Electrodynamic theory}
\fancyhead[ RE,LO]{F. Minotti, L. Bilbao}
\fancyfoot[ RE,LO]{Vibrating Rays Theory}
\fancyfoot[LE,RO]{\thepage}
 
\renewcommand{\headrulewidth}{1pt}
\renewcommand{\footrulewidth}{1pt}

\section{Introduction}\label{sec:8.1}

We deal in this part with the possibility of developing an electrodynamics
theory incorporating Faraday's conception of ``vibrating
rays.'' In so doing some preliminary considerations are in
order. First, if possible novel effects are sought, Galilean relations
between magnitudes in different frames are to be employed, in order to
differentiate the particular frame (in general non-inertial) in which the
speed of the electromagnetic interaction is $c$. Moreover, as we will
presently show, Faraday's idea is better represented by a delayed
action-at-a-distance kind of theory, as it requires the knowledge of the
state of the source at the instant the interaction with another charge takes
place. In this way, we will proceed by analogy with the relativistic
action-at-a-distance theory of Fokker \oldref{[R91]}\cite{fokker1929invarianter}, 
in which the action of a system of point charges is given, from which 
a complete (mechanical and electromagnetic) description of the system 
can be obtained.

We begin with the notion that, according to VRT, the propagation velocity is 
$c$ in the non-rotating, but in general accelerated frame in which the
source is permanently at rest. Employing Galilean relations between
different frames we consequently consider that in the inertial frame in
which the source moves with velocity $\mathbf{V}(t)$, the velocity of
propagation at time $t$ is
\begin{equation*}
c\,\mathbf{n}_{0}+\mathbf{V}\left( t\right)
\end{equation*}
where $\mathbf{n}_{0}$ is a constant unit vector describing the particular
direction of propagation of the interaction in an instantaneously motionless
system relative to the source, with fixed orientation.

If a vibration front leaves the source at time $t^{\prime }$, when the
source is at $\mathbf{X}(t^{\prime })$, and reaches position $\mathbf{x}$ at 
$t$, one has
\begin{equation}
{\displaystyle\int\limits_{t\prime}^{t}}
\left[ c\,\mathbf{n}_{0}+\mathbf{V}\left( \tau \right) \right] 
d\tau =\mathbf{x}-\mathbf{X}\left( t^{\prime }\right)
\label{8.1}
\end{equation}

A direct integration then gives
\begin{equation*}
c\,\mathbf{n}_{0}\left( t-t^{\prime }\right) +\mathbf{X}\left( t\right) -%
\mathbf{X}\left( t^{\prime }\right) =\mathbf{x}-\mathbf{X}\left( t^{\prime
}\right)
\end{equation*}
from which one obtains
\begin{equation*}
\mathbf{n}_{0}=\frac{\mathbf{x}-\mathbf{X}\left( t\right) }{\left\vert 
\mathbf{x}-\mathbf{X}\left( t\right) \right\vert }
\end{equation*}
\begin{equation}
t^{\prime }\ =t-\frac{\left\vert \mathbf{x}-\mathbf{X}\left( t\right)
\right\vert }{c}  \label{8.2}
\end{equation}

Relation \oldref{(8.2)}(\ref{8.2}) thus expresses the delay to be incorporated into
the action-at-a-distance theory, in place of the relativistic one employed
in Fokker's action, which is given as:
\begin{align}
S  & =-\sum_{i}m_{i}c\int\sqrt{c^{2}-\left\vert \mathbf{V}_{i}\left(
t\right)  \right\vert ^{2}}dt\label{8.3}\\
& -%
{\displaystyle\sum\limits_{i}}
\sum_{k\neq i}\frac{Q_{i}Q_{k}}{4\pi \varepsilon _{0}c}\int %
{\displaystyle\int}
\delta \left[ s_{ik}^{2}\right] \left[ c^{2}-\mathbf{V}_{i}\left( t\right) 
\cdot \mathbf{V}_{k}\left( t^{\prime }\right) \right] dtdt^{\prime }
\nonumber%
\end{align}
where
\begin{equation}
s_{ik}^{2}=c^{2}\left( t-t^{\prime }\right) ^{2}-\left\vert \mathbf{X}%
_{i}\left( t\right) -\mathbf{X}_{k}\left( t^{\prime }\right) \right\vert ^{2}
\label{8.4}
\end{equation}

Expressions \oldref{(8.3)}(\ref{8.3}) and \oldref{(8.4)}(\ref{8.4}) are valid for 
point sources with masses $m_{i}$, charges $Q_{i}$, positions $\mathbf{X}_{i}\left(
t\right) $, and velocities $\mathbf{V}_{i}\left( t\right) $.

Note that \oldref{(8.4)}(\ref{8.4}) can be interpreted as indicating that the action
of particle $k$ on particle $i$ is delayed from the ``emission'' time 
$t^{\prime }$ to the ``reception'' time $t$, according to relativistic principles.
It is then clear that a VRT electrodynamics requires the replacement of
expression \oldref{(8.4)}(\ref{8.4}) by that resulting from \oldref{(8.2)}(\ref{8.2})
\begin{equation}
\hat{s}_{ik}^{2}=c^{2}\left( t-t^{\prime }\right) ^{2}-\left\vert \mathbf{X}%
_{i}\left( t\right) -\mathbf{X}_{k}\left( t\right) \right\vert ^{2}
\label{8.5}
\end{equation}
which indicates that even if the action originated in particle $k$ at time $%
t^{\prime }$, the reception time $t$ depends on the position of the emitting
particle $k$ at time $t$. The simple replacement of \oldref{(8.4)}(\ref{8.4}) by 
\oldref{(8.5)}(\ref{8.5}) is however not enough because the resulting action 
is not Galilean invariant. In Appendix 8.A\oldref{sec:8.A} we show with some detail 
the reasoning to derive the appropriate action, which we simply state here as
\begin{eqnarray}
S_{VRT} &=&\sum_{i}\frac{m_{i}}{2}\int \left\vert \mathbf{V}_{i}\left(
t\right) \right\vert ^{2}dt  \notag \\
&&-{\displaystyle\sum\limits_{i}}
\sum_{k\neq i}\frac{Q_{i}Q_{k}}{4\pi \varepsilon _{0}c}\int%
{\displaystyle\int}
\delta \left[ \hat{s}_{ik}^{2}\right] \biggl\{ c^{2}
+\frac{1}{2}\left( \Delta \mathbf{V}_{ik}\left( t\right) 
\cdot \mathbf{n}_{ik}\right) ^{2}
\biggr.  \notag \\
&&\biggl. 
-\left[ \Delta 
\mathbf{V}_{ik}\left( t\right) -\left( \Delta \mathbf{V}_{ik}\left( 
t\right) \cdot \mathbf{n}_{ik}\right) \mathbf{n}_{ik}\right] \cdot 
\mathbf{V}_{k}\left( t^{\prime }\right)
\biggr\} dtdt^{\prime }  
 \label{8.6}
\end{eqnarray}
where $\Delta \mathbf{V}_{ik}\left( t\right) =\mathbf{V}_{i}\left( t\right) -%
\mathbf{V}_{k}\left( t\right) $, and $\mathbf{n}_{ik}=\left( \mathbf{X}%
_{i}\left( t\right) -\mathbf{X}_{k}\left( t\right) \right) /\left\vert 
\mathbf{X}_{i}\left( t\right) -\mathbf{X}_{k}\left( t\right) \right\vert $.

It can be claimed that the term with the parallel component of velocity
differences, $\Delta \mathbf{V}_{ik}\left( t\right) \cdot \mathbf{n}_{ik}$,
in \oldref{(8.6)}(\ref{8.6}) is not really a retarded interaction term as 
it involves only the time $t$ and not $t^{\prime }$. The response to this 
claim is that velocity differences at time $t^{\prime }$ (needed for Galilean 
covariance if the parallel component is included) result in ill-posed equations 
of motion, with a force on particle $i$ at time $t$ depending on its own state
at different times. No other simple expression of the velocities (certainly
not a quadratic one) maintains Galilean covariance of the action, while at
the same time resulting in an experimentally acceptable electrodynamics.
Also, coming back to the ``vibrating rays''
idea, the transverse relative velocity can be related to transverse
excitations traveling at velocity $c$ relative to the emitter. On the other
hand, the relative parallel velocity corresponds to longitudinal tensions or
compressions of the rays, which can have an instantaneous manifestation,
since some kind of instantaneous longitudinal action is of course manifest
in the instantaneous accommodation of the velocity of propagation to the
emitter velocity. In this sense it can be said that expression \oldref{(8.6)}(\ref{8.6})
is compatible with Faraday's conceptions.

Note finally that the symmetry between particles holds only for the full
action. Each pair-interaction term in the action shows the privileged status
of the emitter, as the one that imposes the propagation velocity.

To proceed, as with the relativistic action-at-a-distance theory, it proves
convenient to define ``direct single particle
potentials,'' analogous to the Maxwell field potentials, as
\begin{equation*}
\Phi _{k}\left( \mathbf{X}_{i}\left( t\right) ,t\right) =\frac{Q_{k}c}{4\pi
\varepsilon _{0}}\int \delta \left[ \hat{s}_{ik}^{2}\right] dt^{\prime }
\end{equation*}
\begin{equation}
\mathbf{A}_{k}\left( \mathbf{X}_{i}\left( t\right) ,t\right) =\frac{Q_{k}}{%
4\pi \varepsilon _{0}c}\int \delta \left[ \hat{s}_{ik}^{2}\right] \mathbf{V}%
_{k}\left( t^{\prime }\right) dt^{\prime }  \label{8.7}
\end{equation}
in terms of which the action can be written as
\begin{eqnarray}
S_{VRT} &=&\sum_{i}\frac{m_{i}}{2}\left\vert \mathbf{V}_{i}\left( t\right)
\right\vert ^{2}dt  \notag \\
&&-{\displaystyle\sum\limits_{i}}\sum_{k\neq i}Q_{i}\int 
\biggl\{ \Phi _{k}\left( \mathbf{X}%
_{i}\left( t\right) ,t\right) \left[ 1+\frac{\left( \Delta \mathbf{V}%
_{ik}\left( t\right) \cdot \mathbf{n}_{ik}\right) ^{2}}{2c^{2}}\right]
\biggr.  \label{8.8} \\
&&\biggl. -\mathbf{A}_{k}\left( \mathbf{X}_{i}\left( t\right) ,t\right) \cdot %
\left[ \Delta \mathbf{V}_{ik}\left( t\right) -\left( \Delta \mathbf{V}%
_{ik}\left( t\right) \cdot \mathbf{n}_{ik}\right) \mathbf{n}_{ik}\right]
\biggr\} dt  \notag
\end{eqnarray}

The principle of least action for variations of the trajectory $\mathbf{X}%
_{i}\left( t\right) $ then leads to the equation of motion of particle $i$
allowing to identify the force acting on it. Before giving the expression of
this force the following considerations are in order.

In the process of performing the time integration in expressions 
\oldref{(8.7)}(\ref{8.7}) the delta functions are to be expressed as 
functions of the single variable $t^{\prime }$, which makes explicit the 
appearance of both, retarded and advanced interactions, manifested by 
the two roots of \oldref{(8.5)}(\ref{8.5}) for $t^{\prime }$. Of course 
the same happens with the relativistic version, for which only after one
 includes the ``reaction of the universe,'' as in the absorber theory of 
Wheeler and Feynman \oldref{[R92]}\cite{wheeler1949classical}, 
does the single retarded field remain 
(with the bonus of the ``radiation damping'' being included
automatically). In our case no assumption is necessary for $\Phi _{k}$ as
both, advanced and retarded interactions give the same result. For the case
of $\mathbf{A}_{k}$ the arguments of Wheeler and Feynman are expected to be
applicable because the advanced and retarded interactions for the vector
potential satisfy the same wave equation, as shown in the next subsection.
In this way, we will assume that the absorption process is at work in order
to obtain in the end only retarded interactions. Performing the integration
over $t^{\prime }$ in \oldref{(8.7)}(\ref{8.7}) we obtain
\begin{equation*}
\Phi _{k}\left( \mathbf{X}_{i}\left( t\right) ,t\right) =\frac{Q_{k}}{4\pi
\varepsilon _{0}}\frac{1}{\left\vert \mathbf{X}_{i}\left( t\right) -\mathbf{X%
}_{k}\left( t\right) \right\vert }
\end{equation*}
\begin{equation}
\mathbf{A}_{k}\left( \mathbf{X}_{i}\left( t\right) ,t\right) =\frac{Q_{k}}{%
4\pi \varepsilon _{0}c^{2}}\frac{1}{2}\frac{\mathbf{V}_{k}\left(
t_{adv}\right) +\mathbf{V}_{k}\left( t_{ret}\right) }{\left\vert \mathbf{X}%
_{i}\left( t\right) -\mathbf{X}_{k}\left( t\right) \right\vert }  \label{8.9}
\end{equation}
where $t_{ret}=t-\left\vert \mathbf{X}_{i}\left( t\right) -\mathbf{X}%
_{k}\left( t\right) \right\vert /c$ , and $t_{adv}=t+\left\vert \mathbf{X}%
_{i}\left( t\right) -\mathbf{X}_{k}\left( t\right) \right\vert /c$.

Using expressions \oldref{(8.9)}(\ref{8.9}) in \oldref{(8.8)}(\ref{8.8}), the 
principle of least action for variations of the trajectory $\mathbf{X}_{i}\left( t\right) $
results in
\begin{equation*}
m_{i}\mathbf{\dot{V}}_{i}\left( t\right) =\sum_{k\neq i}\mathbf{F}_{ik}
\end{equation*}
where $\mathbf{F}_{ik}$, the force on the charge $i$ due to charge $k$, is
given as (unless explicitly indicated, the magnitudes are evaluated at the
time $t$)
\begin{eqnarray*}
\hspace{30pt}\mathbf{F}_{ik} \hspace{-5pt} &=&\hspace{-5pt} \frac{Q_{i}Q_{k}}{4\pi \varepsilon_{0}c^{2}}
\frac{\mathbf{n}_{ik}}{R_{ik}^{2}}\left\{ c^{2}+\Delta \mathbf{V}_{ik}\cdot \Delta 
\mathbf{V}_{ik}-\frac{3}{2}\left( \mathbf{n}_{ik}\cdot \Delta \mathbf{V}%
_{ik}\right) ^{2}+R_{ik}\mathbf{n}_{ik}\cdot \Delta \mathbf{\dot{V}}%
_{ik}\right\}  \\
&&\hspace{-70pt}+\frac{Q_{i}Q_{k}}{4\pi \varepsilon _{0}c^{2}}\frac{\mathbf{1}}{R_{ik}}%
\frac{1}{2}\left\{ \left[ \mathbf{\dot{V}}_{k}\left( t_{ret}\right) \cdot 
\mathbf{n}_{ik}\right] \mathbf{n}_{ik}-\mathbf{\dot{V}}_{k}\left(
t_{ret}\right) +\left[ \mathbf{\dot{V}}_{k}\left( t_{adv}\right) \cdot 
\mathbf{n}_{ik}\right] \mathbf{n}_{ik}-\mathbf{\dot{V}}_{k}\left(
t_{adv}\right) \right\}  \\
&&\hspace{10pt}+\frac{Q_{i}Q_{k}}{4\pi \varepsilon _{0}c^{3}}\frac{\mathbf{1}}{R_{ik}}%
\Delta \mathbf{V}_{ik}\times \frac{1}{2}\left\{ \left[ \mathbf{\dot{V}}%
_{k}\left( t_{ret}\right) -\mathbf{\dot{V}}_{k}\left( t_{adv}\right) \right]
\times \mathbf{n}_{ik}\right\} ,
\end{eqnarray*}
where $R_{ik}=\left\vert \mathbf{X}_{i}-\mathbf{X}_{k}\right\vert $.

Applying now arguments of the reaction of an absorbing universe, as in
\oldref{[R92]}\cite{wheeler1949classical}, one concludes that the summation 
of terms involving advanced and retarded times can be expressed as
\begin{equation*}
\sum_{k\neq i}\frac{1}{2}\left[ \delta \mathbf{F}_{ik}\left( t_{adv}\right)
+\delta \mathbf{F}_{ik}\left( t_{ret}\right) \right] =\frac{1}{2}\left[
\delta \mathbf{F}_{ii}\left( t_{ret}\right) -\delta \mathbf{F}_{ii}\left(
t_{adv}\right) \right] +\sum_{k\neq i}\delta \mathbf{F}_{ik}\left(
t_{ret}\right) 
\end{equation*}
where the difference of diverging terms in the square brackets gives a
finite contribution, resulting in the radiation damping force on particle $i$. 
In this way, we write for the final expression of the force of particle $k$
on particle $i$:
\begin{eqnarray}
\mathbf{F}_{ik} &=&\frac{Q_{i}Q_{k}}{4\pi \varepsilon _{0}c^{2}}\frac{%
\mathbf{n}_{ik}}{R_{ik}^{2}}\left\{ c^{2}+\Delta \mathbf{V}_{ik}\cdot \Delta 
\mathbf{V}_{ik}-\frac{3}{2}\left( \mathbf{n}_{ik}\cdot \Delta \mathbf{V}%
_{ik}\right) ^{2}+R_{ik}\mathbf{n}_{ik}\cdot \Delta \mathbf{\dot{V}}%
_{ik}\right\}   \notag \\
&&+\frac{Q_{i}Q_{k}}{4\pi \varepsilon _{0}c^{2}}\frac{\mathbf{1}}{R_{ik}}%
\left\{ \left[ \mathbf{\dot{V}}_{k}\left( t_{ret}\right) \cdot \mathbf{n}%
_{ik}\right] \mathbf{n}_{ik}-\mathbf{\dot{V}}_{k}\left( t_{ret}\right)
\right\}  \notag \\
&&+\frac{Q_{i}Q_{k}}{4\pi \varepsilon _{0}c^{3}}\frac{\mathbf{1}}{%
R_{ik}}\Delta \mathbf{V}_{ik}\times \left[ \mathbf{\dot{V}}_{k}\left(
t_{ret}\right) \times \mathbf{n}_{ik}\right]   \label{8.10}
\end{eqnarray}
while the total force on particle $i$ includes, apart from the sum of terms 
\oldref{(8.10)}(\ref{8.10}) for all $k\neq i$, a ``self-force'' or radiation damping
 force (resulting from the reaction of the universe) given by
\begin{equation}
\mathbf{F}_{i}^{damp}=\frac{Q_{i}^{2}\mathbf{\ddot{V}}_{i}\left( t\right) }{%
6\pi \varepsilon _{0}c^{2}}  \label{8.11}
\end{equation}

The terms in the first line of the right-hand side of \oldref{(8.10)}(\ref{8.10})
correspond to Weber's force (those dependent on velocity result from the
parallel component of $\Delta \mathbf{V}_{ik}$ in \oldref{(8.8)}(\ref{8.8})). The
terms in the other lines come from the contribution of the transverse
component of $\Delta \mathbf{V}_{ik}$. Those in the second line represent
what in Maxwell electrodynamics is the radiative electric field, retarded in
accordance with the principles of VRT. The term in the third line shows the
corresponding radiative magnetic field plus a novel contribution. This novel
force term has in general a component along the ``propagation'' direction 
$\mathbf{n}_{ik}$ and can thus be interpreted as a longitudinal component 
of the radiative electric field, which is smaller in magnitude, by a factor 
$V_{k}/c$, than the transverse component.

Weber's force and its ability to describe electromagnetic interactions, as
compared with Maxwell electrodynamics, are well studied subjects 
\oldref{[R93]}\cite{assis1994weber},
showing an excellent performance of Weber's expression for all non-radiative
processes. It is thus remarkable that Weber's force can be integrated using
Faraday's ideas into an expression including also ``radiative'' terms, which
are missing in the original formalism.

Finally, it is important to verify that the terms added to the original
Weber's expression do not spoil its ability to correctly describe the
experimentally determined laws. In the Appendix 8.B\oldref{sec:8.B} 
we show with some detail that the new terms produce at most negligible 
corrections to the predictions of Weber's theory for non-radiative situations.

\section{Wave equation and Doppler effect}\label{sec:8.2}

The complex formula \oldref{(8.10)}(\ref{8.10}) can formally be expressed 
in terms of derivatives of the particle potentials \oldref{(8.9)}(\ref{8.9}). 
Although that representation is not practical, it tells us that we can analyze the
behavior of those potentials to obtain important clues. For instance, if the
position $\mathbf{X}_{i}\left( t\right) $ of the ``test'' particle 
is taken as an independent variable space
position $\mathbf{x}$, by direct evaluation it can be immediately checked
that the single particle potentials \oldref{(8.9)}(\ref{8.9}) reinterpreted in this
way satisfy the equations
\begin{equation*}
\left[ \frac{1}{c^{2}}\frac{d^{2}}{dt_{k}^{2}}-\nabla ^{2}\right] \Phi
_{k}\left( \mathbf{x},t\right) =\frac{Q_{k}}{\varepsilon _{0}}\delta \left[ 
\mathbf{x}-\mathbf{X}_{k}\left( t\right) \right]
\end{equation*}
\begin{equation}
\left[ \frac{1}{c^{2}}\frac{d^{2}}{dt_{k}^{2}}-\nabla ^{2}\right] \mathbf{A}%
_{k}^{\pm }\left( \mathbf{x},t\right) =\frac{Q_{k}\mathbf{V}_{k}\left(
t\right) }{2\varepsilon _{0}c^{2}}\delta \left[ \mathbf{x}-\mathbf{X}%
_{k}\left( t\right) \right]  \label{8.12}
\end{equation}
where the total time derivative is defined as
\begin{equation*}
\frac{d}{dt_{k}}=\frac{\partial }{\partial t}+\mathbf{V}_{k}\left( t\right)
\cdot \triangledown
\end{equation*}
and $\mathbf{A}_{k}^{\pm }$ is either the retarded or advanced contribution
(this is precisely what allows to use Wheeler and Feynman arguments of an
absorbing universe). In this way, outside the source $k$ its potentials
satisfy the (Galilean) wave equation
\begin{equation*}
\frac{1}{c^{2}}\frac{d^{2}F}{dt_{k}^{2}}-\nabla ^{2}F=0
\end{equation*}

Plane wave solutions for an arbitrarily moving source $k$ can be easily
obtained by checking that a generic function of the form $F\left( \mathbf{k}%
\cdot \mathbf{x}-h\left( t\right) \right) $, with $\mathbf{k}$ a constant wavevector,
satisfies the previous wave equation for arbitrary $\mathbf{X}_{k}\left(
t\right) $ and $F$, if the function $h\left( t\right) $ is given by
\begin{equation*}
\dot{h}\left( t\right) =\mathbf{k}\cdot \mathbf{V}_{k}\left( t\right)
+\left\vert \mathbf{k}\right\vert c
\end{equation*}

The wave angular frequency $\omega $ of $F$ is given as minus the partial
time derivative of the phase $\mathbf{k}\cdot \mathbf{x}-h\left( t\right) $,
which is precisely $\dot{h}\left( t\right) $. So, if for instance we
consider the measurement of the ``radiative'' electric force generated
far away from an harmonically oscillating source $Q_{k}$ we obtain the 
result that the measured frequency of the force corresponds to the Doppler 
shifted frequency, but with the actual source velocity, and not the retarded 
one as in Maxwell electrodynamics.

Note also that the wave group velocity is, consistently with the hypothesis
used to derive the VRT electrodynamics, given by
\begin{equation*}
\mathbf{V}_{g}=\frac{\partial \omega }{\partial \mathbf{k}}=\mathbf{V}%
_{k}\left( t\right) +c\frac{\mathbf{k}}{\left\vert \mathbf{k}\right\vert }
\end{equation*}

\section{Mirror reflected field}\label{sec:8.3}

One essential condition on mirror reflections in VRT is that the reflected
light propagates as ``ray vibrations'' of
the original source and is thus affected by its motion, and not by that of
the image. This effect is indicated by the experimental evidence, for
instance, that in an interferometer the fringe pattern is not affected by
the motion of the source. If the reflected light motion were
``linked'' to the stationary mirror, its
interference with the non-reflected light, which according to VRT follows
the motion of the source, would modify the fringe pattern.

In this way, for a source $Q$ (we leave out here the sub-index $k$ of the
source) at position $\mathbf{X}\left( t\right) $ near a fixed plane mirror
whose normal unit vector is $\mathbf{e}_{0}$ (directed toward the half space
where the source is), the correct boundary condition of zero tangential
electric field (for a plane, perfectly conducting mirror at any point on it
surface) can be obtained by the method of images, with an image of charge $%
-Q $ and reflected position and velocity, and for which to any emitting
direction $\mathbf{n}_{0}$ of the real charge corresponds a direction $%
\mathbf{\bar{n}}_{0}$, mirror image of $\mathbf{n}_{0}$. The interaction
emitted at time $t^{\prime }$ with direction $\mathbf{n}_{0}$ is reflected
by the mirror at time $\tau $ and so propagates further with direction $%
\mathbf{\bar{n}}_{0}$ to reach position $\mathbf{x}$ at $t$
\begin{equation*}
{\displaystyle\int\limits_{t^{\prime}}^{\tau}}
\left[ c\,\mathbf{n}_{0}+\mathbf{V}\left(
t^{\prime \prime }\right) \right] dt^{\prime \prime }+
{\displaystyle\int\limits_{\tau}^{t}}
\left[ c\,\mathbf{\bar{n}}_{0}+\mathbf{V}\left( t^{\prime \prime
}\right) \right] dt^{\prime \prime }=\mathbf{x}-\mathbf{X}\left( t^{\prime
}\right)
\end{equation*}
which gives after integration
\begin{equation}
c\,\mathbf{n}_{0}\left( \tau -t^{\prime }\right) +c\,\mathbf{\bar{n}}%
_{0}\left( t-\tau \right) =\mathbf{x}-\mathbf{X}\left( t\right)  \label{8.13}
\end{equation}
where $\mathbf{n}_{0}\cdot \mathbf{e}_{0}=-\,\mathbf{\bar{n}}_{0}\cdot 
\mathbf{e}_{0}=n_{0\perp }$.

Denoting by $z$ the coordinate normal to the mirror (growing in the
direction of $\mathbf{e}_{0}$) we also have (for the path from the source at 
$Z\left( t^{\prime }\right) $ to the mirror at $z_{m}$)
\begin{equation*}
\mathbf{e}_{0}\cdot 
{\displaystyle\int\limits_{t^{\prime}}^{\tau}}
\left[ c\,\mathbf{n}_{0}+%
\mathbf{V}\left( t^{\prime \prime }\right) \right] dt^{\prime \prime
}=z_{m}-Z\left( t^{\prime }\right)
\end{equation*}
which on integration gives
\begin{equation}
cn_{0\perp }\left( \tau -t^{\prime }\right) =z_{m}-Z\left( \tau \right)
\label{8.14}
\end{equation}

On the other hand, projection of \oldref{(8.13)}(\ref{8.13}) on $\mathbf{e}_{0}$ gives
\begin{equation}
cn_{0\perp }\left( \tau -t^{\prime }\right) -cn_{0\perp }\left( t-\tau
\right) =z-Z\left( t\right)  \label{8.15}
\end{equation}
so that, from \oldref{(8.14)}(\ref{8.14}) and \oldref{(8.15)}(\ref{8.15})
\begin{equation}
n_{0\perp }=\frac{z-Z\left( t\right) }{c\left( 2\tau -t-t^{\prime }\right) }=%
\frac{z_{m}-Z\left( \tau \right) }{c\left( \tau -t^{\prime }\right) }
\label{8.16}
\end{equation}
while, from the inner product of \oldref{(8.13)}(\ref{8.13}) with itself, we have
\begin{equation}
\left( \tau -t^{\prime }\right) ^{2}+\left( t-\tau \right) ^{2}+2\left( \tau
-t^{\prime }\right) \left( t-\tau \right) \left( 1-2n_{0\perp }^{2}\right) =%
\frac{\left\vert \mathbf{x}-\mathbf{X}\left( t\right) \right\vert ^{2}}{c^{2}%
}  \label{8.17}
\end{equation}
where it was used that $\mathbf{n}_{0}\cdot \,\mathbf{\bar{n}}%
_{0}=\left\vert \mathbf{n}_{0}-n_{0\perp }\mathbf{e}_{0}\right\vert
^{2}-n_{0\perp }^{2}=1-2n_{0\perp }^{2}$.

The three equations given by \oldref{(8.16)}(\ref{8.16}) and \oldref{(8.17)}(\ref{8.17}) 
are the complete set to determine $n_{0\perp }$, $t^{\prime }$ and $\tau $, 
given the motion of the source and the observation position and time: $\mathbf{x}$
and $t$.

Unfortunately, the delay $t-t^{\prime }$, to be used instead of \oldref{(8.2)}(\ref{8.2}), 
does not have an analytical expression, and it depends not only on
the source position at $t$, but also at $\tau $ (the latter time is of
course well determined by the system \oldref{(8.16)}(\ref{8.16})-\oldref{(8.17)}(\ref{8.17})).
One useful expression can be obtained using the first equality in \oldref{(8.16)}(\ref{8.16})
together with \oldref{(8.17)}(\ref{8.17}), from which one readily obtains for
the case of non-normal reflection, $n_{0\perp }\neq -1$, that
\begin{equation}
t^{\prime }=t-\frac{\sqrt{\left[ x-X\left( t\right) \right] ^{2}+\left[
y-Y\left( t\right) \right] ^{2}}}{c\sqrt{1-2n_{0\perp }^{2}}}  \label{8.18}
\end{equation}
where of course $n_{0\perp }$ is a function of $t$ and $\mathbf{x}$
determined from relations \oldref{(8.16)}(\ref{8.16}) and \oldref{(8.17)}(\ref{8.17}).

For normal reflection one simply uses both equalities \oldref{(8.16)}(\ref{8.16}) with 
$n_{0\perp }=-1$ to obtain
\begin{equation*}
t^{\prime }=t-2\frac{Z\left( \tau \right) -z_{m}}{c}-\frac{z-Z\left(
t\right) }{c}
\end{equation*}
where $\tau $ is implicitly given by
\begin{equation*}
\tau =t-\frac{Z\left( \tau \right) -z_{m}}{c}-\frac{z-Z\left( t\right) }{c}
\end{equation*}

In any case, by writing in general that for a source $k$ one has
\begin{equation*}
t^{\prime }=t-\frac{D_{k}\left( \mathbf{x},t\right) }{c}
\end{equation*}
the general procedure to obtain the potentials can be carried out as before
remembering also that the ``reflected''
interaction has a sign change. Also, we consider from the beginning only
retarded interactions so that for $\mathbf{A}_{k}$ we take directly twice
the retarded expression, to write
\begin{equation*}
\Phi _{k}\left( \mathbf{X}_{i}\left( t\right) ,t\right) =-\frac{Q_{k}}{4\pi
\varepsilon _{0}}\frac{1}{D_{k}\left( \mathbf{X}_{i}\left( t\right)
,t\right) }
\end{equation*}
\begin{equation}
\mathbf{A}_{k}\left( \mathbf{X}_{i}\left( t\right) ,t\right) =-\frac{Q_{k}}{%
4\pi \varepsilon _{0}c^{2}}\frac{\mathbf{V}_{k}\left[ t-D_{k}\left( \mathbf{X%
}_{i}\left( t\right) ,t\right) /c\right] }{D_{k}\left( \mathbf{X}_{i}\left(
t\right) ,t\right) }  \label{8.19}
\end{equation}

Potentials \oldref{(8.19)}(\ref{8.19}) are then to be used in the action 
\oldref{(8.8)}(\ref{8.8}) to obtain, by the principle of least action under 
variations of $\mathbf{X}_{i}\left( t\right) $, the force of particles $k$ 
on particle $i$ through the reflected interaction. In this case, the directions 
$\mathbf{n}_{ik}$ appearing in \oldref{(8.8)}(\ref{8.8}) are to be interpreted as the
corresponding, post-reflection, $\mathbf{\bar{n}}_{0}$ directions.

\section{Spherical antenna}\label{sec:8.4}

In order to explore the possibility of the experimental verification of the
prediction by VRT of a small longitudinal component of the radiative
electric field, we consider a particular current distribution that generates
only the longitudinal part of the radiative field, a spherical antenna. The
moving charges in this antenna perform radial harmonic oscillations, and we
consider the superposition of $N$ isotropically distributed oscillating
charges, each one characterized by $Q_{k}/N$, $\mathbf{X}_{k}\left( t\right)
=r\left( t\right) \mathbf{e}_{k}$, $\mathbf{V}_{k}\left( t\right) =\dot{r}%
\left( t\right) \mathbf{e}_{k}$, where $\mathbf{e}_{k}$ is the local radial
unit vector. Considering the spherical symmetry we take the observation
point at a distance $R>>r\left( t\right) $ on the $z$ axis, of direction $%
\mathbf{e}_{z}$. By the symmetry of the distribution both, the electric and
magnetic transverse fields given by \oldref{(8.10)}(\ref{8.10}) are zero, and also the
acceleration term in Weber's force (last term in the first line in 
\oldref{(8.10)}(\ref{8.10})) also cancels. Thus, only the superposition of the
non-fully-transverse radiative field $\delta \mathbf{E}_{k}^{rad}$ of each
charge remains. From the last term of \oldref{(8.10)}(\ref{8.10}) we identify $\delta 
\mathbf{E}_{k}^{rad}$ at position $\mathbf{x}$ as
\begin{equation*}
\delta \mathbf{E}_{k}^{rad}=-\frac{Q_{k}}{4\pi \varepsilon _{0}c^{3}}\frac{1%
}{\left\vert \mathbf{x}-\mathbf{X}_{k}\left( t\right) \right\vert }\mathbf{V}%
_{k}\left( t\right) \times \left[ \mathbf{\dot{V}}_{k}\left( t_{ret}\right)
\times \mathbf{n}_{k}\right]
\end{equation*}
with $\mathbf{n}_{k}=\left( \mathbf{x}-\mathbf{X}_{k}\left( t\right) \right)
/\left\vert \mathbf{x}-\mathbf{X}_{k}\left( t\right) \right\vert $.

In this way, with $\mathbf{n}_{k}=$ $\mathbf{e}_{z}$, we have
\begin{equation*}
\mathbf{E}\left( R\mathbf{e}_{z},t\right) =\sum_{k=1}^{N}\delta \mathbf{E}%
_{k}^{rad}=\frac{Q\dot{r}\left( t\right) \ddot{r}\left( t_{ret}\right) }{%
4\pi \varepsilon _{0}c^{3}RN}\sum_{k=1}^{N}\left[ \mathbf{e}_{z}-\left( 
\mathbf{e}_{z}\cdot \mathbf{e}_{k}\right) \mathbf{e}_{k}\right] =\frac{Q\dot{%
r}\left( t\right) \ddot{r}\left( t_{ret}\right) }{6\pi \varepsilon _{0}c^{3}R%
}\mathbf{e}_{z}
\end{equation*}
where it was used that ($\mathbf{I}$ is the identity matrix)
\begin{equation*}
\frac{1}{N}\sum_{k=1}^{N}\mathbf{e}_{k}\mathbf{e}_{k}=\frac{1}{3}\mathbf{I}
\end{equation*}

For an harmonic motion of the form $r\left( t\right) =r_{0}+\Delta \sin
\left( \omega t\right) $ we finally have for the radial field at a distance $%
R$ from the center of the antenna
\begin{equation}
\mathbf{E}\left( R,t\right) =-\frac{Q\Delta ^{2}\omega ^{3}}{6\pi
\varepsilon _{0}c^{3}R}\cos \left( \omega t\right) \sin \left( \omega
t-kR\right) \mathbf{e}_{R}  \label{8.20}
\end{equation}
with $\mathbf{e}_{R}$ the radial unit vector, and $k=\omega /c$.

If this field is to be detected by the currents it induces in a conducting
circuit, one should first note that expression \oldref{(8.20)}(\ref{8.20})
can be rewritten as
\begin{equation*}
\mathbf{E}\left( R,t\right) =-\frac{Q\Delta ^{2}\omega ^{3}}{6\pi
\varepsilon _{0}c^{3}R}\left[ \sin \left( 2\omega t-kR\right) -\sin \left(
kR\right) \right] \mathbf{e}_{R}
\end{equation*}
so that this field induces an oscillating current plus a constant charge
polarization on the detector, the latter not contributing to any electrical
work on the detecting circuit. In this way, the power detected is
proportional to the time average of the square of only the oscillating part,
resulting in power decaying with the distance as $R^{-2}$.

In a real experiment, however, there are also contributions from the fields
reflected from nearby surfaces. Using the ideas of section 8.2 we will
include the radiative terms of the reflected interaction for the spherical
antenna over a conducting infinite plane. To identify the ``reflected'' 
force on a charge at $\mathbf{X}_{i}\left(t\right) $ we make variations 
of the action \oldref{(8.8)}(\ref{8.8}) using the potentials 
\oldref{(8.19)}(\ref{8.19}) and interpreting $\mathbf{n}_{ik}$ as the
corresponding $\,\mathbf{\bar{n}}_{0}$ direction. As only the radiative
terms are looked for the derivation is greatly simplified by the fact that
variations of $\Phi _{k}$\ and $\mathbf{n}_{ik}$, as well as their time
derivatives, generate terms that decay as the square of the distance, that
is, non-radiative terms. The resulting force for a single particle \textit{k}
acting on particle \textit{i}, and decaying as the inverse of the distance
thus results to be%
\begin{eqnarray}
\mathbf{F}_{ik}^{rad} &=&-\frac{Q_{i}Q_{k}}{4\pi \varepsilon
_{0}c^{2}D_{k}\left( \mathbf{X}_{i}\right) }\biggl\{ \left( \mathbf{\bar{n}}%
_{ik}\cdot \Delta \mathbf{\dot{V}}_{ik}\right) \mathbf{\bar{n}}_{ik}+\left[ 
\mathbf{\dot{V}}_{k}\left( t_{ret}\right) \cdot \mathbf{\bar{n}}_{ik}\right] 
\mathbf{\bar{n}}_{ik}-\mathbf{\dot{V}}_{k}\left( t_{ret}\right) \biggr.
 \notag \\
&&\biggl. +\mathbf{\dot{V}}_{k}\left( t_{ret}\right) \cdot \left[ \left( 
\mathbf{\bar{n}}_{ik}\cdot \Delta \mathbf{V}_{ik}\right) \mathbf{\bar{n}}%
_{ik}-\Delta \mathbf{V}_{ik}\right] \frac{\partial D_{k}}{\partial \mathbf{X}%
_{i}}\biggr\}  \label{8.21}
\end{eqnarray}%
where $\mathbf{\bar{n}}_{ik}$ indicates the reflected direction, $%
t_{ret}=t-D_{k}\left( \mathbf{X}_{i}\left( t\right) ,t\right) /c$, and where
unless explicitly noted all magnitudes are evaluated at time $t$.

Proceeding as with the direct interaction with the charges in the antenna,
for a far away charge the direction $\mathbf{\bar{n}}_{ik}$ is the same for
all $k$ (higher order corrections contribute only to non-radiative terms),
so that after summation over all symmetrically distributed charges, only the
terms in the second line of \oldref{(8.21)}(\ref{8.21}) that involve products of 
$\mathbf{\dot{V}}_{k}$ and $\mathbf{V}_{k}$ survive (first order corrections
in the retarded time in the $\mathbf{\dot{V}}_{k}$ appearing in the first
line of \oldref{(8.21)}(\ref{8.21}) also cancel due to the transversality condition).
In this way, the reflected contribution of the electric field at a generic
(far away from the antenna) point $\mathbf{x}$ is
\begin{equation*}
\mathbf{E}_{ref}\left( \mathbf{x},t\right) =-\frac{Q\dot{r}\left( t\right) 
\ddot{r}\left( t_{ret}\right) }{6\pi \varepsilon _{0}c^{3}D\left( \mathbf{x}%
\right) }\frac{\partial D}{\partial \mathbf{x}}
\end{equation*}
where $D\left( \mathbf{x}\right) /c$ measures the distance from the center
of the antenna to the position $\mathbf{x}$ along the reflected path, so
that $t_{ret}=t-D\left( \mathbf{x}\right) /c$. Written explicitly for the
harmonic charge motion, the reflected field is
\begin{equation}
\mathbf{E}_{ref}\left( \mathbf{x},t\right) =\frac{Q\Delta ^{2}\omega ^{3}}{%
6\pi \varepsilon _{0}c^{3}D\left( \mathbf{x}\right) }\cos \left( \omega
t\right) \sin \left( \omega t-kR\right) \frac{\partial D}{\partial \mathbf{x}%
}  \label{8.22}
\end{equation}

The measured field is then the purely oscillating part of the superposition
of \oldref{(8.20)}(\ref{8.20}) and \oldref{(8.22)}(\ref{8.22}), noting also that 
$D\left( \mathbf{x}\right) $ is simply the distance form the center of the image
antenna to the observation point.

\begin{figure}[htb]
\begin{center}
\includegraphics[scale=0.5]{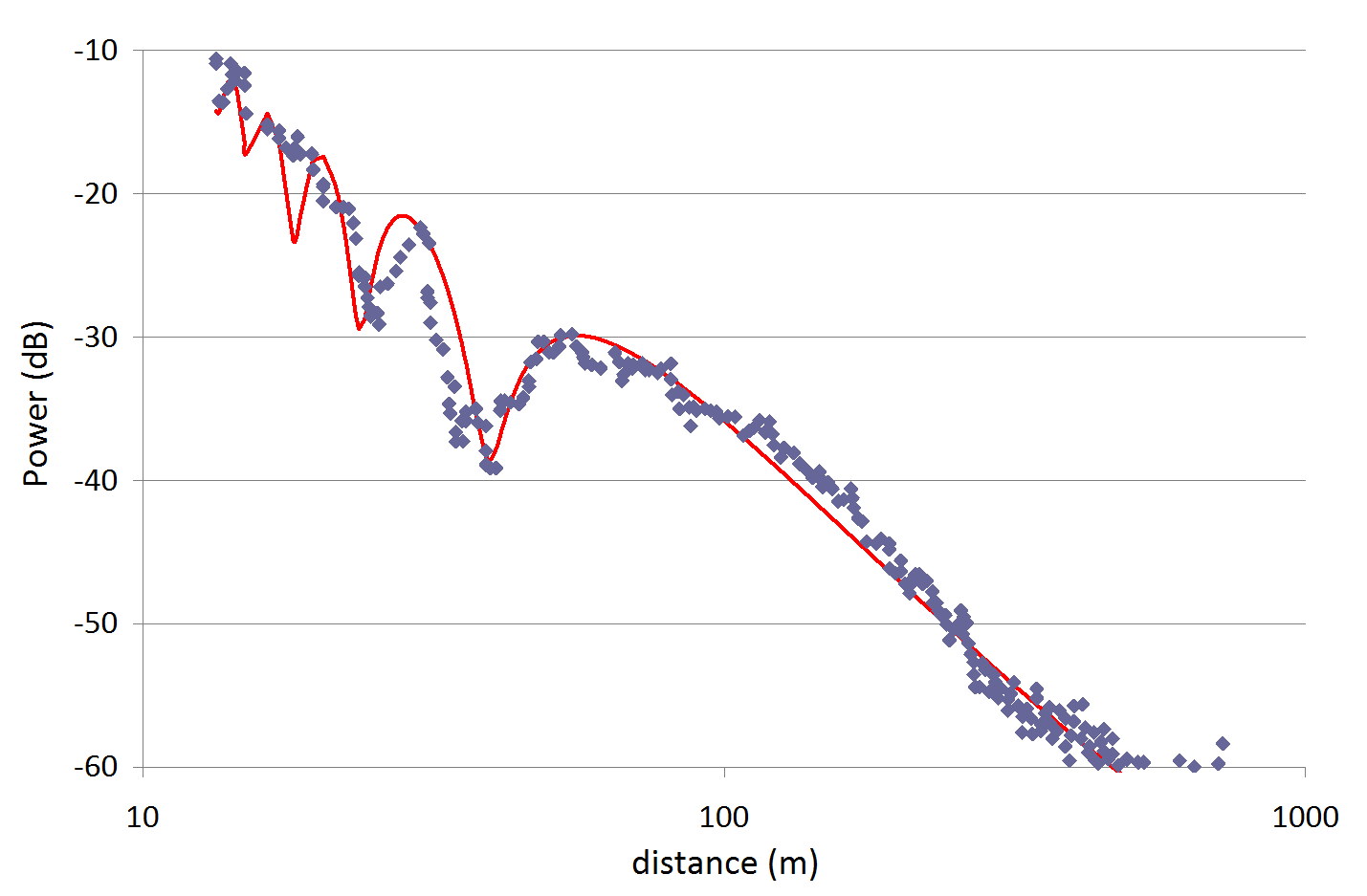}
\caption{Power of longitudinal electric field radiated from spherical
antenna for the conditions of Monstein and Wesley experiment. Blue squares:
experimental values, and solid red line: theoretical values using VRT.}
\label{fig:8.1}
\end{center}
\end{figure}

As mentioned above, the measured power is proportional to the time average
of the oscillating part of the superposition of \oldref{(8.20)}(\ref{8.20}) and 
\oldref{(8.22)}({8.22}), which we represent in figure \oldref{8.1}\ref{fig:8.1}
for conditions similar to the experiment by Monstein and Wesley 
\oldref{[R94]}\cite{monstein2002observation}. 
The best fit, adjusting the amplitude of the reflected field,\oldref{(8.22)}
(\ref{8.22}), and the height of the receiving antenna, was obtained
with an amplitude of the reflected field of 70\% the incident amplitude (about 50\% of energy reflected,
similar to the theoretical model of \oldref{[R94]}\cite{monstein2002observation}), 
and using a vertical support of the antenna 20\% shorter than the reported height 
of the antenna. The fit is statistically significant, $p<10^{-4}$, $R^{2}=0.970$,
meaning that the measurements support the model given by the superposition of
\oldref{(8.20)}(\ref{8.20}) and \oldref{(8.22)}(\ref{8.22}).

\newpage
\section*{Appendix 8.A. Derivation of VRT action}\label{sec:8.A}
\addcontentsline{toc}{section}{Appendix 8.A. Derivation of VRT action}

To determine a correct Galilean covariant action we start with the notion that
in VRT the frame of the emitting particle is a privileged one, so it is
natural to refer the velocity of the interacting particle $i$ to that frame
(at the time $t$ of reception) to write, instead of the Fokker action
\oldref{(8.3)}(\ref{8.3}), the modified action
\begin{eqnarray}
S &=&\sum_{i}\frac{m_{i}}{2}\int \left\vert \mathbf{V}_{i}\left( t\right)
\right\vert ^{2}dt  \label{8.A.1} \\
&&
- {\displaystyle\sum\limits_{i}}
\sum_{k\neq i}\frac{Q_{i}Q_{k}}{4\pi \varepsilon _{0}c}%
\int 
{\displaystyle\int}
\delta \left[ \hat{s}_{ik}^{2}\right] \left[ c^{2}-\left[ \mathbf{%
V}_{i}\left( t\right) -\mathbf{V}_{k}\left( t\right) \right] \cdot \mathbf{V}%
_{k}\left( t^{\prime }\right) \right] dtdt^{\prime }  \notag
\end{eqnarray}
where the single particle contribution is now given in Galilean form, and 
\oldref{(8.5)}(\ref{8.5}) is used in the argument of the delta function. The
interaction term of this expression is still not Galilean invariant due to
the presence of $\mathbf{V}_{k}\left( t^{\prime }\right) $. In order to
obtain a fully Galilean action we observe the following. The integration
over $t^{\prime }$ in \oldref{(8.A.1)}(\ref{8.A.1}) of the term involving 
$\mathbf{V}_{k}\left( t^{\prime }\right) $ gives a result proportional to
\begin{equation}
{\displaystyle\int}
\delta \left[ \hat{s}_{ik}^{2}\right] \mathbf{V}_{k}\left( t^{\prime
}\right) dt^{\prime }=\frac{1}{2c}\frac{\mathbf{V}_{k}\left( t_{+}\right) +%
\mathbf{V}_{k}\left( t_{-}\right) }{\left\vert \mathbf{X}_{i}\left( t\right)
-\mathbf{X}_{k}\left( t\right) \right\vert }  \label{8.A.2}
\end{equation}
where $t_{\pm }$ are the two roots of \oldref{(8.5)}(\ref{8.5}). In 
this way, if the action \oldref{(8.A.1}(\ref{8.A.1}) is expressed in a 
different inertial frame, moving relatively to the original one with velocity 
$\mathbf{V}_{0}$, a new term appears of the form
\begin{equation*}
{\displaystyle\sum\limits_{i}}
\sum_{k\neq i}\frac{Q_{i}Q_{k}}{4\pi \varepsilon _{0}c^{2}}%
{\displaystyle\int}
\frac{\left[ \mathbf{V}_{i}\left( t\right) -\mathbf{V}_{k}\left(
t\right) \right] \cdot \mathbf{V}_{0}}{\left\vert \mathbf{X}_{i}\left(
t\right) -\mathbf{X}_{k}\left( t\right) \right\vert }dt
\end{equation*}
in which the integrand cannot be reduced to a total time derivatives.
However, it is a simple matter to see that if only the component of $\mathbf{%
V}_{i}\left( t\right) -\mathbf{V}_{k}\left( t\right) $ transverse to $%
\mathbf{X}_{i}\left( t\right) -\mathbf{X}_{k}\left( t\right) $ is used in %
\oldref{(8.A.1)}(\ref{8.A.1}), then the term generated in the action by a 
change of inertial frame corresponds to integrals in $t$ of a total time 
derivative, thus not contributing to the variation (with fixed end points) 
of the action. Indeed, if we replace $\mathbf{V}_{i}\left( t\right) -\mathbf{V}
_{k}\left( t\right) $ in \oldref{(8.A.1)}(\ref{8.A.1}) by its transverse component
\begin{equation}
\Delta \mathbf{V}_{ik}\left( t\right) -\left( \Delta \mathbf{V}_{ik}\left(
t\right) \cdot \mathbf{n}_{ik}\right) \mathbf{n}_{ik}  \label{8.A.3}
\end{equation}
where $\Delta \mathbf{V}_{ik}\left( t\right) =\mathbf{V}_{i}\left( t\right) -
\mathbf{V}_{k}\left( t\right) $ and $\mathbf{n}_{ik}=\left[ \mathbf{X}%
_{i}\left( t\right) -\mathbf{X}_{k}\left( t\right) \right] /\left\vert 
\mathbf{X}_{i}\left( t\right) -\mathbf{X}_{k}\left( t\right) \right\vert $ ,
the terms generated in the action by a Galilean change of frame are now

\begin{eqnarray*}
&&{\displaystyle\sum\limits_{i}}
\sum_{k\neq i}\frac{Q_{i}Q_{k}}{4\pi \varepsilon _{0}c^{2}} 
{\displaystyle\int}
\frac{\left[ \Delta \mathbf{V}_{ik}\left( t\right) -\left( \Delta 
\mathbf{V}_{ik}\left( t\right) \cdot \mathbf{n}_{ik}\right) \mathbf{n}_{ik}
\right] \cdot \mathbf{V}_{0}}{\left\vert \mathbf{X}_{i}\left( t\right) -
\mathbf{X}_{k}\left( t\right) \right\vert }dt \\
&=&
{\displaystyle\sum\limits_{i}}
\sum_{k\neq i}\frac{Q_{i}Q_{k}}{4\pi \varepsilon _{0}c^{2}%
}\int \frac{d}{dt}\left( \mathbf{n}_{ik}\cdot \mathbf{V}_{0}\right) dt
\end{eqnarray*}
thus not contributing to the variation of the action.

However, the simple replacement of $\Delta \mathbf{V}_{ik}\left( t\right) $
by its transverse component \oldref{(8.A.3)}(\ref{8.A.3}) in \oldref{(8.A.1)}(\ref{8.A.1}) 
is not enough, for the component parallel to $\mathbf{n}_{ik}$ needs to be
included in order to obtain a consistent theory (as shown in the main text).
Quadratic expressions involving these parallel components can only be
included if they are evaluated only at the interaction time $t$ (as further
discussed in the main text) to give the final expression \oldref{(8.6)}(\ref{8.6})
in the main text.

\newpage
\section*{Appendix 8.B. Weber's quasistationary electrodynamics laws are not affected by
the radiative terms of VRT}\label{sec:8.B}
\addcontentsline{toc}{section}{Appendix 8.B. Weber's quasistationary electrodynamics laws are not affected by
the radiative terms of VRT}

For the case of relatively slowly varying currents in nearby circuits,
retardation effects can be neglected, and so the radiative electric force
results in an expression similar to the term depending on acceleration in
Weber's force. Since the latter accounts for Faraday's law of induction
between closed circuits, it is important to make sure that the new terms do
not modify in an unacceptable way that result. To check this we consider a
closed filamentary circuit $C_{k}$ in which the circulating charges $Q_{k}$
give rise to a neutral current $I_{k}$. Since the current is neutral, each
element $d\mathbf{X}_{k}$ of the circuit contains equal amounts of positive
and negative charge (the positive one denoted as $\delta Q_{k}^{+}$), which
have velocities $\mathbf{V}_{k}^{+}$ and $\mathbf{V}_{k}^{-}$, respectively.
These velocities have in general a component along the circuit element $d%
\mathbf{X}_{k}$, responsible for the current $I_{k}$, plus a component due
to possible motion of the element $d\mathbf{X}_{k}$, resulting from rigid
roto-translations and also deformations of the circuit. However, as the
velocity due to the motion of $d\mathbf{X}_{k}$ is the same for charges of
both signs, this component does not contribute to the current. We can thus
in general write that $I_{k}d\mathbf{X}_{k}=\delta Q_{k}^{+}\left( \mathbf{V}%
_{k}^{+}-\mathbf{V}_{k}^{-}\right) $, noting that, due to what was just
said, the velocity difference has component only along $d\mathbf{X}_{k}$.

Now, if we consider an element $d\mathbf{X}_{i}$ located at $\mathbf{X}_{i}$%
, and belonging to a nearby circuit $C_{i}$, the electromotive force induced
in that portion of the circuit by the radiative electric terms (in the
non-retarded situation considered) due to the charges in the element $d%
\mathbf{X}_{k}$, located at $\mathbf{X}_{k}$, is given by
\begin{equation}
d\varepsilon =\frac{1}{4\pi \varepsilon _{0}c^{2}}\frac{\delta Q_{k}^{+}}{%
\left\vert \mathbf{X}_{i}-\mathbf{X}_{k}\right\vert }d\mathbf{X}_{i}\cdot
\left\{ \left[ \left( \mathbf{\dot{V}}_{k}^{+}-\mathbf{\dot{V}}%
_{k}^{-}\right) \cdot \mathbf{n}_{ik}\right] \mathbf{n}_{ik}-\left( \mathbf{%
\dot{V}}_{k}^{+}-\mathbf{\dot{V}}_{k}^{-}\right) \right\}  \label{8.B.1}
\end{equation}

As with the velocities, the acceleration of the charges in $d\mathbf{X}_{k}$
has a component due to the acceleration of the element $d\mathbf{X}_{k}$
itself, common to all charges, plus the component corresponding to the
motion of charges along $d\mathbf{X}_{k}$. The latter can be divided into
that along $d\mathbf{X}_{k}$, $\mathbf{\dot{V}}_{k\parallel }^{\pm }$, which
is related to the time variation of the current $I_{k}$, and a perpendicular
component of direction $\mathbf{n}_{\perp }$ due to the local radius of
curvature $R_{c}$ of the circuit. We can thus write%
\begin{eqnarray*}
\delta Q_{k}^{+}\left( \mathbf{\dot{V}}_{k}^{+}-\mathbf{\dot{V}}%
_{k}^{-}\right) &=&\delta Q_{k}^{+}\left( \mathbf{\dot{V}}_{k\parallel }^{+}-%
\mathbf{\dot{V}}_{k\parallel }^{-}\right) +\frac{\delta Q_{k}^{+}}{R_{c}}%
\left( \mathbf{V}_{k\parallel }^{+}\cdot \mathbf{V}_{k\parallel }^{+}-%
\mathbf{V}_{k\parallel }^{-}\cdot \mathbf{V}_{k\parallel }^{-}\right) 
\mathbf{n}_{\perp } \\
&=&\delta Q_{k}^{+}\left( \mathbf{\dot{V}}_{k\parallel }^{+}-\mathbf{\dot{V}}%
_{k\parallel }^{-}\right) +\frac{\delta Q_{k}^{+}}{R_{c}}\left[ \left( 
\mathbf{V}_{k\parallel }^{+}-\mathbf{V}_{k\parallel }^{-}\right) \cdot
\left( \mathbf{V}_{k\parallel }^{+}+\mathbf{V}_{k\parallel }^{-}\right) %
\right] \mathbf{n}_{\perp } \\
&=&\dot{I}_{k}d\mathbf{X}_{k}+I_{k}\left[ d\mathbf{X}_{k}\cdot \left( 
\mathbf{V}_{k\parallel }^{+}+\mathbf{V}_{k\parallel }^{-}\right) \right] 
\frac{\mathbf{n}_{\perp }}{c}
\end{eqnarray*}%
If $\mathbf{n}_{\parallel }$ designates the unit vector along the circuit $%
C_{k}$, we can write
\begin{equation*}
\left[ d\mathbf{X}_{k}\cdot \left( \mathbf{V}_{k\parallel }^{+}+\mathbf{V}%
_{k\parallel }^{-}\right) \right] \frac{\mathbf{n}_{\perp }}{c}=\left[ 
\mathbf{n}_{\parallel }\cdot \left( \mathbf{V}_{k\parallel }^{+}+\mathbf{V}%
_{k\parallel }^{-}\right) \right] d\mathbf{n}_{\parallel }=\left(
V_{k\parallel }^{+}+V_{k\parallel }^{-}\right) d\mathbf{n}_{\parallel }
\end{equation*}
to have
\begin{equation}
\delta Q_{k}^{+}\left( \mathbf{\dot{V}}_{k}^{+}-\mathbf{\dot{V}}%
_{k}^{-}\right) =\dot{I}_{k}d\mathbf{X}_{k}+I_{k}\left( V_{k\parallel
}^{+}+V_{k\parallel }^{-}\right) d\mathbf{n}_{\parallel }  \label{8.B.2}
\end{equation}
where $V_{k\parallel }^{\pm }$ are the longitudinal components of the drift
velocities, and $d\mathbf{n}_{\parallel }$ is the change in the longitudinal
unit vector associated to the displacement $d\mathbf{X}_{k}$ along the
circuit.

With all this we can finally write that
\begin{eqnarray}
d\varepsilon &=&\frac{1}{4\pi \varepsilon _{0}c^{2}}\frac{\dot{I}_{k}}{%
\left\vert \mathbf{X}_{i}-\mathbf{X}_{k}\right\vert }\left[ \left( \mathbf{n}%
_{ik}\cdot d\mathbf{X}_{k}\right) \left( \mathbf{n}_{ik}\cdot d\mathbf{X}%
_{i}\right) -d\mathbf{X}_{k}\cdot d\mathbf{X}_{i}\right]  \label{8.B.3} \\
&&+\frac{1}{4\pi \varepsilon _{0}c^{2}}\frac{I_{k}\left( V_{k\parallel
}^{+}+V_{k\parallel }^{-}\right) }{\left\vert \mathbf{X}_{i}-\mathbf{X}%
_{k}\right\vert }\left[ \left( \mathbf{n}_{ik}\cdot d\mathbf{n}_{\parallel
}\right) \left( \mathbf{n}_{ik}\cdot d\mathbf{X}_{i}\right) -d\mathbf{n}%
_{\parallel }\cdot d\mathbf{X}_{i}\right]  \notag
\end{eqnarray}

Following \oldref{[R93]}\cite{assis1994weber} we can now 
show that the integral of each of both lines in
the right hand side of this equation is zero when extended to the closed
circuit $C_{i}$ (needed to determine the electromotive force induced in that
whole circuit by the charge motions in the single element $d\mathbf{X}_{k}$).
Indeed, we consider the general integral
\begin{equation}
\oint_{C_{i}}\frac{\left( \mathbf{n}_{ik}\cdot d\mathbf{X}_{k}\right) \left( 
\mathbf{n}_{ik}\cdot d\mathbf{X}_{i}\right) }{\left\vert \mathbf{X}_{i}-%
\mathbf{X}_{k}\right\vert }=\int_{S_{i}}\left[ \nabla _{i}\times \frac{%
\left( \mathbf{n}_{ik}\cdot d\mathbf{X}_{k}\right) \mathbf{n}_{ik}}{%
\left\vert \mathbf{X}_{i}-\mathbf{X}_{k}\right\vert }\right] \cdot d\mathbf{S%
}_{i}  \label{8.B.4}
\end{equation}
where Stokes theorem was used to write the right hand side of \oldref{(8.B.4)}(\ref{8.B.4}). 
Since $\mathbf{n}_{ik}=\nabla _{i}\left\vert \mathbf{X}_{i}-\mathbf{%
X}_{k}\right\vert $ we have by direct derivations that
\begin{equation*}
\nabla _{i}\times \frac{\left( \mathbf{n}_{ik}\cdot d\mathbf{X}_{k}\right) 
\mathbf{n}_{ik}}{\left\vert \mathbf{X}_{i}-\mathbf{X}_{k}\right\vert }%
=\nabla _{i}\left[ \frac{\mathbf{n}_{ik}\cdot d\mathbf{X}_{k}}{\left\vert 
\mathbf{X}_{i}-\mathbf{X}_{k}\right\vert }\right] \times \mathbf{n}_{ik}=%
\frac{d\mathbf{X}_{k}}{\left\vert \mathbf{X}_{i}-\mathbf{X}_{k}\right\vert
^{2}}\times \mathbf{n}_{ik}=\nabla _{i}\times \frac{d\mathbf{X}_{k}}{%
\left\vert \mathbf{X}_{i}-\mathbf{X}_{k}\right\vert }
\end{equation*}
from which, using again Stokes theorem, we have
\begin{equation}
\oint_{C_{i}}\frac{\left( \mathbf{n}_{ik}\cdot d\mathbf{X}_{k}\right) \left( 
\mathbf{n}_{ik}\cdot d\mathbf{X}_{i}\right) }{\left\vert \mathbf{X}_{i}-%
\mathbf{X}_{k}\right\vert }=\oint_{C_{i}}\frac{d\mathbf{X}_{k}\cdot d\mathbf{%
X}_{i}}{\left\vert \mathbf{X}_{i}-\mathbf{X}_{k}\right\vert }  \label{8.B.5}
\end{equation}

In this way, the integral of the first line of \oldref{(8.B.3)}(\ref{8.B.3}) is zero
when extended to the closed circuit $C_{i}$.

Noting that in the previous derivation the vector $d\mathbf{X}_{k}$ played
only the role of an arbitrary vector, constant in the integration along $%
C_{i}$, we need simply to replace $d\mathbf{X}_{k}$ by $d\mathbf{n}%
_{\parallel }$ in \oldref{(8.B.5)}(\ref{8.B.5}) to see that also the integral 
along $C_{i}$ of the second line is zero.

In the same way we see that the result is rather general, as we could have
started directly from \oldref{(8.B.1)}(\ref{8.B.1}) and use relation \oldref{(8.B.5)}(\ref{8.B.5}) 
with $\delta Q_{k}^{+}\left( \mathbf{\dot{V}}_{k}^{+}-\mathbf{\dot{V}%
}_{k}^{-}\right) $ instead of $d\mathbf{X}_{k}$. Even the condition of
neutral current is not necessary as we could have as well used $\delta
Q_{k}^{+}\mathbf{\dot{V}}_{k}^{+}+\delta Q_{k}^{-}\mathbf{\dot{V}}_{k}^{-}$,
with and arbitrary amount of negative charge $\delta Q_{k}^{-}$ in the
circuit element $d\mathbf{X}_{k}$.

It is still necessary to evaluate the effect of the remaining novel terms.
In the same situation considered for the radiative electric force, the
electromotive force induced by those terms is given as
\begin{eqnarray}
d\varepsilon ^{\prime } &=&\frac{\delta Q_{k}^{+}}{4\pi \varepsilon _{0}c^{3}%
}\frac{d\mathbf{X}_{i}}{\left\vert \mathbf{X}_{i}-\mathbf{X}_{k}\right\vert }%
\cdot \biggl\{ \mathbf{V}_{i}\times \left[ \left( \mathbf{\dot{V}}_{k}^{+}-%
\mathbf{\dot{V}}_{k}^{-}\right) \times \mathbf{n}_{ik}\right] \biggr.
 \notag \\
&&\biggl. -\mathbf{V}_{k}^{+}\mathbf{\times }\left[ \mathbf{\dot{V}}%
_{k}^{+}\times \mathbf{n}_{ik}\right] +\mathbf{V}_{k}^{-}\mathbf{\times }%
\left[ \mathbf{\dot{V}}_{k}^{-}\times \mathbf{n}_{ik}\right] \biggr\} \label{8.B.6}
\end{eqnarray}

It is clear, from a comparison with \oldref{(8.B.1)}(\ref{8.B.1}), which has the same
order of magnitude as that in Weber's electrodynamics, that the magnitude of 
\oldref{(8.B.6)}(\ref{8.B.6}) is a factor $U_{d}/c$ smaller, with $U_{d}$ the drift
velocity of the charges. It is then clear that any contribution from this
term is negligible in any common situation.

Finally, always in the situation in which retarded effects can be neglected,
we must consider the effect of the terms added to the original Weber's force
on the derivation of Amp\`{e}re's force between neutral current elements. It
is clear that the radiative electric force does not have any effect on a
neutral element, and the same happens with its non-transverse part. The only
possible added force comes from the radiative magnetic force, and to
consider it we must include the motion of charges in the element $d\mathbf{X}
_{k}$ (electrically neutral, and with positive amount of charge $\delta
Q_{i}^{+}$), to write the additional force on $d\mathbf{X}_{i}$ due to $d
\mathbf{X}_{k}$ as
\begin{eqnarray*}
\delta \mathbf{F}_{ik}^{\prime }=\frac{\delta Q_{i}^{+}\delta Q_{k}^{+}}{4\pi
\varepsilon _{0}c^{3}}\frac{1}{\left\vert \mathbf{X}_{i}-\mathbf{X}_{k}\right\vert }
\biggl\{ 
\mathbf{V}_{i}^{+}\times \left[ \mathbf{\dot{V}}_{k}^{+}\times \mathbf{n}_{ik}\right] 
+\mathbf{V}_{i}^{-}\times \left[\mathbf{\dot{V}}_{k}^{-}\times \mathbf{n}_{ik}\right]
\biggr. \\
\biggl. 
-\mathbf{V}_{i}^{+}\times \left[ \mathbf{\dot{V}}_{k}^{-}\times \mathbf{n}_{ik}\right]
-\mathbf{V}_{i}^{-}\times \left[ \mathbf{\dot{V}}_{k}^{+}\times \mathbf{n}_{ik}\right]
 \biggr\} 
\end{eqnarray*}
which after some rearrangement and use of vector identities can be recast as
\begin{eqnarray*}
\delta \mathbf{F}_{ik}^{\prime }=\frac{\delta Q_{i}^{+}\delta Q_{k}^{+}}{4\pi
\varepsilon _{0}c^{3}}\frac{1}{\left\vert \mathbf{X}_{i}-\mathbf{X}%
_{k}\right\vert }
\biggl\{ 
\left[ \left( \mathbf{V}_{i}^{+}-\mathbf{V}%
_{i}^{-}\right) \cdot \mathbf{n}_{ik}\right] \left( \mathbf{\dot{V}}_{k}^{+}-
\mathbf{\dot{V}}_{k}^{-}\right) 
\biggr. \\
\biggl. 
-\left[ \left( \mathbf{V}_{i}^{+}-\mathbf{V}%
_{i}^{-}\right) \cdot \left( \mathbf{\dot{V}}_{k}^{+}-\mathbf{\dot{V}}
_{k}^{-}\right) \right] \mathbf{n}_{ik}
\biggr\}
\end{eqnarray*}
Since $I_{i}d\mathbf{X}_{i}=\delta Q_{i}^{+}\left( \mathbf{V}_{i}^{+}-
\mathbf{V}_{i}^{-}\right) $, using also \oldref{(8.B.2)}(\ref{8.B.2}), we can write
\begin{eqnarray}
\delta \mathbf{F}_{ik}^{\prime } &=&\frac{1}{4\pi \varepsilon _{0}c^{3}}
\frac{I_{i}}{\left\vert \mathbf{X}_{i}-\mathbf{X}_{k}\right\vert }\biggl\{
\left( d\mathbf{X}_{i}\cdot \mathbf{n}_{ik}\right) \left[ \dot{I}_{k}d
\mathbf{X}_{k}+I_{k}\left( V_{k\parallel }^{+}+V_{k\parallel }^{-}\right) d
\mathbf{n}_{\parallel }\right] \biggr. \notag \\
&&\biggl. -d\mathbf{X}_{i}\cdot \left[ \dot{I}_{k}d\mathbf{X}_{k}+I_{k}\left(
V_{k\parallel }^{+}+V_{k\parallel }^{-}\right) d\mathbf{n}_{\parallel }
\right] \mathbf{n}_{ik}\biggr\}   \label{8.B.7}
\end{eqnarray}

As can be seen, this force depends in a rather complex way on the time
variation of the current $I_{k}$, as well as on the curvature and drift
velocities of the charges. However, as the magnitude of Amp\`{e}re's force
between the considered current elements is about
\begin{equation*}
\delta \mathbf{F}_{Ampere}=\frac{1}{4\pi \varepsilon _{0}c^{3}}\frac{
I_{i}I_{k}}{\left\vert \mathbf{X}_{i}-\mathbf{X}_{k}\right\vert ^{2}}
dX_{i}dX_{k}
\end{equation*}
we see that \oldref{(8.B.7)}(\ref{8.B.7}) is much smaller, by factors about $U_{d}/c$
and $\tau _{c}/\tau _{I}$, where $U_{d}$ is the drift velocity of charges in
the conductor (multiplied by the distance between current elements and
divided by the radius of curvature of the conductor at the point $\mathbf{X}
_{k}$ considered), while $\tau _{c}$ is the time it takes light to travel
the distance between current elements, and $\tau _{I}$ is the characteristic
time of variation of the current $I_{k}$. Both contributions are clearly
negligible in any practical situation.


%% file: 2014arxiv_v14_3_9.tex


\chapter{Propagation of light emitted from sources with different
speed}\label{ch:9}

\fancyhf{}

\fancyhead[LE,RO]{9. Propagation of light from different sources}
\fancyhead[RE,LO]{L. Bernal, L. Bilbao}
\fancyfoot[RE,LO]{Vibrating Rays Theory}
\fancyfoot[LE,RO]{\thepage}
 
\renewcommand{\headrulewidth}{1pt}
\renewcommand{\footrulewidth}{1pt}

\section{Introduction}\label{sec:9.1}

Although VRT is radically different from SRT, it is not easy to design an
experiment that brings out the differences between them. The present work is
aimed to measure the propagation of light produced by different macroscopic
moving sources relative to a detector. After more than a century of
relativity there is no experiment that directly compares the speed of
propagation of light produced by extensive macroscopic sources moving with
different speeds under adequate conditions in order to discern between VRT
and SRT.

It should be noted that many of the experiments relating to mobile sources
relate, in fact, to moving images of stationary or moving sources, produced
by transparent media or movable mirrors, see, for example,\oldref{[R37]}\oldref{[R38]} 
\cite{babcock1964determination,beckmann1965test}. Even
the famous experience of Ives and Stilwell \oldref{[R39]}\cite{ives1938experimental} 
suffers from this defect of interpretation. Observations made with moving 
mirrors bear no relation to those with moving sources, and these may produce 
different consequences. In Chapter \oldref{1}\ref{ch:1} we mentioned that a 
moving source exhibits time dilatation (i.e., twin paradox) while a moving image, 
produced by a moving mirror, does not. Therefore, a moving image experiment 
cannot be considered as a moving source experiment.

\section{Main features of VRT}\label{sec:9.2}

In order to design an experiment that distinguishes VRT from SRT, one must
first find the main differences between theories.

In short, the interpretation of VRT is:

\begin{list}{(\arabic{qcounter})}{\usecounter{qcounter}}
\item Light propagates in vacuum with constant velocity $c$ only in a
non-rotating system fixed to the source, no matter how the source moves.
\item The wave equation is not invariant under a Galilean
transformation. It only applies in the proper frame of the source, in the
same way that the mechanical wave equations can only be used in the system
attached to the transmission material.
\item The travel time of a pulse is independent of the reference system
and it can be calculated as the distance between the source and detector
(both at time of reception) divided by $c$.
\item The speed of light in any system is the sum of $c$ plus the
average speed of the source between the epoch of emission and the epoch of
reception.
\end{list}

This can be summarized by saying that in the case of VRT, the speed of light
as measured by an observer is $c$ plus the speed of the source at the
instant of detection. Indeed, an experiment to test VRT requires that at the
instant of detection a simultaneous measurement of the speed of light and of
the speed of the source is performed. This point is crucial and so far there
is not known experiment in which these values were measured simultaneously.
For example, in the experiment by Alvager and others \oldref{[R4]} 
\cite{alvager1964measuring}, where high-energy
particles striking a beryllium target produce pions moving at $.99975c$,
which, in turn, decays generating two photons moving in opposite direction,
the speed of light is measured at more than 60 meters from the beryllium
target, while the speed of the pions were measured at time of emission (i.e.
about 200 ns before the detection of photons). No simultaneous measurement
of the source velocity at the time of photon detection is reported.

Finally, some words on the Extinction Theorem. Usually, it is considered
that a transparent material absorbs and re-emits radiation, resulting in
that the speed of light is linked to the media and not to the original
source, so that it would seem impossible to prove VRT, except in situations
of high vacuum. However, as Tolman quoted in 1910 \oldref{[R95]}\cite{tolman1910second} 
``the possibility that an original difference in velocity would be destroyed when 
the light reached the neighborhood of the Earth is not entirely excluded.
Nevertheless, the experiments of Fizeau and Michelson seem to show that the
presence of air or other transmitting medium would not completely destroy
such a difference.''

In addition, according to Dingle, one obvious objection to the usual way of
interpreting the extinction theorem is that if one states that the source of
the light that emerges from the material medium are the atoms or molecules
of that medium, then the frequency spectrum should be that of the medium and
not that of the original source. However, it is not so. For example, if you
see the light of a mercury lamp through a glass, the spectrum observed is
that of the mercury lamp and not that of the glass, so if the speed of light
is related to the glass particles (considered the new source of the
emission), why does the spectrum correspond to the original source?
Arguably, the extinction (in the sense that the information on the source
speed is lost) will occur when the spectrum of the light source is lost.
Under VRT the Extinction Theorem should be reformulated.

In short, a method that measures the speed of light, must also measure the
speed of the source at the epoch of detection. Therefore, in addition to the
experience proposed by Dingle, variants can be found using other phenomena
associated with the propagation of light, such as the Doppler effect and the
Sagnac experiment \oldref{[R66]}\cite{sagnac1913comptes}.

In the case of optical Doppler effect, there are very few situations in
which there is an independent measurement of the shift in frequency and of
the velocity of the source. There are even situations where the measurements
differ from the predictions of the theory of relativity, being the most
famous, the so-called ``Pioneer anomaly'' \oldref{[R5]}\cite{anderson1998indication} 
as it was mentioned in Chapter \oldref{6}\ref{ch:6}.
This effect could be an indication that the Doppler formulas need to be
modified (at least for macroscopic sources), hence the interest of the
experiment proposed in this section.

The Sagnac effect \oldref{[R66]}\cite{sagnac1913comptes} is also present 
in other experiments such as Michelson and Gale (which allowed measuring 
the angular velocity of Earth's rotation)\oldref{[R69]}\oldref{[R70]}
\cite{michelson1925effect1,michelson1925effect2} 
or the use of ``ring laser''\oldref{[R53]}\oldref{[R54]}
\cite{macek1964measurement,bilger1972fresnel}, 
or global-scale experiment of Allan et a\oldref{[R96]}\oldref{[R97]}
\cite{allan1985around,allan1985accuracy} 
or the configuration of the optical fiber train by Wang et al.
\oldref{[R71]}\cite{wang2003modified}. 
The time lag difference between the two theories
is of second order in $v/c$ which is virtually undetectable. However, if we
consider each branch separately the time lag is different depending on the
used model. In the case of SRT there are two interpretations of this
phenomenon: that the delay is evenly distributed throughout the circuit
(see, for example, \oldref{[R96]}\cite{allan1985around}), or that locally 
there is no delay, and the delay only appears when considering coordinates 
along the global circuit \oldref{[R73]}\cite{wucknitz2004sagnac}. In
contrast, in the case of VRT the time lag is not uniform throughout the
circuit, the maximum difference is produced in the more distant branches
from the source while no time lag is produced in a leg where the source is
located (see Chapter \oldref{5}\ref{ch:5}). Thus, measuring the delay 
in a branch, using two different sources located in different corners, 
there will be a difference of first order in $v/c$.

\begin{figure}[h]
\begin{center}
\includegraphics[scale=1]{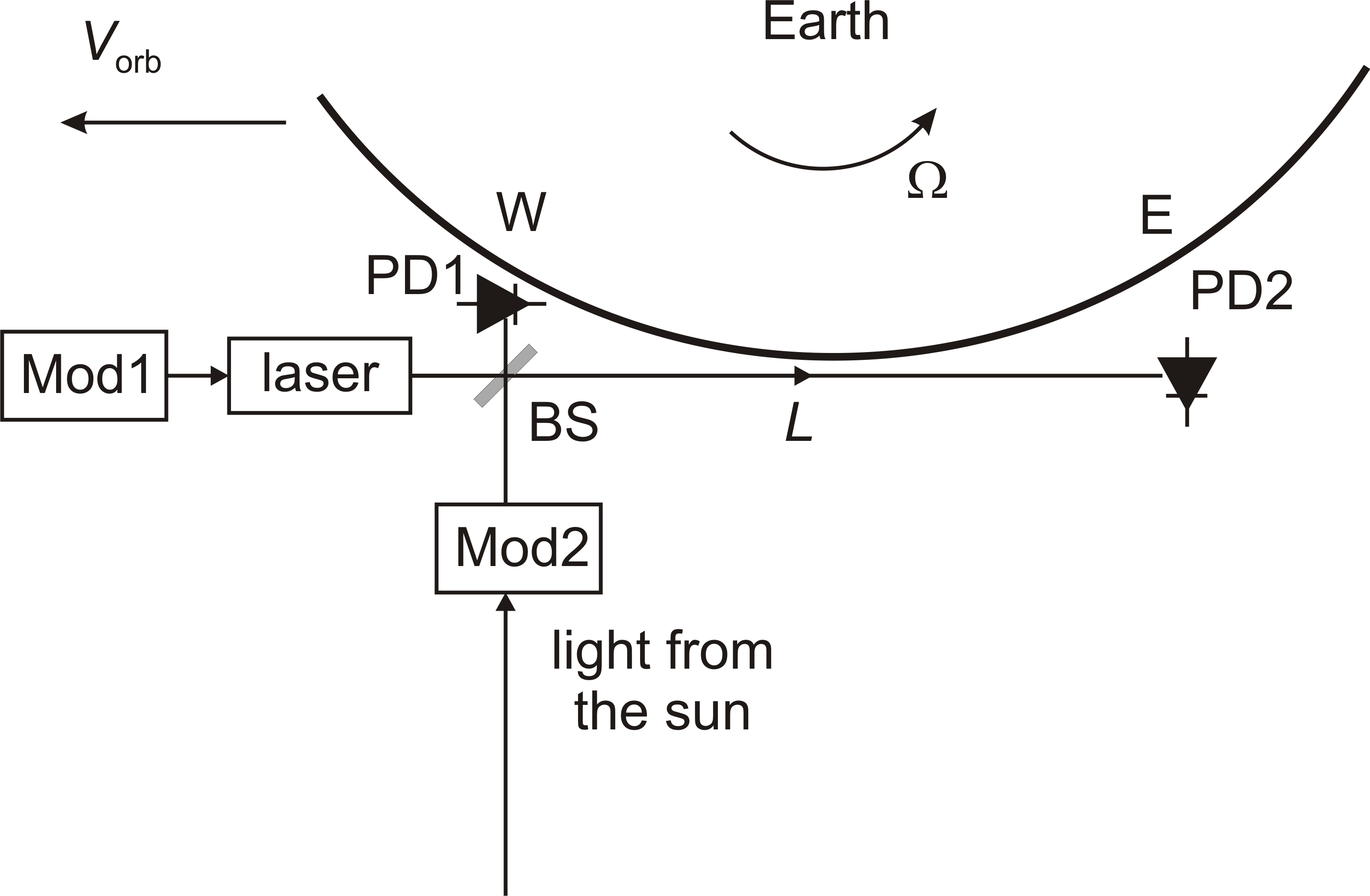}
\caption{Basic description of the experiment. Amplitude modulated light
from the Sun and a laser are combined and simultaneous time-of-flight
measurement is performed between two photodetectors.}
\label{fig:9.1}
\end{center}
\end{figure}

In the experiment we have carried out, the speed of light from the Sun and a laser are
simultaneously measured. Consider a triangle where the Sun is located in a
vertex, a laser and a detector (near detector, PD1) in another corner, and a
second detector (far detector, PD2) is located in the remaining vertex (see
figure \oldref{9.1}\ref{fig:9.1}). An amplitude modulator is included in 
the path of the Sun, while a modulated source feeds the laser. A time 
of flight measurement is performed in the branch PD1-PD2, oriented 
West-East. According to SRT there will be no time lag between the arrival 
time of the laser light and the sunlight, while according to VRT sunlight 
must reach the far detector before the laser light does. At solar noon, 
during December or June solstice, the arm will be aligned with Earth's 
orbital motion, which allows us to estimate the time lag as (see next paragraph)
\begin{equation}
\Delta t_{\text{sun-laser}}=-\frac{Lv_{\text{orb}}}{c^{2}}  \label{9.1}
\end{equation}
where $L$ is the length of the branch and $v_{\text{orb}}$ ($\approx 30$ km/s) 
the orbital velocity of Earth. Six hours earlier or later (because of the
Earth's rotation), the branch will be oriented perpendicular to orbital
speed and the time lag will be zero.

\section{Measurement method}\label{sec:9.3}

The method is based in measuring the time-of-flight (TOF) delays between a
near and a far photodectector for light from two different sources. The
expected time lag lies in the ps range, therefore pulsed signals cannot be
used because a very broad bandwidth oscilloscope would be needed. Therefore,
we used amplitude modulated signals at fixed frequency in combination with a
deep memory oscilloscope (4,000,000 record length). Using
``cross-correlation'' or similar techniques it is possible to measure the
relative delay between signals with a precision in the ps range.

Assuming that there are two detectors separated by a distance L in the
direction West-East, as shown in figure \oldref{9.1}\ref{fig:9.1}, 
according to SRT the transit time of light for Sun and laser are equal, 
i.e. no time lag is expected beyond dispersion effects in the air.

In contrast, according to VRT there will be a time lag. The laser signal
transit time (being the laser at fixed distance from the detectors), will be
\begin{equation}
\Delta t_{\text{laser}}=\frac{nL}{c}  \label{9.2}
\end{equation}
where $n$ is the refraction index of the medium (the same delay is predicted
by SRT). In order to estimate the transit time of the sunlight, we assume
that the measurement is performed at noon, during December or June solstice
in a West-East direction. According to Chapter \oldref{3}\ref{ch:3} the transit 
time of sunlight will be (the calculation is performed in a system fixed to the Sun)
\begin{equation}
\Delta t_{\text{sun}}=\frac{L}{c^{\prime }+v_{\text{orb}}}  \label{9.3}
\end{equation}
where
\begin{equation}
c^{\prime }=\frac{c+v_{\text{orb}}}{1+\left( n-1\right) \frac{c+v_{\text{orb}}}{c}}-v_{\text{orb}}  \label{9.4}
\end{equation}
and the tangential velocity due to Earth rotation has been neglected. Thus,
to first order in $v_{\text{orb}}$,
\begin{equation}
\Delta t_{\text{sun-laser}}=-\frac{Lv_{\text{orb}}}{c^{2}}+O\left( v_{\text{orb}}^{2}\right)  \label{9.5}
\end{equation}
or
\begin{equation}
\frac{\Delta t_{\text{sun-laser}}}{c}=-0.33\text{ ps/m}  \label{9.6}
\end{equation}
which is independent of the index of refraction (neglecting dispersion in
the medium). As it was mentioned above, six hour earlier or later the delay
will be zero, therefore the measurement relies on the variation of the time
lag during the day.

\begin{figure}[h]
\begin{center}
\includegraphics[scale=0.5]{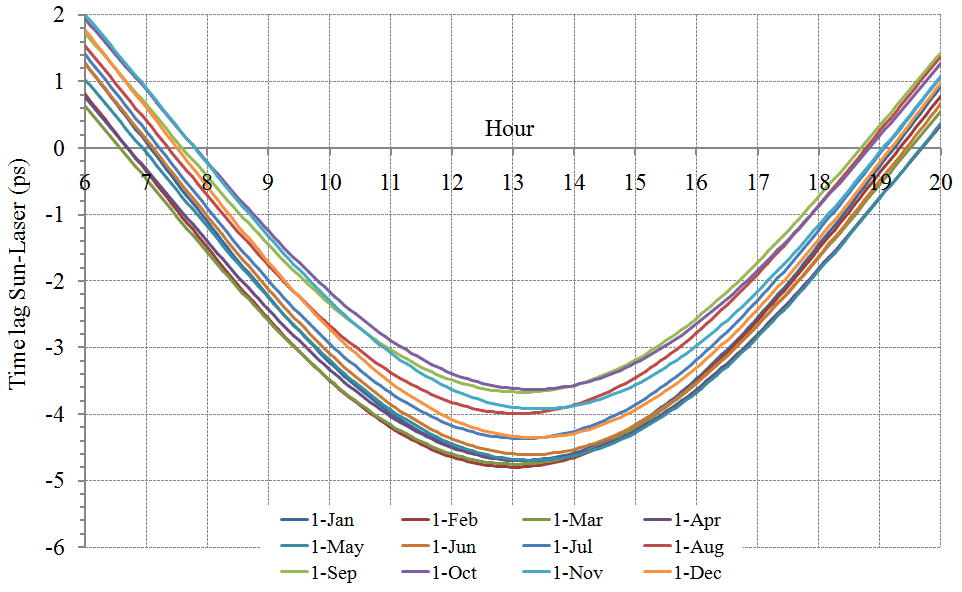}
\caption{Predicted diurnal variation of the time lag during the year
according to VRT (civil time, UTC-03:00). The near detector (PD1) is located
at 38${{}^\circ}$00'54''S, 57${{}^\circ}$33'26''W, and the far detector
(PD2) is 14.3 m apart (13.6 m East, 4.0 m North, and 1.5 m lower).}
\label{fig:9.2}
\end{center}
\end{figure}

From the orbital motion of Earth, the latitude and the longitude of both
detectors, and the date of the measurement, it is possible to calculate the
predicted delay as a function of the hour of the day. As an example, in
figure \oldref{9.2}\ref{fig:9.2} we show the variation of the time lag 
during a day, at different dates along the year, under the condition of 
experimental sets \#3 to \#8 described below.

For a practical use, a very good daily approximation of the theoretical time
lag is
\begin{equation}
\Delta t_{\text{sun-laser}}=A_{0}+A_{1}\cos \left( 2\pi \left(
h-h_{0}\right) /T\right)  \label{9.7}
\end{equation}
where $h_{0}$, $A_{0}$, and $A_{1}$ depend on the configuration, latitude,
longitude, and the day of the year, $h$ is the hour of the day, and $T=24$.
The parameters for the year 2011 are given in Table \oldref{9.1}\ref{table:9.1}.

\begin{table}[h]
\captionsetup{width=.6\textwidth}
\centering
\caption{Parameters for the year 2011 predicted by VRT as described in
figure \oldref{9.2}\ref{fig:9.2}}
\begin{tabular}{cccc}
Date & $A_{0}$ & $A_{1}$ & $h_{0}$ \\
& (ps) & (ps) & (hour) \\
01-Jan & -0.095 & 4.594 & 1.151 \\
01-Feb & -0.359 & 4.434 & 1.010 \\
01-Mar & -0.507 & 4.242 & 1.036 \\
01-Apr & -0.530 & 4.168 & 1.195 \\
01-May & -0.413 & 4.274 & 1.305 \\
01-Jun & -0.185 & 4.420 & 1.268 \\
01-Jul & 0.081 & 4.444 & 1.147 \\
01-Aug & 0.331 & 4.319 & 1.069 \\
01-Sep & 0.496 & 4.165 & 1.140 \\
01-Oct & 0.527 & 4.156 & 1.314 \\
01-Nov & 0.420 & 4.338 & 1.421 \\
01-Dec & 0.197 & 4.544 & 1.345
\end{tabular}
\label{table:9.1}
\end{table}

Finally, note that due to the dispersion in air, the different wavelength of
the laser and the Sun may produce a time lag. Using the equation developed
by Ciddor \oldref{[R98]}\cite{ciddor1996refractive}, the time lag between 
700 nm and 400 nm light is
\begin{equation}
\frac{\Delta t_{\text{700nm-400nm}}}{c}\approx -0.023\text{ ps/m}
\label{9.8}
\end{equation}

For a green laser (532 nm) or a red laser (650 nm) and assuming a peak at
500 nm for sunlight, the time lag due to dispersion will be at most one
order of magnitude smaller than the expected time lag according to VRT.
Therefore, dispersion in air can be neglected.

\section{Brief description of the experimental scheme}\label{sec:9.4}

The used laser was a diode laser, modulated electronically in amplitude to
frequencies around 20 MHz. The amplitude modulator for the Sun was required
to be achromatic, non-polarized and passive. So we have used three different
methods.

\begin{figure}[h]
\begin{center}
\includegraphics[scale=0.25]{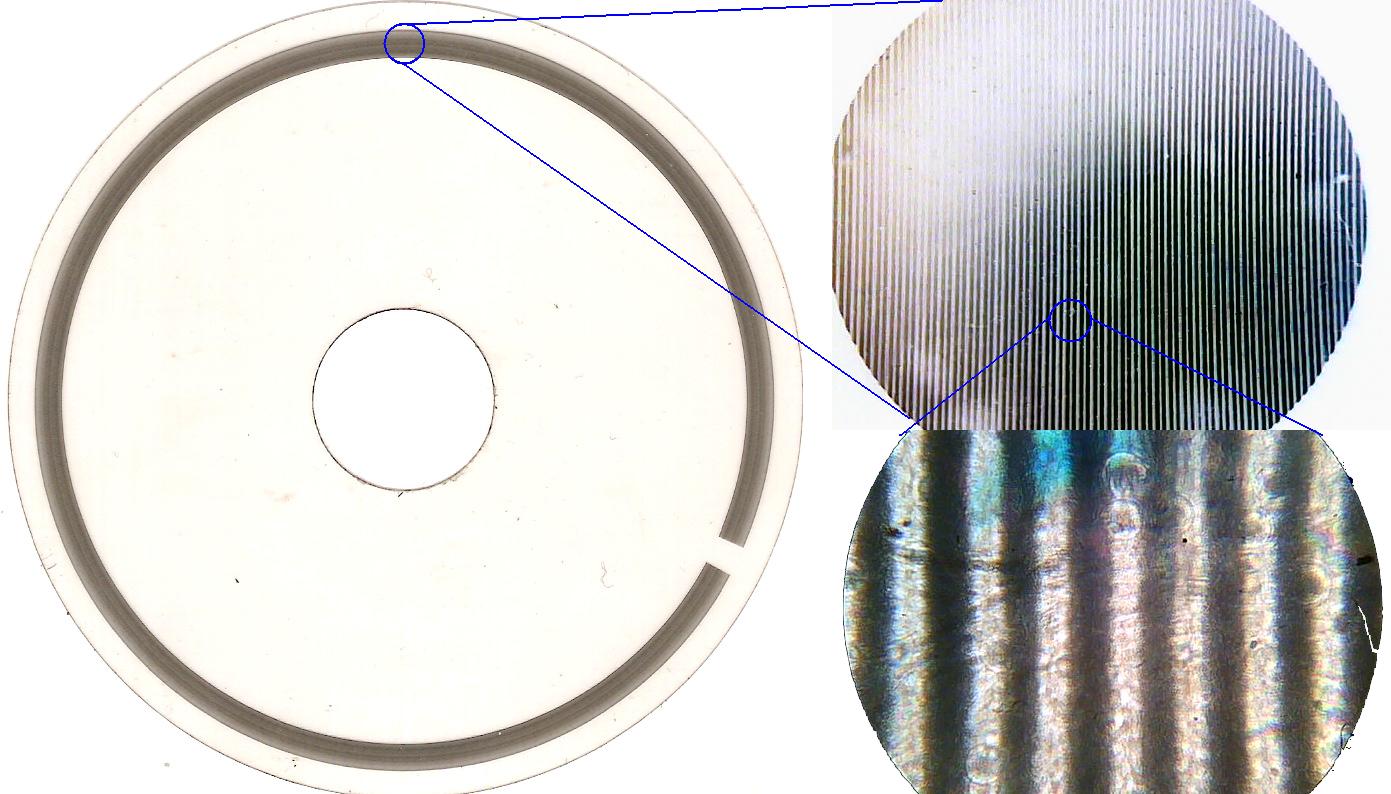}
\captionsetup{width=.8\textwidth}
\caption{Film used to modulate the sunlight in amplitude. The wheel
diameter is 89.5 mm and the spacing between black bands is 7 $\mu$m. 
At 500 Hz a 20 MHz modulation is obtained.}
\vspace{-12pt}
\label{fig:9.3}
\end{center}
\end{figure}

\begin{figure}[h]
\centering
\begin{minipage}{.5\textwidth}
  \centering
  \includegraphics[scale=0.25]{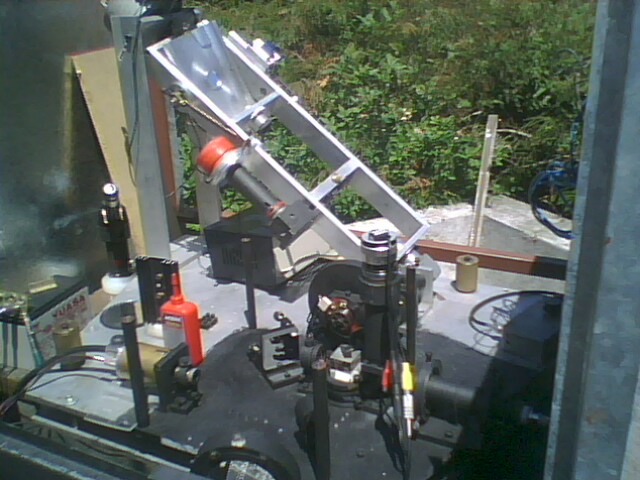}
\end{minipage}%
\begin{minipage}{.5\textwidth}
  \centering
  \includegraphics[scale=0.25]{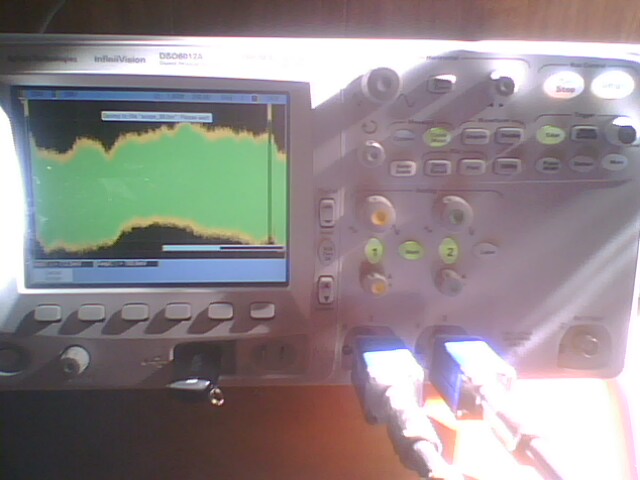}
\end{minipage}
\par
\medskip
\noindent
\begin{minipage}[t]{.5\textwidth}
  \centering
  \captionsetup{width=.9\textwidth}
  \captionof{figure}{Front view of the wheel and solar heliostat.}
  \label{fig:9.4}
\end{minipage}%
\hfill
\begin{minipage}[t]{.5\textwidth}
  \centering
  \captionsetup{width=.9\textwidth}
  \captionof{figure}{Typical signal of the superposition of sunlight modulated with the
wheel at $(20.605\pm 0.019)$ MHz and laser light modulated at $(19.993975\pm
0.000008)$ MHz. In yellow the near photodiode and in green the far detector.}
  \label{fig:9.5}
\end{minipage}
\end{figure}

The first one was a mechanical screening method consisting of a rotating
transparent film with opaque stripes uniformly arranged near the edge as
shown in figure \oldref{9.3}\ref{fig:9.3} (the modulation method is similar 
in principle to the toothed wheel of Fizeau). The strips are distributed 
uniformly in a circle of 89.5 mm diameter. A gap of 4 mm width was left 
in the film intended to produce a reference mark on the oscilloscope. 
There are 39,500 strips separated by 7 $\mu$m. Working at 500 Hz 
an amplitude modulated frequency of 20 MHz was achieved (although 
low-amplitude, around 10\% of the continuous signal). The wheel was 
located at the output of a heliostat, as shown in figure \oldref{9.4}\ref{fig:9.4}
(a sample signal is shown in figure \oldref{9.5}\ref{fig:9.5}). 

\begin{figure}[h]
\centering
\begin{minipage}{.5\textwidth}
  \centering
  \includegraphics[scale=0.25]{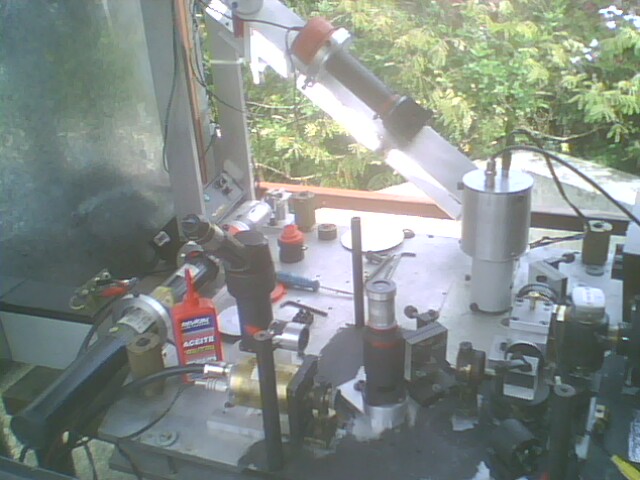}
\end{minipage}%
\begin{minipage}{.5\textwidth}
  \centering
  \includegraphics[scale=0.25]{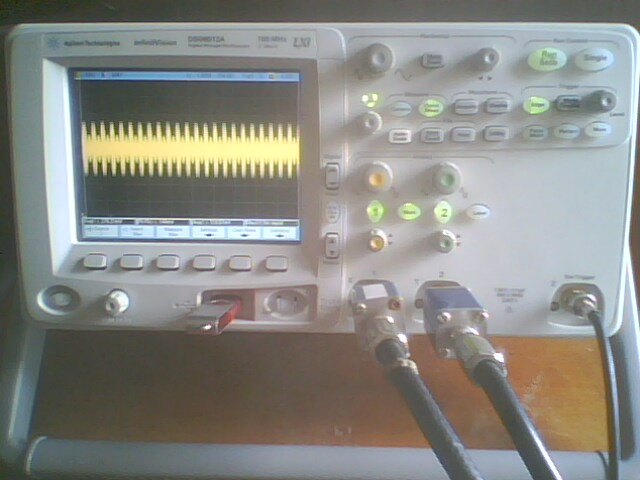}
\end{minipage}
\par
\medskip
\noindent
\begin{minipage}[t]{.5\textwidth}
  \centering
  \captionsetup{width=.9\textwidth}
  \captionof{figure}{Front view of the acousto-optic modulator and solar heliostat.}
  \label{fig:9.6}
\end{minipage}%
\hfill
\begin{minipage}[t]{.5\textwidth}
  \centering
  \captionsetup{width=.9\textwidth}
  \captionof{figure}{Typical signal of the
superposition of modulated sunlight with the acousto-optic modulator at $
(20.47303\pm 0.00002)$ MHz and modulated laser light at $(20.48903\pm
0.00002)$ MHz. PD1 in yellow and PD2 in green.}
  \label{fig:9.7}
\end{minipage}
\end{figure}

The second method was a modulated Fabry-Perot, and the third method, was an
acousto-optic modulator driven by a piezoelectric modulated at about 20 MHz,
mounted in replacement of the wheel, see figure \oldref{9.6}\ref{fig:9.6} 
(a sample signal is shown in figure \oldref{9.7}\ref{fig:9.7}).

Both modulated signals (laser and Sun) were combined using a beamsplitter.
This combined signal was divided into two beams directed to two photodiodes
(the near one and the far one). The signals of the photodiodes were sent via
RG-232 cables to a 2 channel, 4,000,000 record length digitizer. The time
lag ($\tau $) is defined as
\[
\tau =\Delta t_{\text{sun}}-\Delta t_{\text{laser}}=\left( t_{\text{sun}}-t_{\text{laser}}\right) _{\text{CH2}}-\left( t_{\text{sun}}-t_{\text{laser}}\right) _{\text{CH1}}
\]

Therefore the time lag can be interpreted as the time difference between Sun
and laser in channel 2 minus the difference between Sun and laser in channel
1. The idea of using difference between signals that travel the same path
was to keep alterations due to temperature, pressure, etc., as low as
possible (see Appendix 9.A\oldref{sec:9.A}. for further details on calibration).

The analysis of the measurements was performed using different mathematical
techniques. Basically, first, the signals from the Sun and the laser were
separated, then a cross-correlation between the near and far detectors was
performed, and finally, a delay is obtained for both the laser and for the
Sun, from which the time lag was obtained.

The use of a deep memory digitizer allows us to reach ps time resolution. As
a rule of thumb the statistical precision of the measurement of a time delay
is determined by the time base resolution divided by the square root of the
number of peaks registered in each measurement. For example, using a 2
Gsa/s, 2 channel, 4,000,000 record length oscilloscope, the precision (under
a low noise to signal ratio) may be as low as 1.4 ps for a 30 MHz signal.

The time lag is measured as a function of the hour of the day and compared
with the predictions of VRT (under SRT no time lag is expected). Since
measurements can only be performed on sunny days, the number of data points
strongly depended on weather conditions. Further, the method of the wheel
allowed the acquisition of less data points than the other methods because
some temporal gaps were needed during measurements in order to avoid
excessive heating of the wheel while working at 500 Hz in air.

\section{Measurements}\label{sec:9.5}

The first tests were conducted in August 2008 near the winter solstice in
the southern hemisphere, and other 7 different sets of measurements were
performed until November 2011. In each of those sets different modulation
methods, different frequencies, and different optical configurations were
used, as shown in Table \oldref{9.2}\ref{table:9.2}. Set \#7 (June 2011) 
produced low quality results due to the ashes of volcano Puyehue (whose 
eruption started on June 4, 2011) that affected the quality of the sunlight.

\begin{landscape}
\begin{table}
\centering
\caption{Details of the time lag measurement between sunlight and a laser.
The location of the second photodiode (PD2) is relative to the first one
(PD1), while $d$ is the line of sight distance. The sunlight was amplitude
modulated via either a toothed wheel, a Fabry-Perot (FP), or an
acousto-optic (AO) modulator at a frequency $f$. The laser was amplitude
modulated using a sinusoidal power supply having a frequency close or equal
to $f$.}
\begin{tabular}{ccccccc}
Set \# & Date &  & Location & $d$ (m) &  $f$ (MHz)  &  Data points \\
1 & 9-Aug-08 & PD1 & 38${{}^\circ}$05'35''S, 57${{}^\circ}$36'10''W & 102.0 & 13 & 56 \\
&          & PD2 & 100.6 m E, 16.9 m S & & Wheel & \\
2 & 22-Dec-08 / 28-Dec-08 & PD1 & 38${{}^\circ}$05'32''S, 57${{}^\circ}$36'20''W & 91.6 & 13 & 522 \\
&          & PD2 & 91.6 m E & & Wheel & \\
3 & 22-Nov-09 / 17-Dec-09 & PD1 & 38${{}^\circ}$00'54''S, 57${{}^\circ}$33'26''W & 14.3 & 20 & 319 \\
& & PD2 & 13.6 m E, 4.0 m N, 1.5 m below & & Wheel & \\
4 &  28-Mar-10 & PD1 & 38${{}^\circ}$00'54''S, 57${{}^\circ}$33'26''W & 14.3 & 2.5 & 37 \\
& & PD2 & 13.6 m E, 4.0 m N, 1.5 m below & & FP & \\
5 & 19-Sep-10 / 10-Oct-10 & PD1 & 38${{}^\circ}$00'54''S, 57${{}^\circ}$33'26''W & 14.3 & 24 & 425 \\
& & PD2 & 13.6 m E, 4.0 m N, 1.5 m below & & AO & \\
6 & 16-Feb-11 / 2-Mar-11 & PD1 & 38${{}^\circ}$00'54''S, 57${{}^\circ}$33'26''W & 14.3 & 20.4 & 922 \\
& & PD2 & 13.6 m E, 4.0 m N, 1.5 m below & & AO & \\
7 & 13-Jun-11 / 1-Jul-11 & PD1 & 38${{}^\circ}$00'54''S, 57${{}^\circ}$33'26''W & 14.3 & 28 & 719 \\
& & PD2 & 13.6 m E, 4.0 m N, 1.5 m below & & AO & \\
8 & 3-Oct-11 / 12-Nov-11 & PD1 & 38${{}^\circ}$00'54''S, 57${{}^\circ}$33'26''W & 14.3 & 28 & 3,677 \\
& & PD2 & 13.6 m E, 4.0 m N, 1.5 m below & & AO &
\end{tabular}
\label{table:9.2}
\end{table}
\end{landscape}

Also calibration measurements were performed using a second laser beam
instead of the sunlight (called ``virtual Sun''). A complete description of
the all data sets will produce an excessively long paper, therefore we will
show the main results and conclusions, leaving technical details for a
separated technical report to be completed.

\subsection{Measurements with $L=102$ m.}\label{sub:9.5.1}

\begin{figure}[h]
\centering
\begin{subfigure}{.5\textwidth}
  \centering
  \includegraphics[scale=0.2]{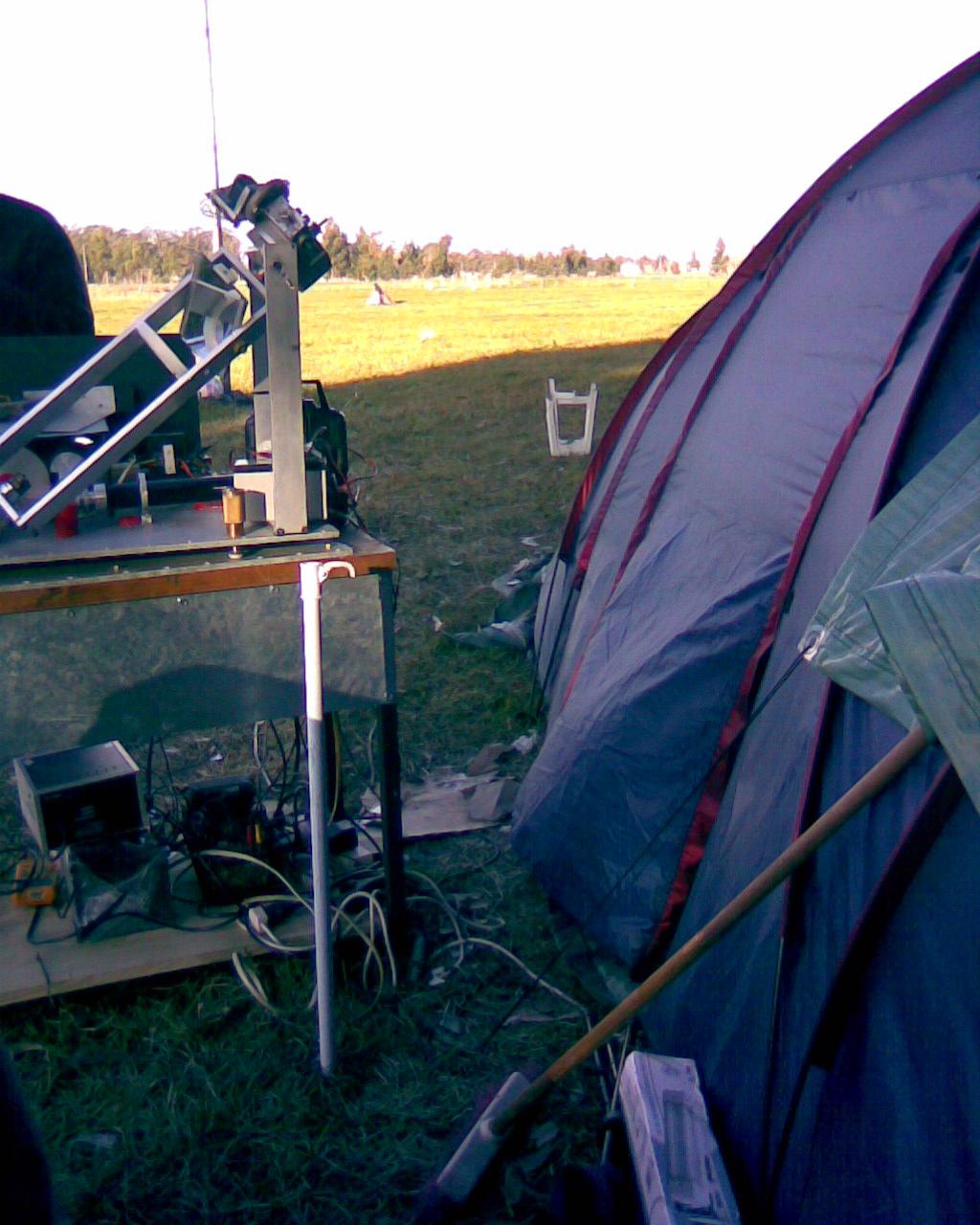}
  \caption{}
  \label{fig:9.8.a}
\end{subfigure}%
\begin{subfigure}{.5\textwidth}
  \centering
  \includegraphics[scale=0.2]{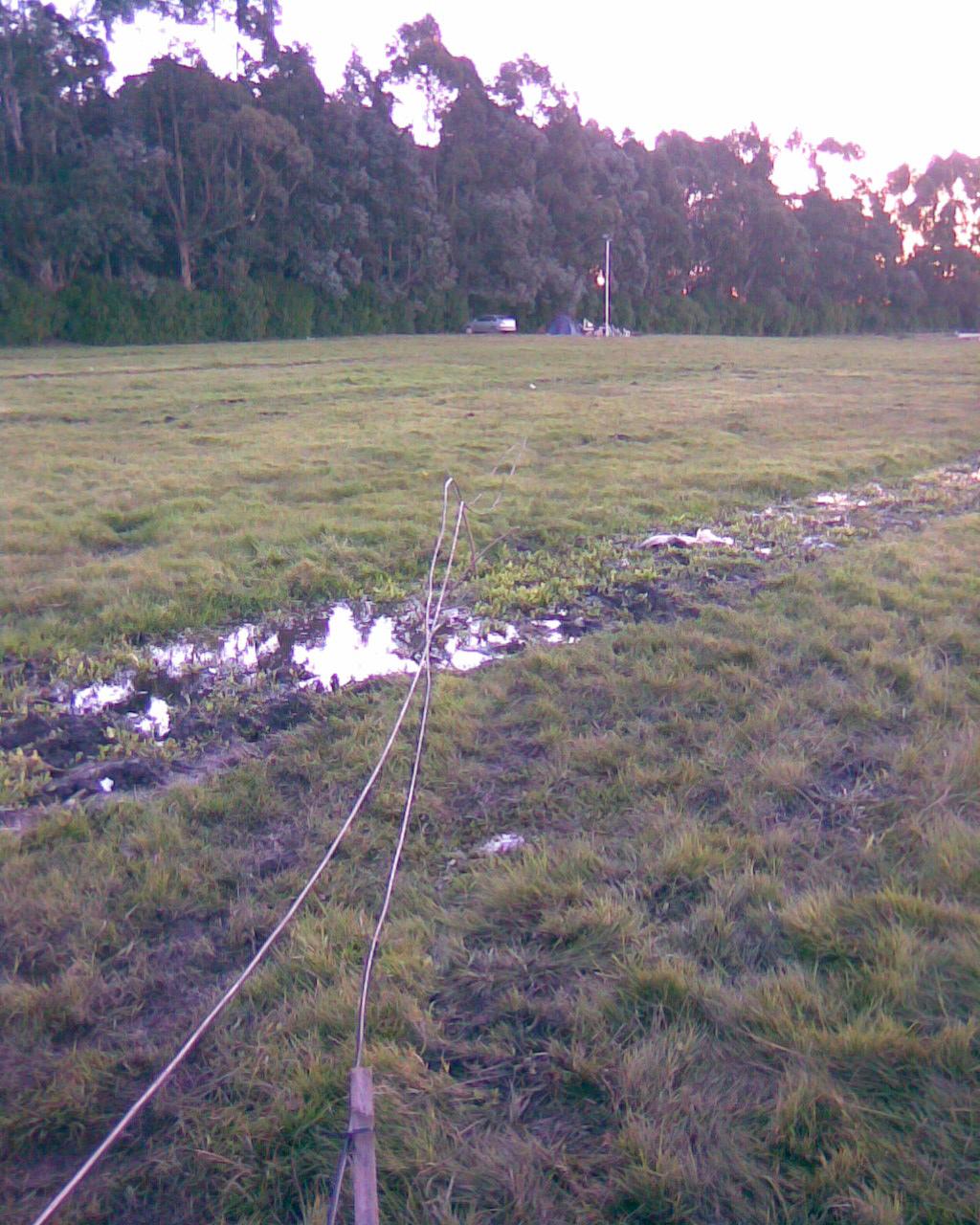}
  \caption{}
  \label{fig:9.8.b}
\end{subfigure}
\caption{Measurements with $L=102$ m (August 2008). Left: view to the
East (the far detector is hardly seen in the middle, upper part) from the
main base (heliostat+near detector). Right: view of the main base (to the
West) from the far detector.}
\label{fig:9.8}
\end{figure}

During August 2008 measurements were performed with the far detector located
100.6 m East, and 16.9 m South from the near detector (see figure
\oldref{9.8}\ref{fig:9.8}). A rotating wheel produced a 13 MHz modulated signal.

\begin{figure}[h]
\begin{center}
\includegraphics[scale=0.5]{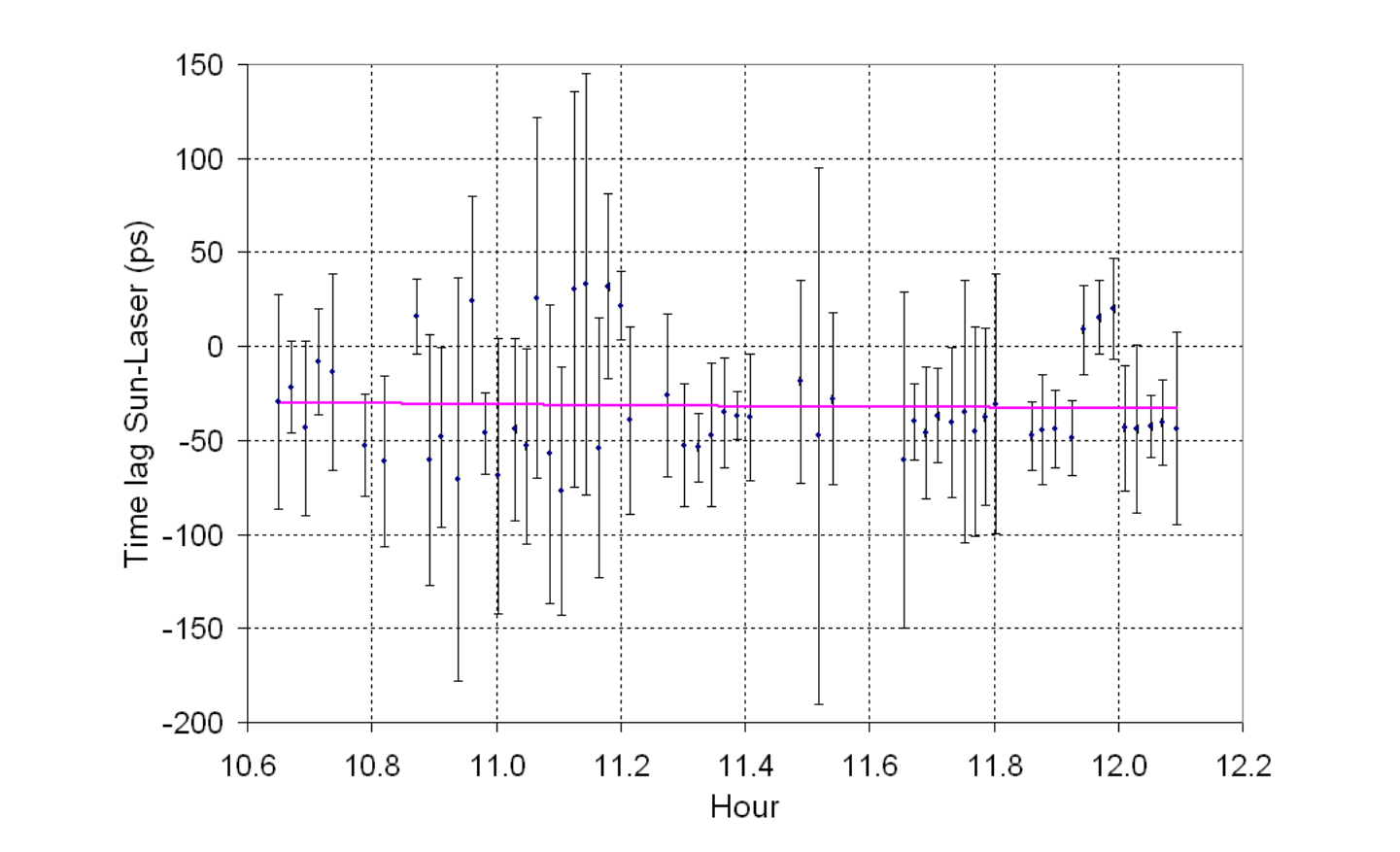}
\caption{August 2008, set \#1. Time lag Sun-laser as a function of the
hour of the day. Experimental results (black circles) and VRT prediction
(magenta full line). SRT predicts a null time lag.}
\label{fig:9.9}
\end{center}
\end{figure}

Fitting the data with a cosine (see figure \oldref{9.9}\ref{fig:9.9}) produces 
statistically non-significant results ($p=0.83$). Anyway results were promising 
since they suggested a variation along the day. Also, they were important in
understanding the sources of errors and bias (see Appendix 9.A\oldref{sec:9.A}).

\subsection{Measurements with $L=91.6$ m.}\label{sub:9.5.2}

\begin{figure}[htb]
\begin{center}
\includegraphics[scale=0.5]{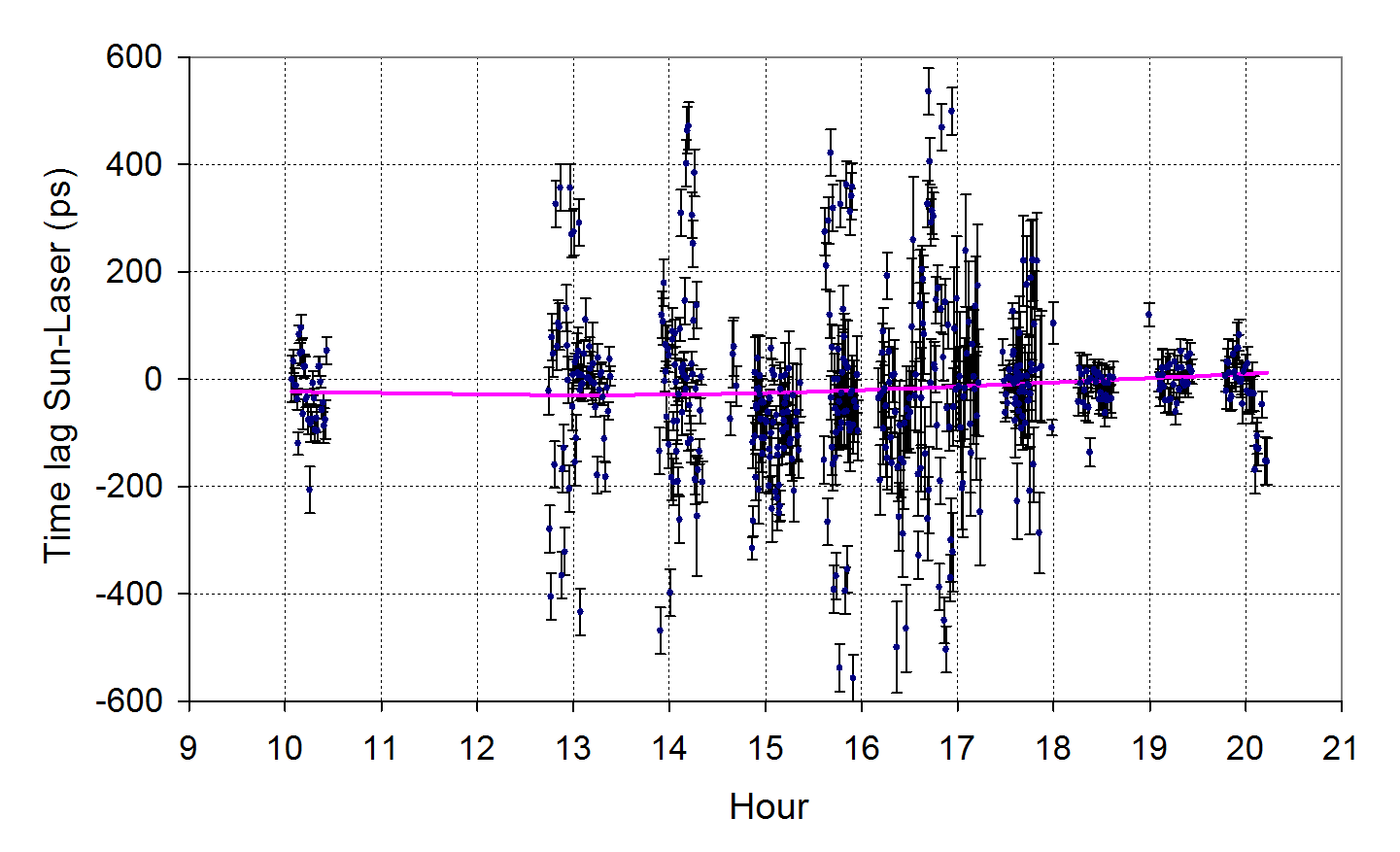}
\caption{December 2008, set \#2. Time lag Sun-laser as a function 
of the hour of the day. Experimental results (black circles) and VRT prediction
(magenta full line). SRT predicts a null time lag.}
\label{fig:9.10}
\end{center}
\end{figure}

During December 2008 the far detector was located 91.6 m East from the near
detector. The measured time lag is shown in figure \oldref{9.10}\ref{fig:9.10}. 
The parameters predicted by VRT are $A_{0}=0.012$ ps, $A_{1}=30.867$ ps, 
and $h_{0}=0.828$ h. Although results agree quite well with the VRT model, 
they are statistically non-significant ($p=0.59$). The standard deviation was 
165 ps, much higher than the previous set. There are many reasons for this, 
among them: a) the far detector was connected to channel 2 with a 100 m long
RG-232 cable, while the near detector was connected by a meter long cable,
b) measurements were performed during various days, at different hours with
a large variation in solar intensity, and ambient temperature, c) the
rotational frequency of the wheel had not been controlled with high
precision, d) the optical system was modified during the measurements. These
facts further indicated the problems resulting from unbalanced paths between
channel 1 and channel 2 at the ps level (see Appendix 9.A\oldref{sec:9.A}).

\subsection{Measurements with $L=14.3$ m.}\label{sub:9.5.3}

\begin{figure}[h]
\centering
\begin{subfigure}{.5\textwidth}
  \centering
  \includegraphics[scale=0.3]{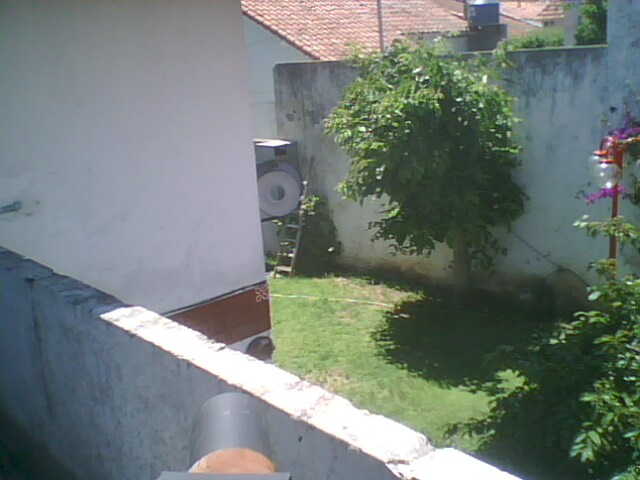}
  \caption{}
  \label{fig:9.11.a}
\end{subfigure}%
\begin{subfigure}{.5\textwidth}
  \centering
  \includegraphics[scale=0.3]{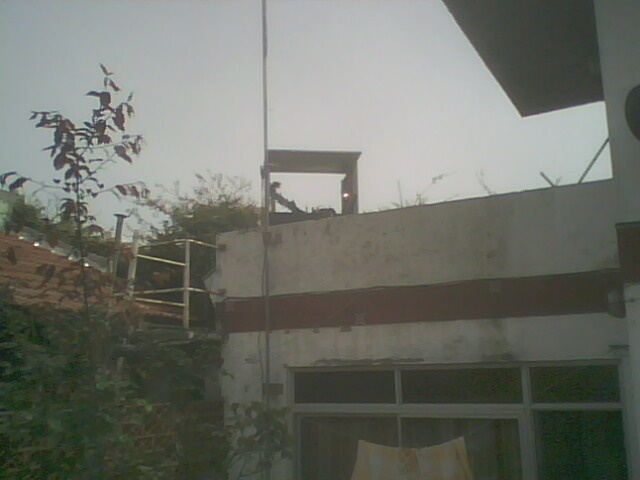}
  \caption{}
  \label{fig:9.11.b}
\end{subfigure}
\caption{Measurements with $L=14.3$ m. Left: view of the far detector
from the main base (heliostat+near detector). Right: view of the main base
from the far detector.}
\label{fig:9.11}
\end{figure}

In order to keep a more controlled environment the distance was reduced to
14.3 m and the channels were balanced using both a 50 m long RG-232 cable
(see figure \oldref{9.11}\ref{fig:9.11}). The sunlight was modulated via a 
wheel, a Fabry-Perot or an Acousto-optic modulator, as described in 
Table \oldref{9.2}\ref{table:9.2}.

The offset was measured on a periodic basis by using of a second laser in
the sunlight path (``virtual Sun''), performed before or after the main
measurements. This calibration gave an overall offset of $(-7.2\pm 0.3)$ ps,
that was added to the results. This is valid for we were trying to study the
time lag variation along the day, i.e., a relative time lag. The offset
correction increases the error of $A_{0}$ but not those of $A_{1}$ and $
h_{0} $ which are strongly linked to VRT. Actually, due to the small
variation of the parameters (i.e., temperature, frequency and intensity) the
analysis of experimental results may include a procedure for discounting the
systematic effects from the calibration, as discussed in Appendix 
9.A\oldref{sec:9.A}. On the contrary, in our analysis there are 
no any other corrections applied to the observed values besides the
offset described above.

Data acquisition was controlled via a computer program that kept the
received intensity of the modulated sunlight and laser as constant as
possible. Also frequency was sensed and adjusted online during the
measurements. None of the data has been discarded. The effect of a few poor
quality data (mainly when solar light was dimmed in late afternoon) did not
significantly affect the results.

We found a small drift from one set to another, due to the use of different
optical systems. This fact inhibits us to make a precise (to the ps range)
measurement of the absolute time lag. At the level of ps time the Sun-laser
time lag depends on different variables. We identified the dependence on
temperature, on modulation frequency of laser and Sun, and on intensity of
the signal of laser and Sun. Although we tried to reduce these factors as
much as possible, an offset was still present, mainly produced by the
different optical configuration.

\begin{figure}[htb]
\begin{center}
\includegraphics[scale=0.5]{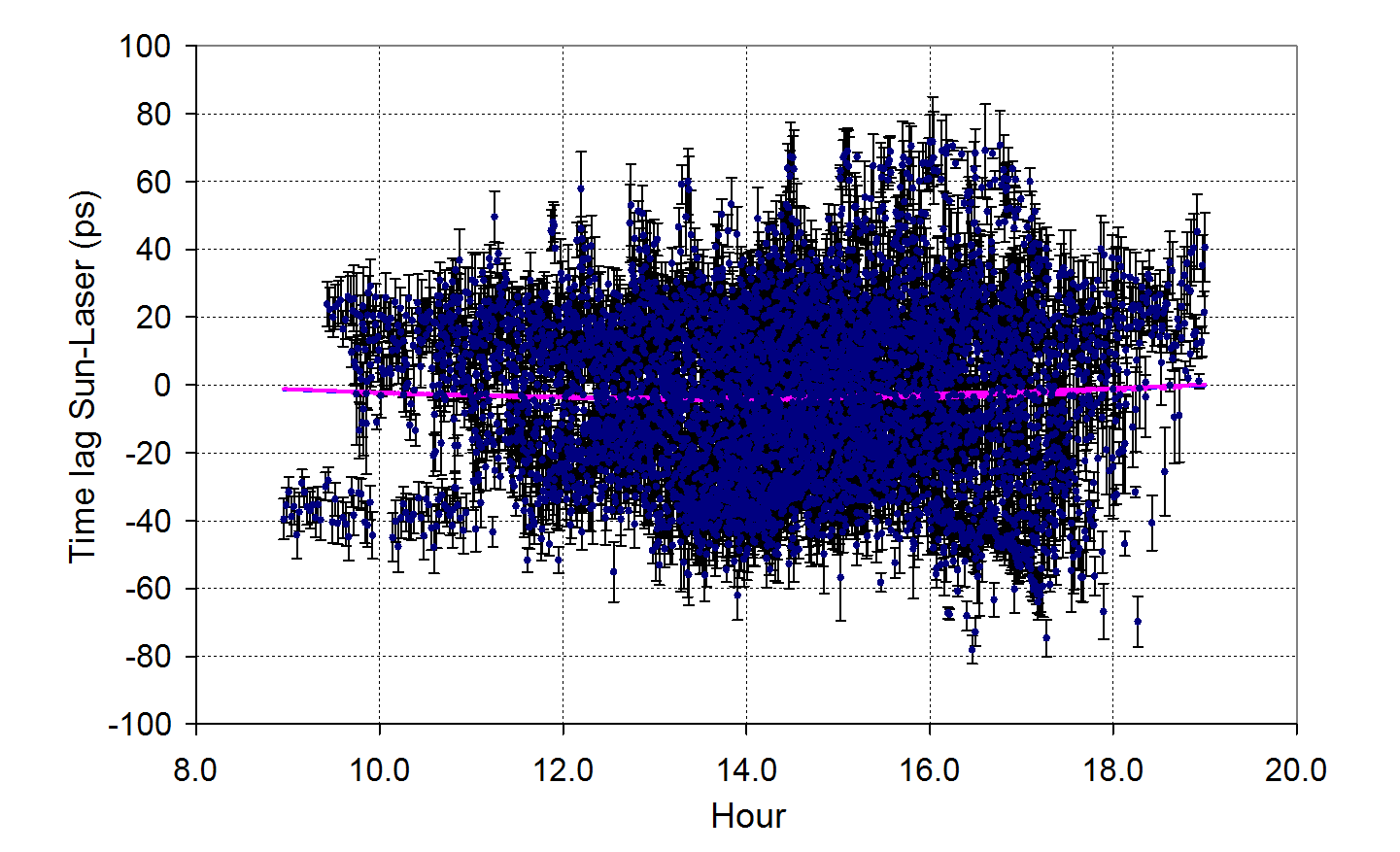}
\caption{$L=14.3$ m, set \#3 to \#8, 6,099 data points. Time lag
Sun-laser as a function of the hour of the day. Experimental results (black
circles), VRT prediction (magenta full line) and best cosine fit (yellow
full line, hardly seen behind the VRT prediction). SRT predicts a null time
lag.}
\label{fig:9.12}
\end{center}
\end{figure}

In figure \oldref{9.12}\ref{fig:9.12} we plot the measured time lag from 
Nov. 22, 2009 to Nov. 12, 2011 (sets \#3 to \#8). A total of 6,099 data 
points were recorded. The mean values of the parameters predicted by 
VRT are $A_{0}=0.433$ ps, $A_{1}=4.345$ ps, and $h_{0}=1.410$ h. 
The best cosine fit gives $A_{0}=(1.8\pm 2.0)$ ps ($p$-value $=0.367$), 
$A_{1}=(5.8\pm 2.2)$ ps ($p$-value $=0.0077$), 
$h_{0}=(1.7\pm 0.5)$ h ($p$-value $=0.0014$). 
Both $A_{1}$ and $h_{0}$ are statistically different from zero at 0.01 significance level.


As expected, due to the large size of the sample, a very small $p$-value is obtained from the $F$-test. 
Further, the dependence of the time lag with the hour of the day is largely hidden by 
the large experimental errors. Therefore, we cannot assure whether or not the effect 
of the speed of the source is responsible for the observed variation. 
However, the fact that  $A_{1}$ is different from zero means that an actual dependence of the time lag 
with the hour of the day has been observed.


\section{Conclusions}\label{sec:9.6}

Within the experimental error, the experiment has detected a statistically
significant sinusoidal variation of the time lag along the day of the same
order of magnitude as that predicted by VRT.

\begin{figure}[htb]
\begin{center}
\includegraphics[scale=0.5]{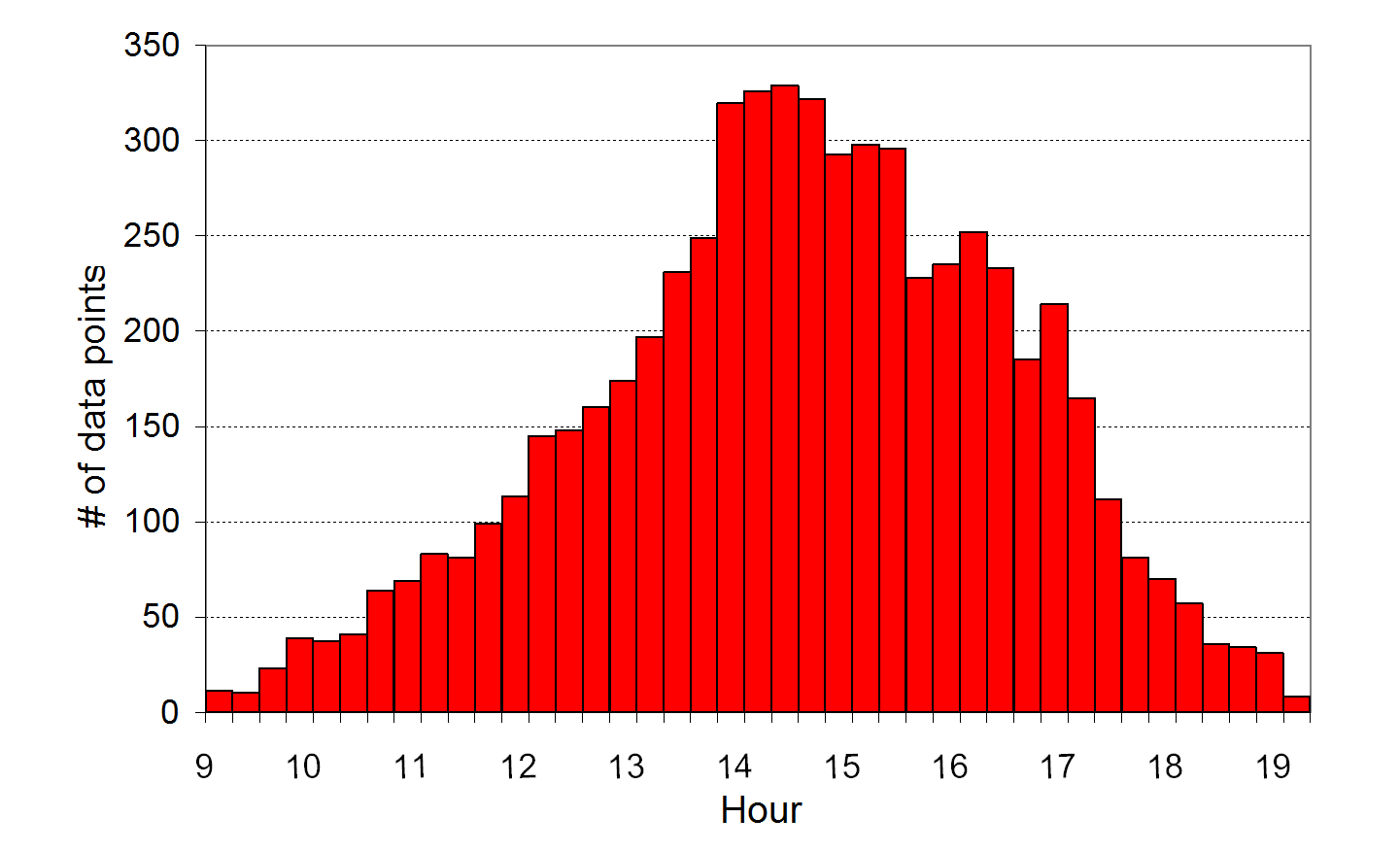}
\caption{Set \#3 to \#8, 6,099 data points. Total number of data points
measured every 15 minutes. The maximum is around 2:30 pm (civil time,
UTC-03:00).}
\label{fig:9.13}
\end{center}
\end{figure}

Note that the minimum of the measured time lag is at (1:40$\pm $0:30) pm
(civil time, UTC-03:00) in accordance with VRT predictions (1:24 pm). Of
course, other non-considered effects may produce a time lag. For example, a
chromatic variation of the sunlight may produce a different time response of
the detector (dispersion in air may be neglected, as stated above), but the
minimum or maximum should be expected at solar noon, that is at 12:50 pm
civil time. Another possibility is a statistical bias for the data was not
acquired uniformly during the day (see figure \oldref{9.13}\ref{fig:9.13}). 
In this case the minimum or maximum should be related to the hour of 
maximum data point acquisition, which is 2:30 pm civil time. Finally, 
temperature variations may produce a minimum or maximum around the 
time of maximum temperature, that is, around 3 or 4 pm civil time. None 
of these effects lie inside the measured minimum time lag error interval.

\begin{figure}[htb]
\begin{center}
\includegraphics[scale=0.5]{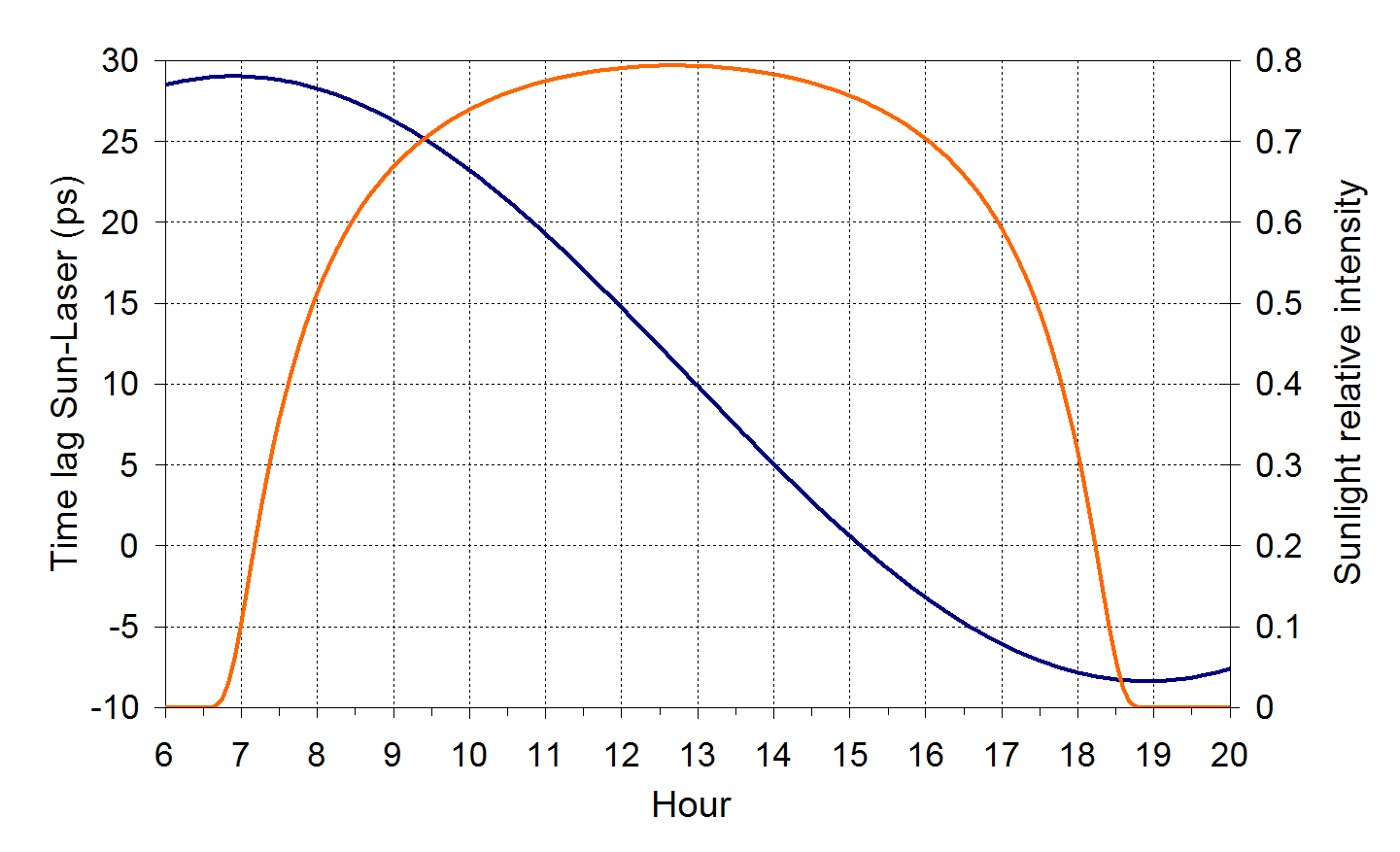}
\caption{VRT time lag prediction for a south-north orientation, with L =
100 m, during vernal equinox (blue full line). Sunlight intensity (orange
full line) is relative to that of the Sun in the zenith.}
\label{fig:9.14}
\end{center}
\end{figure}

In order to distinguish between VRT and those possible effects, one
possibility is to use a different orientation, for example south-north. In
figure \oldref{9.14}\ref{fig:9.14} we show the delay predicted by VRT 
for the vernal equinox (southern hemisphere) of 2011 using a south-north 
orientation and $L=100$ m. The corresponding parameters are 
$A_{0}=10.34$ ps, $A_{1}=19.69$ ps, and $h_{0}=6.906$ h.

Since the absolute delay with high precision will be very difficult to
measure with the available detectors (including the oscilloscope)
measurement may focus on the variation of the time lag along the day.
According to the current orientation (roughly west-east) the maximum effect
should be obtained close to noon when both intensity and chromaticity of
sunlight exhibit a symmetric behavior, therefore another direction should be
preferable. An alternative is to use a south-north orientation for which VRT
predicts a anti-symmetric behavior around noon, see figure \oldref{9.14}\ref{fig:9.14}. 
Although the amplitude of the effect is reduced relative to the west-east
orientation, the time derivative is maximum around noon (5 ps/h for $L=100$
m) when the maximum sunlight intensity occurs. This has also the advantage
that intensity, chromatic, and temperature effects will exhibit a very
different behavior as a function of time, and should be easily
discriminated. Finally, a distance of about 100 m should be desirable, as
long as ambient conditions can be controlled (for example, using an
evacuated tube between detectors, and keeping thermal insulations for both
detectors and cables).

\newpage
\section*{Appendix 9.A. Characterization of the detecting system}\label{sec:9.A}
\addcontentsline{toc}{section}{Appendix 9.A. Characterization of the detecting system}

The measured time lag depends on different factors: optical system,
detectors, cables, oscilloscope, and data processing. Most of the time delay
variations are due to the transmission of the signals in the coaxial cables.
Variations of intensity or temperature will produce fluctuations in the
results.

Another source of error comes from the fact that the delay depends on the
solid angle seen by the detectors. Since the optical systems of the near and
far detectors are different, it is almost impossible to balance the delay in
each channel, therefore, a bias remains in the results. A better design
would contemplate this problem, for example, by placing the heliostat in the
middle of the 2 detectors and using the same optics for both channels. In
our case this was impossible to implement, therefore the offset was deducted
from the results. Note that we were trying to detect variations of the time
lag along the day, rather than a precise absolute measurement.

There may be other source of errors that couldn't be identified, such as
chromatic variations on the sunlight, micro turbulence in the air, etc.

\begin{figure}[htb]
\begin{center}
\includegraphics[scale=0.5]{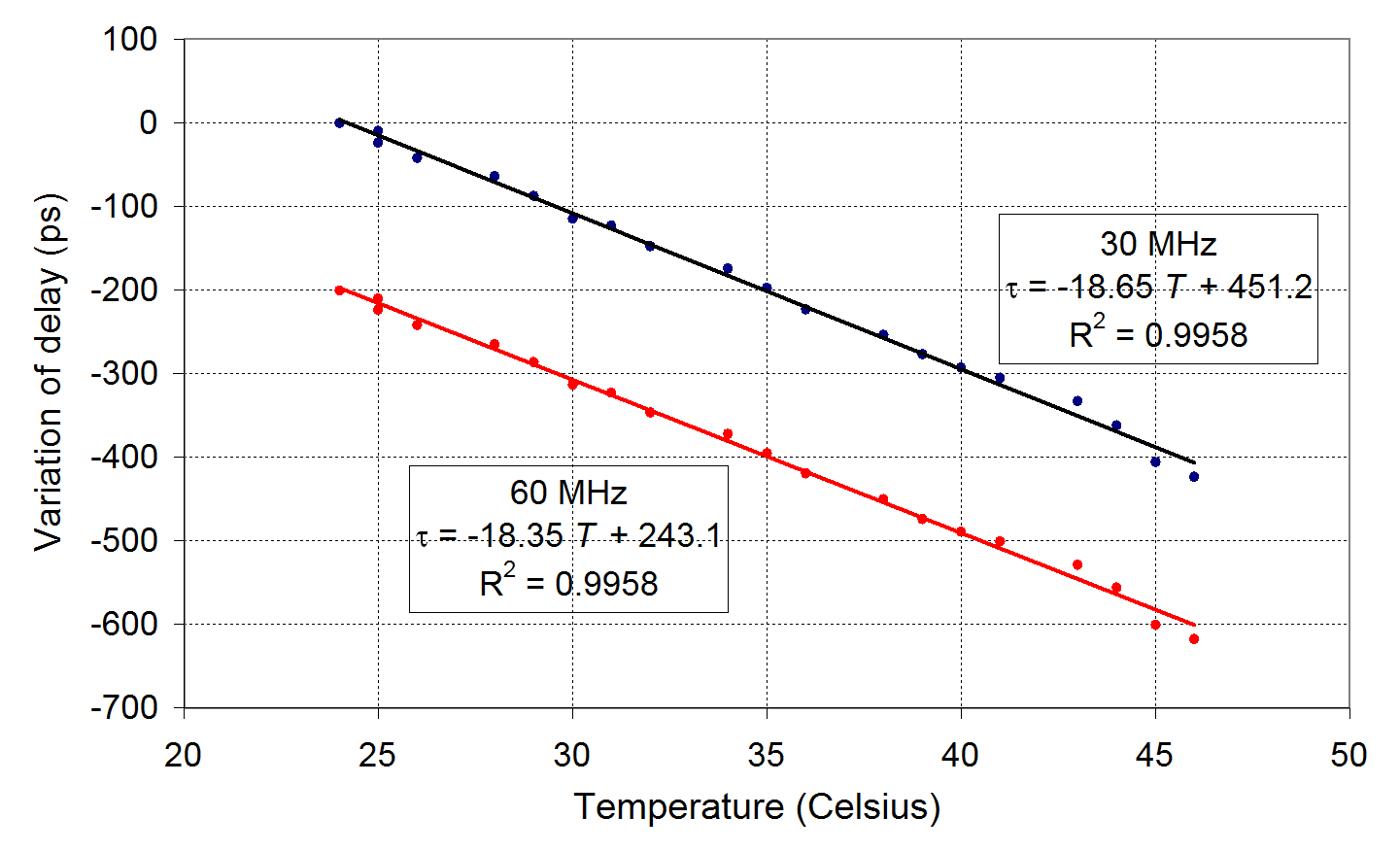}
\caption{Variation of the delay of a signal in a channel as a function
of temperature for two different frequencies.}
\label{fig:9.A.1}
\end{center}
\end{figure}

\begin{figure}[htb]
\begin{center}
\includegraphics[scale=0.5]{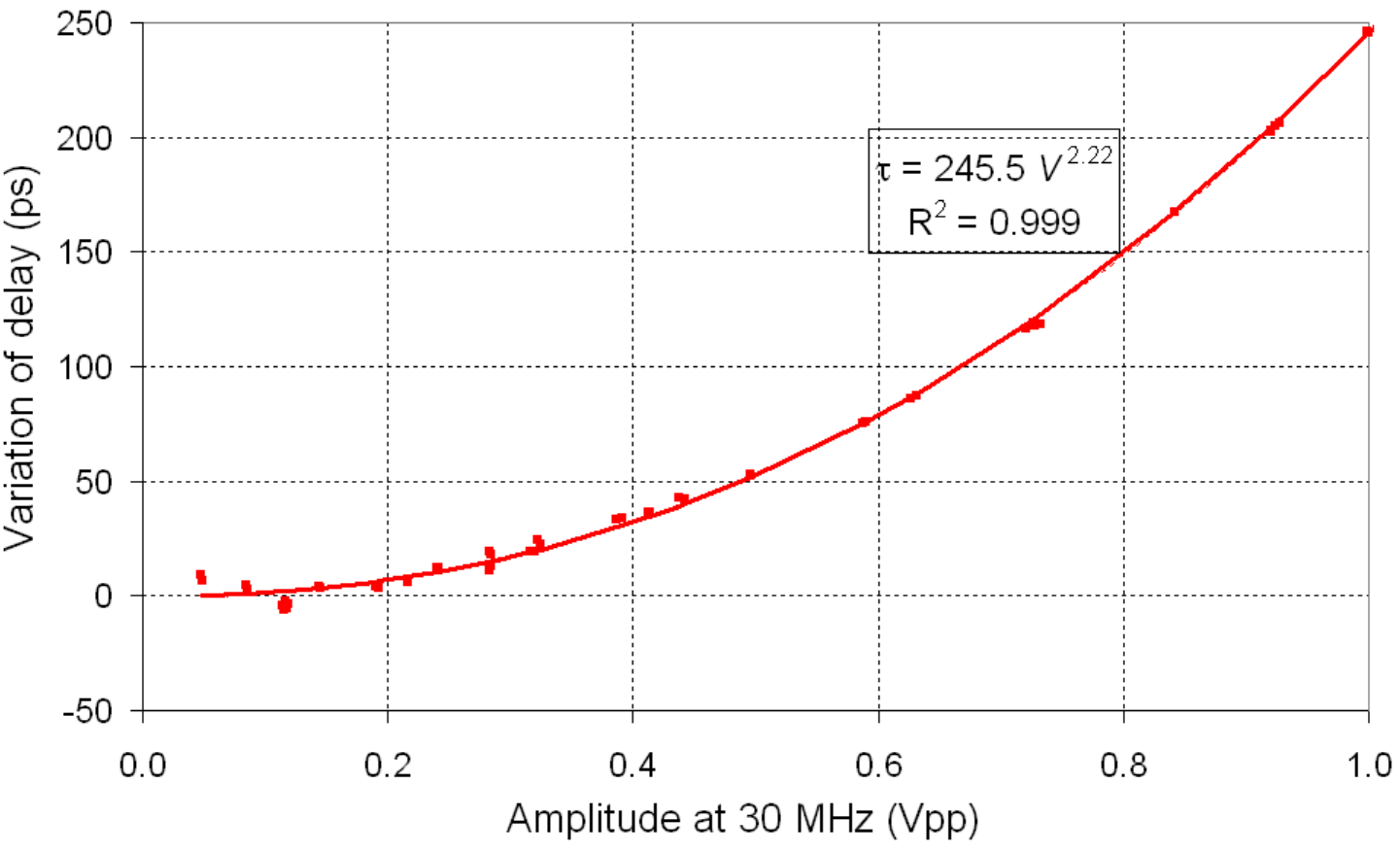}
\caption{Variation of the delay of a signal in a channel as a function
of amplitude at 30 MHz.}
\label{fig:9.A.2}
\end{center}
\end{figure}

\begin{figure}[htb]
\begin{center}
\includegraphics[scale=0.5]{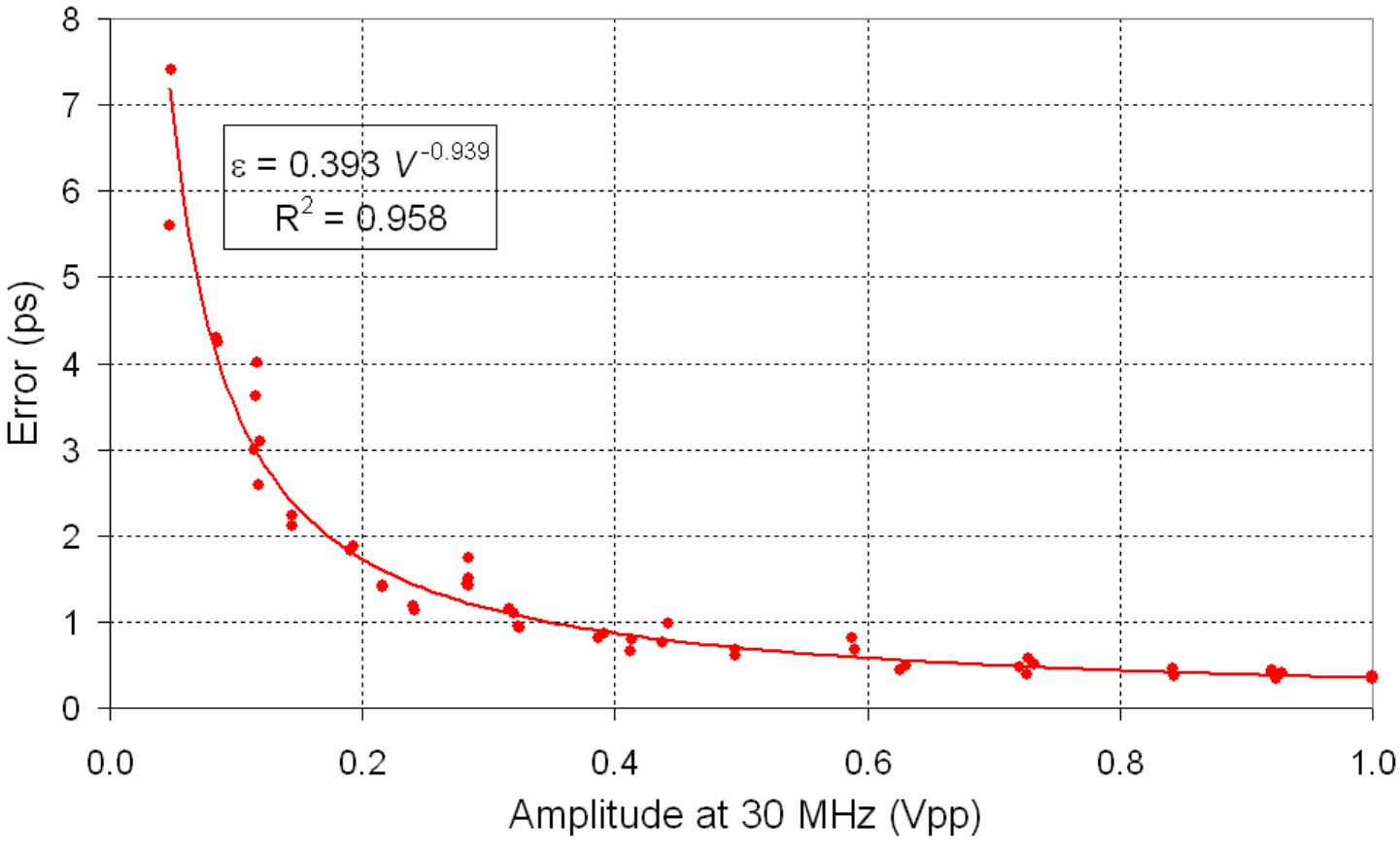}
\caption{Error of the delay of a signal in a channel as a function of
amplitude at 30 MHz.}
\label{fig:9.A.3}
\end{center}
\end{figure}

The use of coaxial cables for picosecond timing has been studied elsewhere
\oldref{[R99]}\oldref{[R100]}\cite{kalliomaki2006applicability,corey2006phase}. 
The measured time delay of an individual signal on a given
channel depends on signal amplitude, frequency, and temperature. Other
parameters like humidity or air pressure may also affect the measurements in
a lesser form. The amplitude of the signal modifies also the quality of the
data. Low signals give more error than larger signals do. The error is
inversely proportional to the amplitude. In figures \oldref{9.A.1}\ref{fig:9.A.1}, 
\oldref{9.A.2}\ref{fig:9.A.2}, and \oldref{9.A.3}\ref{fig:9.A.3}
we show some sample characterization.

Note, however, that measurements are relative (not absolute) since 
the time lag is defined as
\begin{eqnarray*}
\tau &=&\Delta t_{\text{sun}}-\Delta t_{\text{laser}} \\
&=&\left( t_{\text{sun CH2}}-t_{\text{sun CH1}}\right) -\left( t_{\text{laser CH2}}-t_{\text{laser CH1}}\right) \\
&=&\left( t_{\text{sun}}-t_{\text{laser}}\right) _{\text{CH2}}-\left( t_{\text{sun}}-t_{\text{laser}}\right) _{\text{CH1}}
\end{eqnarray*}
where each parenthesis in the last line is calculated in a single channel
making a zero correction when frequency and amplitude are identical for the
Sun and laser signals. Also, zero correction should be applied when the
offset of channel 2 equals the offset of channel 1, for example, same
frequency and amplitude ratios in both channels. Thus the global offset is
expected to be a small difference between the offset on each channel (which
individually may be larger).

Unfortunately it is impossible to keep the amplitude constant along a day
and for all measurements during a year. The amplitude and frequency ratio
between Sun and laser depend on the superposition method, the optical
system, and the hour of the day. In fact, sunlight changes its intensity
during a day.

When superimposing simultaneously both signals a different frequency is
mandatory in order to discriminate the signals. A difference of 1 to 32 kHz
was used for frequencies of 20 MHz and above, thus an offset should be
discounted to the measurements. Other methods, like chopping techniques
between Sun and laser signals do no suffer from this problem, and they are
to be preferred.

Data analysis can be performed by correcting the measured time lag according
to the frequency and amplitude of each signal, and the ambient temperature.
Instead, we preferred to develop a computer controlled acquisition system
designed to capture the signal when it met different conditions, like
intensity (absolute and relative) within a narrow band, frequencies within
the allowed values, etc. in order to avoid large variations that may lead to
large dispersion of the time lag. Then we used the raw data; with no
corrections of any kind applied to the observed values besides a global
offset.

Another approach, used in our analysis, was to assume that due to the large
number of observations, the computed time lag will be distributed according
to the normal distribution centered in the actual value.


%% file: 2014arxiv_v14_2_9c.tex


\chapter*{Conclusions}\label{ch:c}

\addcontentsline{toc}{chapter}{Conclusions}

\fancyhf{}
\fancyhead[LE,RO]{Conclusions}
\fancyhead[RE,LO]{L. Bilbao, L. Bernal, F. Minotti}
\fancyfoot[RE,LO]{Vibrating Rays Theory}
\fancyfoot[LE,RO]{\thepage}
 
\renewcommand{\headrulewidth}{1pt}
\renewcommand{\footrulewidth}{1pt}

In this work we have presented observational evidence favoring a dependence
of the speed of light on that of the source, in the manner implied in
Faraday's ideas of ``vibrating rays.''\ 

It is remarkable and very suggestive that, as derived from Faraday's
thoughts, simply relating the velocity of light and the corresponding
Doppler effect with the velocity of the source at the time of detection, is
enough to quantitatively and qualitatively explain a variety of spacecraft
anomalies.

Under VRT the manifestation of the movement of the source in the speed of
light is more subtle than the naive $c+kv$ hypothesis ($k$ is a constant, 
$0\le k\le 1$) usually used to test their dependence \cite{brecher1977speed}.
Thus, it is also of fundamental importance the fact that, from the
experimental point of view, it is very difficult to detect differences
between VRT and SRT in usual experiments and observations. For example,
stellar aberration (Chapter \ref{ch:2}), Fresnel drag (Chapter \ref{ch:3}),
Sagnac experiment (Chapter \ref{ch:5}) or GPS (Chapter \ref{ch:7}) give
non-measurable differences between models. 

The measured spacecraft anomalies, although small, exhibit a non-random
pattern. Notice that a slight rotation of the orbital plane may produce similar
effects in both Doppler and range as those expected from VRT. In other
words, VRT signature in spacecraft tracking may be hidden by a rotation of
the orbital plane. 

Due to maneuvres or perturbations, the orbital parameters are continuosly 
adjusted to fit the measurements. If SRT is valid and the orbits are adjusted
using SRT, then, the actual trajectory is retrieved, and no anomalies should
be present. But if VRT is valid, the fit of the orbits using SRT will give a ghost 
orbit that differs from the actual one by a small rotation of its plane. 
This rotation may hide part of first order effects, as it was discussed in 
Section \ref{sub:6.1.2}, but it cannot completely remove the first order term. 
Also second order terms remain. The remanent of the first order term seems 
to produce the anual term (Section \ref{sub:6.1.3}), while the flyby 
anomaly is the manifestation of a second order difference (Section \ref{sec:6.2}). 
The comparison of the measured anomalies and the difference between SRT 
and VRT gives a statistically significant fit, with a high goodness of the fit.

In Section \ref{sec:6.3} we have shown that the range disagreement  strongly 
supports the dependence of the speed of light with that of the source.
In this case, two different orbital adjustment would be needed by the 
DSN and the SSN due to their different propagation speed. In consequence, 
it will be impossible to simultaneously eliminate the first order difference, 
as it seems to happen with the range disagreement.

A simultaneous time-of-flight measurement of the speed of 
light from two different sources with different velocities has never been
performed. In Chapter \ref{ch:9} we described an attempt in this direction. 
Although the experiment has detected a statistically significant sinusoidal 
variation of the time lag of the same order of magnitude as that predicted 
by VRT, the experimental error was large enough to prevent reaching a 
definitive answer. In this sense, we recall Fox's words regarding the 
possibility of conducting an experiment
on the propagation of light relative to the motion of the source: 
``{\it Nevertheless if one balances the overwhelming odds against such 
an experiment yielding anything new against the overwhelming importance 
of the point to be tested, he may conclude that the experiment should 
be performed}'' \cite{fox1962experimental}.

Finally, it is worth mentioning that a formulation of electromagnetism
compatible with Faraday's conception is possible, as shown in
Part \ref{ch:8}, which is also compatible with the known electromagnetic 
phenomena. The most remarkable fact of this new formalism is the 
simultaneous presence of instantaneous (static terms) and delayed 
(radiative terms) interactions (i.e., local and nonlocal phenomena in 
the same interaction).

We believe that, given the above evidence, a conscientious
experimental research is needed to settle the question of the dependence of
the speed of light on that of its source as predicted by Vibrating Rays
Theory, and that has been observed during the 1998 NEAR flyby. 
